\documentstyle[12pt,epsf]{article}
\newcommand{\beq}{\begin{equation}}   
\newcommand{\eeq}{\end{equation}}
\newcommand{\ra}{\rightarrow}
\newcommand{\gsim}{\lower.7ex\hbox{$
\;\stackrel{\textstyle>}{\sim}\;$}}
\newcommand{\lsim}{\lower.7ex\hbox{$
\;\stackrel{\textstyle<}{\sim}\;$}}

\newcommand {\th}{\theta}
\newcommand {\bt}{\bar{\theta}}
\newcommand {\la}{\lambda}
\newcommand {\La}{\Lambda}
\newcommand {\bl}{\bar{\lambda}}
\newcommand {\ga}{\gamma}
\newcommand {\dg}{\dot{\gamma}}
\newcommand {\ep}{\epsilon}
\newcommand {\bep}{\bar{\epsilon}}
\newcommand {\al}{\alpha}
\newcommand {\be}{\beta}
\newcommand {\da}{\dot{\alpha}}
\newcommand {\db}{\dot{\beta}}
\newcommand {\eps}{\epsilon}

\begin{document}
\begin{titlepage}
\renewcommand{\thefootnote}{\fnsymbol{footnote}}

\begin{center} \Large
{\bf Theoretical Physics Institute}\\
{\bf University of Minnesota}
\end{center}
\begin{flushright}
TPI-MINN-97/09-T\\
UMN-TH-1535-97\\
hep-th/9704114
\end{flushright}
\vspace{.3cm}
\begin{center}

{\LARGE Non-Perturbative Dynamics in Supersymmetric Gauge 
Theories}

\vspace{1.5cm}

Extended version of lectures given at International School of Physics 
``Enrico 
Fermi",  Varenna, Italy, July 3 - 6, 1995, Institute of Nuclear Science, 
UNAM, 
Mexico, April 11 - 17, 1996, and Summer School in High-Energy 
Physics and 
Cosmology,  10 - 26 July, 1996, ICTP, Triest, Italy.

\end{center}
\vspace*{.3cm}
\begin{center} {\Large 
M. Shifman} \\
\vspace{0.4cm}
{\it  Theoretical Physics Institute, Univ. of Minnesota,
Minneapolis, MN 55455}
\end{center}

\vspace*{.2cm}
\begin{abstract}

I give an introductory review of recent, 
fascinating developments in supersymmetric gauge theories. 
I explain pedagogically the 
miraculous properties of  supersymmetric gauge
dynamics allowing one to obtain exact solutions
in many instances. Various dynamical
regimes emerging in supersymmetric Quantum Chromodynamics
 and its generalizations are discussed.
I emphasize those features that have a chance of survival in
QCD and those which are drastically different 
in supersymmetric and non-supersymmetric gauge theories.

Unlike most of the recent reviews focusing almost entirely on the 
progress in extended supersymmetries (the Seiberg-Witten solution 
of
$N=2$ models), these lectures are mainly devoted to $N=1$ theories.
The primary task is extracting lessons for non-supersymmetric 
theories.

\end{abstract}

\end{titlepage}

\newpage

{\tiny\tableofcontents}

\newpage

\section{Lecture 1. Basic Aspects of Nonperturbative Gauge 
Dynamics}

\renewcommand{\theequation}{1.\arabic{equation}}
\setcounter{equation}{0}

\subsection{Introduction}

All fundamental interactions established in nature are described by 
non-Abelian gauge theories. The standard model of the electroweak
interactions belongs  to this class. In this model,
the coupling constant is weak, and its dynamics is fully controlled 
(with the  possible 
 exception of  a few, rather exotic problems, like  baryon number 
violation at high energies).

Another  important example of the non-Abelian gauge theories is 
Quantum Chromodynamics (QCD). This 
theory has been  under  intense scrutiny for over two decades, yet 
remains  mysterious.  Interaction in QCD 
becomes strong at  large distances. What is even worse, 
 the degrees of freedom appearing in the Lagrangian (microscopic 
variables -- colored quarks and gluons in the case at hand)  are not  
those degrees of freedom that show up as physical asymptotic states 
(macroscopic degrees of freedom -- colorless hadrons). Color is 
permanently confined. What are the dynamical reasons of this 
phenomenon?

Color confinement is believed to take place even in pure 
gluodynamics, i.e. with no dynamical quarks. Adding massless
quarks produces   another surprise. The chiral symmetry of the 
quark sector, present at the Lagrangian level, is spontaneously 
broken (realized nonlinearly) in the physical amplitudes. Massless 
pions are the remnants
of the spontaneously broken chiral symmetry.
What can be said, theoretically, about the pattern of the
spontaneous breaking of the chiral symmetry?

Color confinement and  the
spontaneous breaking of the chiral symmetry
are the two most sacred questions of  strong non-Abelian dynamics;
and the progress of  understanding  them is painfully slow.
At  the end of the 1970s Polyakov showed that in
3-dimensional compact electrodynamics (the so called
Georgi-Glashow model, a  primitive relative of QCD) color 
confinement does indeed take place \cite{polyakov}. 
Approximately at the same time a qualitative picture of how this
phenomenon could actually happen in 4-dimensional QCD was 
suggested by
Mandelstam \cite{mandelstam} and 't Hooft \cite{thooft}.
Some insights, though quite limited, were provided by 
models of the various degree of fundamentality, and by numerical
studies on  lattices. This is, basically, all we had before
1994, when a significant breakthrough was achieved
in understanding  both issues in supersymmetric (SUSY) gauge 
theories.

Unlike the Georgi-Glashow model in three dimensions mentioned 
above, which is 
quite a distant relative of QCD, four-dimensional supersymmetric 
gluodynamics
and supersymmetric gauge theories with matter come much closer
to genuine QCD. Moreover, the dynamics of these theories is rich and 
interesting by itself, which accounts for  the attention they have 
attracted
  in the last two or three years. Although
the development is not yet complete, the lessons are promising,
and definitely deserve thorough studies.
Several topics which I consider to be most interesting are discussed 
below in 
this lecture course. Before submerging into supersymmetry proper, 
however,
it is worth reiterating the main  general ideas which are 
the key players
in this range of questions: the Meissner and the dual Meissner 
effects,
monopoles, Abelian projection of QCD, and so on. The first part
is a brief review of these issues intended mostly to refresh 
the memory
and to provide a representative list of pedagogical literature.
We will start an 
excursion into supersymmetric gauge theories in  Sect. 2, and 
gradually proceed from simpler topics to 
more 
complicated ones. The simplest supersymmetric non-Abelian model 
is SUSY 
gluodynamics. Simultaneously, it happens to be the closest 
approximation to
QCD (without light quarks). Although there was essentially no 
progress 
towards
the solution of this theory the seeds of the miraculous properties of 
the
supersymmetric gauge dynamics are clearly visible. I will explain 
how some 
exact results (the first example ever in four-dimensional strongly 
coupled field 
theory!) can be derived. These results will become a part of our tool 
kit used 
in revealing various dynamical scenarios in SUSY gauge theories with 
matter.

In Sect. 3 we 
will open a fascinating world of supersymmetric $SU(N_c)$ QCD, a 
world
populated by a variety of unusual regimes governed by
nonperturbative 
supersymmetric dynamics. Here, among other rarities,  
 we will find confinement without spontaneous chiral symmetry 
breaking, with spontaneous breaking of the baryon 
number; the so called $s$-confinement, with additional composite 
massless fields not related to  Goldstone modes of the spontaneously 
broken global symmetries. We will discover a conformal window --
a set of pairs of theories with different gauge groups but
identical global symmetries that are dual (equivalent) to each other 
as far as  infrared behavior is concerned. The infrared 
asymptotics of these theories is (super)conformal. Outside the 
conformal window we will encounter dual pairs, with one theory 
coupled superstrongly and another free in the infrared domain. 
The gauge bosons of the latter can be considered as composite 
superstrongly bound states in the former theory. 

Section 4 is a very brief travel guide to the supersymmetric 
gauge theories with other gauge groups. New phenomena we will 
encounter are the so called oblique confinement and triality --
the infrared equivalence of three distinct theories. One of them is in 
the Higgs
phase, another in the confinement phase, and the third one
is in the oblique confinement phase. 

Finally, in Sect. 5, supersymmetry is explicitly (softly) broken by the 
gluino and squark masses.   
Ideally, we would like to send these masses to infinity,
evolving towards non-supersymmetric theories, without loosing our 
calculational abilities. Unfortunately, once  the gluino and squark 
masses become large enough, the calculational abilities are lost.
We have to settle for 
small perturbations of the supersymmetric solution. 
By exploring the dynamical properties of the
theory obtained in this way one hopes to get qualitative insights 
about what happens in the limit of infinitely heavy squarks and 
gluino.

This review is an extended version of my lecture notes.
 The pedagogical style of presentation is preserved, where possible.
Simple and general issues are discussed
first, providing a necessary background
for more advanced theoretical constructions and
conclusions. Occasional remarks intended for expert readers will
slip, though; they may be ignored in the first reading. 

\subsection{Phases of  gauge theories (Abelian version)}

Quantum electrodynamics (QED) was historically the first gauge 
theory studied in detail. Although from the modern perspective it 
seems to be  a very simple model, with no mysteries, it can exhibit  
at least three different types of behavior.
Let us consider supersymmetric version, SQED. The Lagrangian of the 
model
is \cite{GL}
$$
{\cal L} = -\frac{1}{4} F_{\mu\nu}^2 + \bar\psi i\not\!\! D\psi - 
m\bar\psi\psi + 
$$
$$
(D_\mu \phi )^\dagger (D_\mu \phi ) - m^2\phi^\dagger\phi
+ (D_\mu \chi )^\dagger (D_\mu \chi ) - m^2\chi^\dagger\chi -
$$
$$
\frac{e^2}{2} (\phi^\dagger\phi - \chi^\dagger\chi )^2+
$$
\beq
\frac{i}{2}\bar\lambda \not\! \partial\lambda
+ e \left\{ \bar\lambda \frac{1+\gamma^5}{2}\psi \phi^\dagger +
\bar\lambda \frac{1-\gamma^5}{2}\psi \chi +\mbox{h.c.}\right\}\, ,
\label{SQEDL}
\eeq
where $\phi$ and $\chi$ are complex scalar fields with the charges
$+e$ and $-e$, respectively, ({\em selectrons}), and $\lambda$
is the {\em photino} (Majorana) field. The first line in Eq. 
(\ref{SQEDL})
is just conventional quantum electrodynamics of photons and 
electrons, the second line gives the kinetic and the mass terms
of electron's superpartners, selectrons, the third line is the selectron 
self-interaction, and, finally, the fourth line presents the photino 
kinetic term and photino's interactions. As we will see later, 
supersymmetry guarantees that the overall form of the Lagrangian is
preserved under quantum corrections -- no new counterterms 
appear.

Assume first that the electron and selectron mass (they are the 
same)
does not vanish, and $\alpha \equiv e^2/4\pi \ll 1$. For $m\neq 0$
the vacuum state of the theory is unique; it corresponds to
the vanishing expectation values of the scalar fields, 
$\langle \phi \rangle = \langle \chi \rangle = 0$. Apart from the fact 
that some of the charged particles have spin zero, the theory is very 
much like QED. If heavy (static) probe charges are introduced,
their interaction is just Coulomb, with the potential
proportional to 
$$
V(R) \sim \frac{\alpha (R)}{R}
$$
where $R$ is the distance between the probe charges.
Classically $\alpha$ is constant, of course, but  quantum 
renormalization makes $\alpha$ run. The behavior of $\alpha$ is 
determined by the well-known Landau formula. At large distances 
$\alpha$ decreases logarithmically; if $m$ is finite, however, the 
logarithmic
fall off is frozen at $R\sim m^{-1}$; the corresponding critical value of 
$\alpha$ is $\alpha_* = \alpha (R=m^{-1})$. The potential between 
the
distant 
static charges is
\beq
V(R) \sim \frac{\alpha_*}{R}\, , \,\,\,\, R\ra \infty\, .
\label{CLRP}
\eeq
The dynamical regime with this type of the long-distance behavior
is referred to as the {\em Coulomb phase}. In the case at hand we 
deal 
with the Abelian Coulomb phase. Similar behavior,
Eq. (\ref{CLRP}),  can take place in 
the non-Abelian gauge theories as well.
Non-Abelian gauge theories with the long-range potential
(\ref{CLRP}) are said to be in the non-Abelian Coulomb phase.

What happens if the mass parameter $m$ is set equal to zero?
Note, that in SUSY theories, if the bare value of $m$ is fine-tuned to 
vanish, it will remain at zero with all quantum corrections.

The most drastic change is evident at first glance, after examining 
the
third line in Eq. (\ref{SQEDL}). The minimum of the potential energy
is achieved now not only at the vanishing values of the
scalar fields but, rather, on a one-dimensional complex manifold.
Indeed, $\phi$ can be arbitrary complex number; if $\phi
=\chi$ the potential energy obviously vanishes. The continuous 
degeneracy of the classical minimal-energy state (the so called 
{\em vacuum valley}) is a rather typical feature of supersymmetric 
theories with matter. We will return to the in-depth discussion
of this aspect later.
 
Quantum-mechanically one can say that the expectation value 
$\langle\phi \chi\rangle\neq 0$ may take  arbitrary complex value; 
here $\phi \chi$ is a convenient 
gauge invariant product parametrizing the vacuum valley.
If   the scalar fields $\phi$ and $ \chi$ take constant non-vanishing 
values in the vacuum, the Higgs phenomenon takes place
\cite{Higgs}: the gauge symmetry ($U(1)$ in the case at hand) is 
spontaneously broken.

What does one mean by saying that the gauge symmetry is 
spontaneously broken? The gauge symmetry, in a sense,
is not a symmetry at all -- rather, it is a description of $x$ physical
degrees of freedom in terms of $x+y$ variables; $y$ variables are 
redundant; the corresponding degrees of freedom are physically 
unobservable. 
In other words, only a subspace of all field space
($\psi , \bar\psi, \phi, \phi^\dagger ,  \chi, \chi^\dagger , A_\mu$
in the model under consideration), corresponding to gauge 
non-equivalent points,
describe physically observable degrees of freedom.

Let us first switch off the electric charge, $e=0$.
Then the  Lagrangian (\ref{SQEDL}) is invariant under the global 
phase 
rotations, $\phi \ra e^{i\alpha}\phi $, $\chi \ra e^{-
i\alpha}\chi$, $\psi \ra e^{i\alpha}\psi $. The condensation of 
the scalar fields breaks this 
invariance. But invariance of the model is not lost.
Under the phase transformation one vacuum goes into another, 
physically equivalent. Say, if we start from the  vacuum 
characterized
by a real value of the order parameter $\phi$ and $\chi$, in the 
``rotated" one the order
parameter  is complex. The spontaneous breaking of any global 
symmetry leads to a set of degenerate (and physically equivalent) 
vacua.

If we now switch on the photons, ($e\neq 0$), the degeneracy 
associated with the spontaneous breaking of the global symmetry is 
gone. All states related by the phase rotation are gauge-equivalent, 
and only one of them should be left in the Hilbert space of the 
theory. In other words, one
can always 
{\em choose} the vacuum values of $\phi$ and $\chi$ 
to be real. This is nothing but the gauge condition. Thus,
spontaneous breaking of the gauge symmetry does not imply, 
generally speaking, the existence of a degenerate set of vacua,
as is the case with the global symmetries. Then, what does 
it mean, after all?

By inspecting Lagrangian (\ref{SQEDL}) it is not difficult to see
that if $\phi$ and $\chi$ have non-vanishing (and constant) values 
in the vacuum, the spectrum of the theory does not contain massless 
vector particles at all. The photon acquires mass, $M_V^2
=4e^2 |v|^2$, where $v^2 = \langle \phi\chi\rangle$, through the 
mixing
with the ``phases" of the fields $\phi$ and $\chi$. In the 
supersymmetric model considered, we ``cook" in this way a massive 
vector field 
and a massive (real) scalar field, both with masses $M_V$,
and a massless complex scalar field, out of massless photon
and two massless complex scalar fields. (All these boson fields are 
accompanied by their fermion superpartners, of course).

This regime is referred to as the {\em Higgs phase}. One
massless  scalar field is eaten up by the photon field in the 
process of the transition to the Higgs phase. In the Higgs phase the 
electric charge
is screened by the vacuum condensates. If we put a probe (static) 
electric charge in the vacuum, the Coulomb potential $\sim 1/R$ it 
induces 
at short distances (i.e. distances less than $M_V^{-1}$) gives place to 
the 
Yukawa potential $\sim \exp (-M_VR)/R$ at  distances larger
than $M_V^{-1}$. The gauge coupling runs,
according the standard Landau formula, 
 only at distances shorter than $M_V^{-1}$, and is frozen at
$M_V^{-1}$.

There is one single point in the vacuum valley, the origin,
(i.e. $\langle \phi\chi\rangle =0$)
where   the gauge symmetry is
unbroken. The long-range force due to massless photons
is not screened by the vacuum condensates of the scalar fields. 
A different type of screening does  occur, however, due to quantum 
effects.
Indeed, the photon propagator is dressed by the virtual pairs of 
electrons and selectrons. This dressing results in the running of the 
effective charge $\alpha (R)$,
\beq
\alpha (R) \sim \frac{1}{\ln R}\, .
\label{massless}
\eeq
Unlike the massive case, where this running is frozen
at $R=m^{-1}$, in the theory with $m=0$ (and $\langle 
\phi\chi\rangle =0$) the logarithmic fall off (\ref{massless}) 
continues 
indefinitely:
at asymptotically large $R$ the effective coupling becomes 
asymptotically small. 

Thus, the  asymptotic limit of massless QED is a  free photon
(and photino) plus massless matter fields whose charge
is completely screened. The theory does not have localized 
asymptotic states and no mass shell at all, no $S$ matrix in the usual 
sense of this word.
Still, it is well-defined in finite volume. 

 This phase of the theory 
is referred to as a {\em free phase}. Sometimes it is also called {\em 
the
Landau zero-charge phase}. Strictly speaking, the model is 
ill-defined
at short distances where the effective coupling grows and finally hits 
the Landau pole. To make it self-consistent, at short distances
it must be embedded into an asymptotically free theory. This is not 
difficult to achieve. The Georgi-Glashow model gives an example of 
such an embedding.

Summarizing, even in the  simplest Abelian example we encounter 
three
different phases, or dynamical regimes: the Coulomb phase,
the Higgs phase and the free (Landau) phase, depending on the 
values
of parameters of the model and the choice of the vacuum state
(in the case of the vanishing mass parameter, $m=0$). All these 
regimes are attainable in non-Abelian 
models too. The non-Abelian gauge theories are richer, however,
since they admit one more dynamical regime,
{\em confinement of color}, a famous property of QCD
which attracted so much attention but still 
defies  analytic solution. 

\subsection{Non-Abelian Higgs model; monopoles}

Interactions of quarks and leptons at the fundamental level are 
described by non-Abelian gauge theories. The standard model
of electroweak interactions is, probably, the most well-studied
non-Abelian gauge theory in the Higgs phase. Apart from a few 
exotic phenomena (e.g. the baryon number violation at high
energies, for reviews see \cite{Mattis}) all processes in this
model occur in the weak coupling regime, and are well-understood.
The dynamical content is almost exhausted by perturbation theory;
very small nonperturbative corrections are due to instantons.

Since this model is so well-studied and familiar to everybody, it 
makes more sense to discuss another example of a non-Abelian Higgs 
phenomenon, the Georgi-Glashow model \cite{GeG}. This example is 
instructive and more relevant to the discussion below since this 
model exhibits magnetic monopoles \cite{THP}. A nice review
of the Georgi-Glashow model, with special emphasis on this particular 
aspect, magnetic monopoles and dyons, can be found in Ref.
\cite{RAJ}. 

The gauge group of the model is $SU(2)$, so it has three gauge 
bosons. The matter sector includes one real scalar field 
$\phi^a$
in the adjoint representation of $SU(2)$ (i.e. $a=1,2,3$). The 
Lagrangian has the form
\beq
{\cal L} = -\frac{1}{4}G_{\mu\nu}^a G_{\mu\nu}^a
+ \frac{1}{2} (D_\mu\phi^a) (D_\mu\phi^a)
-\frac{1}{4}\lambda (\phi^a\phi^a -v^2)^2
\label{GGL}
\eeq
where $\lambda$ is a scalar coupling constant, and $v$ is a constant 
of dimension of mass determining the vacuum expectation value
of the $\phi$ field. The so-called Bogomol'nyi-Prasad-Sommerfield 
(BPS) limit \cite{Bog,PrS}
\beq
\lambda \ra 0\, , \,\,\, v \,\, \mbox{fixed}\, ,
\label{BPS}
\eeq
is most relevant for our purposes. In this limit the scalar 
self-interaction disappears from the Lagrangian and the equations of 
motion.
The only remnant of the  scalar 
self-interaction term is the boundary condition for the field $\phi$
at spatial infinity. Indeed, requiring the energy of field 
configurations  to be finite we single out only those for which
$$
\phi^a\phi^a \ra v^2 \,\,\, \mbox{at} \,\,\, |\vec x | \ra\infty\, .
$$
The BPS limit naturally emerges in many supersymmetric theories.

If $v\neq 0$ the minimum of the classical energy is achieved
for $\phi^a\phi^a = v^2$. In the weak coupling regime,
when the gauge coupling constant $g$ is small, the quantum  vacuum 
of the 
model is characterized by a nonvanishing expectation value of 
$\phi^2$. The vacuum field can always be chosen as follows
\beq
\phi^3 = v\, , \,\,\, \phi^{1,2} = 0\, .
\label{GGcond}
\eeq
It is not difficult to check that the gauge fields with the color indices
1,2 propagating in the condensate (\ref{GGcond}) acquire masses
$M_V = gv$, and become $W$ bosons, while the gauge field
with the color index 3 remains massless. The gauge transformations 
corresponding to rotations around the third axis leave the condensate 
(\ref{GGcond}) intact.
Correspondingly, the gauge group $SU(2)$ is spontaneously broken
down to $U(1)$; $A^3$ plays the role of the photon of the 
$U(1)$ gauge theory. The particle spectrum of the theory, apart from 
this ``photon", consists of one neutral
massless scalar $\phi^3$ (neutral  with respect to the $U(1)$ group), 
and two massive vector particles,
$W^\pm = (2)^{-1/2} (A^1\pm iA^2)$, with the charges $\pm 1$, 
a rather typical pattern of the non-Abelian Higgs phenomenon.
The 1,2 components of the $\phi$ field are eaten up:
they became the longitudinal components of $W$'s. 

Since the residual gauge symmetry is $U(1)$,
while at high energies (when the spontaneous symmetry breaking is 
inessential) we deal with the full original $SU(2)$, which is a compact 
group, the low-energy electrodynamics obtained after the 
spontaneous breaking of $SU(2)$ down to $U(1)$ is actually compact.
Topological arguments then prompt us \cite{THP,RAJ}
that this model has topologically stable
localized finite-energy configurations with a non-vanishing magnetic 
charge, magnetic monopoles. I cannot go into details regarding
these objects, referring the interested reader to a vast literature 
devoted to the subject of the 't Hooft-Polyakov monopoles
(recent review papers \cite{Harv} contain a representative list
of references). Here I will only sketch how the monopole mass can be 
calculated using a limited information coded in the asymptotics of 
the corresponding fields.

The Lagrangian (\ref{GGL}) in the BPS limit implies that the energy 
$E$ of any static field configuration (in the $A_0=0$ gauge) can be 
written as
$$
E= \int d^3 x \left( \frac{1}{4} G_{ij}^a G_{ij}^a +\frac{1}{2}
D_k\phi^a D_k\phi^a\right) =
$$
\beq
\int d^3 x\frac{1}{4}\left( G_{ij}^a-\epsilon_{ijk}D_k\phi^a \right)^2 +
\int d^3 x\frac{1}{2}\epsilon_{ijk} G_{ij}^a D_k\phi^a \, .
\label{B1}
\eeq
Now, the second term is actually an integral over a full derivative; it 
reduces identically to a two-dimensional integral over the large 
sphere,
\beq
\int d^3 x\frac{1}{2}\epsilon_{ijk} G_{ij}^a D_k\phi^a
= \int d^3 x\partial_k \left( \frac{1}{2}\epsilon_{ijk} G_{ij}^a \phi^a
\right)
=\int_{S_2} d\sigma_k \left( \frac{1}{2}\epsilon_{ijk} G_{ij}^a \phi^a
\right)
\, ,
\label{B2}
\eeq
where $ d\sigma_k$ is the area element. In deriving Eq. (\ref{B2})
it was taken into account that
$$
D_\mu \tilde G_{\mu\nu} =0\, , \,\,\,
\tilde G_{\mu\nu} = 
(1/2)\epsilon_{\mu\nu\alpha\beta}G_{\alpha\beta}\, .
$$
It is not difficult to show that the surface integral in
Eq. (\ref{B2}) is nothing but the flux of the magnetic field through 
the large sphere,  proportional to the 
topological charge.
In this way we arrive at the following expression for the 
energy
\beq
E = \frac{4\pi \kappa  v}{g} +\int d^3 x\frac{1}{4}\left( G_{ij}^a-
\epsilon_{ijk}D_k\phi^a \right)^2
\label{B3}
\eeq
where $\kappa$ is the  topological  charge of the configuration 
considered (see below),
related to the magnetic charge $m$,
$$
m=\frac{4\pi\kappa }{g}\, .
$$
The second term is obviously positive-definite.
Thus, in the sector with the given $\kappa$
\beq
E \geq \frac{4\pi \kappa v}{g}\, ;
\label{B4}
\eeq
the equality is achieved if and only if
\beq
 G_{ij}^a =\epsilon_{ijk}D_k\phi^a \, .
\label{B5}
\eeq
Equation (\ref{B5}) is called the Bogomol'nyi condition. All states
satisfying this condition are called the BPS-saturated states;
the 't Hooft-Polyakov monopole belongs to this class.
The mass of the monopole is, thus, unambiguously related to its 
magnetic charge,
$M_{\rm mon} =  |m| v $. 

It remains to be added that a topologically non-trivial
solution of the Bogomol'nyi condition corresponding to
$\kappa=1$ (the one-monopole solution) has a ``hedgehog" form,
\beq
\phi^a (\vec x) = \frac{x^a}{r } F(r)\, ,
\,\,\, 
A_i^a (\vec x) = \epsilon_{aij}\, \frac{x^j}{ r } \, W(r)\, ,
\,\,\,  A_0^a (\vec x) = 0\, ,
\label{B6}
\eeq
where 
$$ 
r= |\vec x|
$$
and 
the asymptotics of the functions $F$ and $W$ are
as follows:
\beq
F\ra v\, , \,\,\, W \ra \frac{1}{g r }\,\,\,
\mbox{at}\,\,\, r \ra\infty\, , \,\,\, F,W\ra 0\,\,\, 
\mbox{at}\,\,\, \vec x \ra 0\, .
\label{B7}
\eeq
Substituting the {\em ansatz} (\ref{B6}) in the Bogomol'nyi condition
we check that it goes through, and get two coupled differential 
equations
for the invariant functions $F,W$.
The solutions of these equations satisfying the boundary conditions
(\ref{B7}) are \cite{THP}
\beq
F = \frac{1}{gr}\left( \frac{ gr v}{{\rm tanh} (gr v)} - 1\right)\, 
,\,\,\, 
W = -\frac{1}{ gr }\left( \frac{ gr v}{{\rm sinh} (gr v)} - 
1\right)\, .
\label{B8}
\eeq
The gauge-invariant definition of the electromagnetic field tensor is
\beq
F_{\mu\nu} = \hat\phi^aG_{\mu\nu}^a -\frac{1}{g}
\epsilon^{abc} \hat\phi^a D_\mu \hat\phi^b D_\nu \hat\phi^c\, , 
\,\,\,\,
\hat\phi^a =\phi^a/\sqrt{ \phi^b\phi^b }\, .
\label{B9}
\eeq  
The topological current, whose conservation is obvious, has the form
\beq
K_\mu =\frac{1}{8\pi}\epsilon^{\mu\nu\alpha\beta}\epsilon^{abc}
\partial_\nu  \hat\phi^a \partial_\alpha  \hat\phi^b \partial_\beta  
\hat\phi^c\, .
\eeq
The corresponding charge, $\kappa = \int d^3 x K_0$, counts the 
windings
of the mapping of the two-dimensional large sphere in the 
configurational space
onto $S_2$, the group space of $SU(2)$. 
The reader is invited to check this statement, using the definition
of the topological current.

Using Eq. (\ref{B9}) and 
 the solution (\ref{B8}) it is not difficult to see that
\beq
\frac{1}{2}\epsilon_{\mu\nu\alpha\beta}\partial^\nu 
F^{\alpha\beta} =\frac{4\pi}{g} K_\mu\, ,
\eeq
and, hence, 
the magnetic charge of the 
't Hooft-Polyakov monopole
$$
m \equiv \int_{S_2} d\sigma_i B_i
$$
is indeed equal to $4\pi /g$.  The magnetic charge quantization 
condition is, thus, 
\beq
gm = 4\pi \, .
\label{mchqu}
\eeq
The magnetic charge quantum seemingly is twice larger than for 
the Dirac monopole \cite{DirM}. (Note, however, that Eq. 
(\ref{mchqu}) coincides with the Schwinger quantization condition
\cite{SchM}.) This is due to the fact that the electric charge of the 
$W$ bosons, $g$, is not the minimal one, in principle.
It is conceivable that the matter fields in the fundamental (doublet) 
representation are added in the Georgi-Glashow model.
Then, their charge with respect to the $U(1)$ is $g/2$,
and the product of the minimal electric charge and the
magnetic charge of the monopole is $2\pi$, as required
by Dirac's argument. 

Shortly after the discovery of the monopoles in the Georgi-Glashow 
model it was pointed out
\cite{JZ} that the same model also has {\em dyon} solutions --
localized field configurations carrying both, the magnetic and electric 
charges.  The monopole solution (\ref{B6}), (\ref{B8})
carries no electric charge, since the fields are time independent and 
$A_0=0$; hence, $E_i = -F_{0i} \equiv 0$. One can modify the 
hedgehog {\em ansatz} (\ref{B6}), keeping its static nature but 
allowing for $A_0\neq 0$. In this way one can obtain \cite{JZ} in the 
BPS limit
an analytic solution in which both integrals
$$
\frac{1}{4\pi} \int_{S_2} d\sigma_i B_i
\,\,\, \mbox{and} \,\,\, \frac{1}{4\pi} \int_{S_2} d\sigma_i E_i
$$
are non-vanishing. These objects, dyons, also supposedly play a role
in some mechanisms ensuring color confinement. 

\subsection{Phases of gauge theories (non-Abelian version)}

The Georgi-Glashow model discussed above teaches us that
in the non-Abelian case the microscopic variables in the Lagrangian
(gauge bosons, adjoint matter fields) do not necessarily coincide with 
those quanta we can observe (massive $W$ bosons, magnetic 
monopoles). In the weak coupling regime the relation between
the microscopic and macroscopic degrees of freedom is pretty
transparent, though.
Theoretical situation in QCD  is far from being so cloudless. 
The QCD Lagrangian is well established. Thus, we know that
at short distances the microscopic variables are colored quarks and 
gluons.
The macroscopic degrees of freedom are strongly bound states
whose analysis cannot be carried out perturbatively or in the 
semiclassical approximation. 
Empirically we know
that asymptotic states are colorless hadrons. Thus, if the quarks have 
fractional charges ($2/3$ and $-1/3$), all observable asymptotic 
states have integer 
charges
and are built from quark-antiquark pairs (mesons) or
three quarks (baryons). Here we encounter for the first time in our 
brief excursion a new phase of the gauge 
theory, the {\em confining 
phase}.

Consider pure gluodynamics, i.e. the theory of gluons, with no 
dynamical quarks. 
The Lagrangian has the familiar form
\beq
{\cal L} = -\frac{1}{4g_0^2} G_{\mu\nu}^a G_{\mu\nu}^a\, ,
\eeq
where $g_0^2$ is the gauge coupling at the ultraviolet cut-off.
Although this coupling is dimensionless, actually the true
parameter characterizing interactions in the theory is the scale
$\Lambda$ related to  $g_0^2$ as follows
\beq
\Lambda = M_0 \exp \left[ \left(-\frac{8\pi^2}{\beta_0 g_0^2}\right)
\right] \left( \frac{8\pi^2}{ g_0^2} \right)^{\beta_1/\beta_0^2}
\eeq
where $\beta_0$ and $\beta_1$ are the first and the second 
coefficients in the Gell-Mann-Low 
function ($\beta_0=11$ in QCD). 
Unlike the Higgs-phase standard model, QCD is strongly coupled
at momenta of order of $\Lambda$;
all interesting dynamical features of this theory reflect
the strong-coupling dynamics. The intricacies of this dynamics are 
such that
the microscopic degrees of freedom -- gluons -- must disappear at 
distances larger than $\Lambda^{-1}$, giving place to macroscopic 
degrees of freedom, hadrons. A qualitative picture of 
how and why  this process might take place is believed to be known. 

If we place  two heavy (static) color charges at a large distance from 
each other they create a chromoelectric field
which is believed to form a flux tube between the charges. This flux 
tube of the confining non-Abelian theory substitutes the dispersed 
Coulomb field one observes between the static charges in 
electrodynamics (in the Coulomb phase). 
The flux tube is a string-like one-dimensional object,
with the cross section $\sim \Lambda^{-2}$, and constant string 
tension $\sigma \sim \Lambda^{2}$. Then the interaction energy of 
two
static charges grows linearly with the distance between them, 
$V(R) =\sigma R$, and they can never be separated asymptotically, 
since this separation costs infinite energy. The formal signature of 
this regime is the area law for the Wilson loop.

Do we have any precedents of such a behavior -- constant force, 
linearly rising potential -- in the dynamical systems studied 
previously?

There exists one example known for a long time. Let us return to 
supersymmetric electrodynamics, Eq. (\ref{SQEDL}). The gauge 
symmetry is $U(1)$, and if $\phi \neq 0$ the theory is in the Higgs 
phase. A non-relativistic analog of this theory is nothing but the 
Ginzburg-Landau model of superconductivity, describing the 
Bose-condensation of the Cooper electron pairs in the vacuum state.
An electrically charged order parameter develops a non-vanishing  
expectation value. 
The vector quanta acquire mass, and the electric potential becomes
short-range. The magnetic fields are repelled completely from the 
domain where the condensate develops, the famous Meissner effect.
Assume, however, that two static magnetic charges 
(magnetic monopoles) are placed by hand inside this domain.
Since the magnetic flux is conserved, the magnetic field cannot 
vanish everywhere. A strong  repulsion it experiences in the vacuum 
medium results in formation of narrow flux tubes connecting
the magnetic charges. The flux tubes are solutions of the classical 
equations of motion corresponding
to the overall $2\pi$ change of the phase of the field $\phi$
when one makes a full rotation around a line connecting the 
magnetic charges. To avoid singularity the value of the $\phi$ field
in the center of the tube must vanish. These solutions 
in the Ginzburg-Landau theory were found by Abrikosov
(Abrikosov vortices) \footnote{A remark for more educated readers: 
mathematically, the existence of the topologically stable vortices is 
due to the fact that $\pi_1 (U(1)) = {\bf Z}$.}. It is not difficult to 
calculate the energy of the vortex per unit length -- far away from 
the sources it is constant.
In other words, the energy between two magnetic charges in the 
superconducting medium grows linearly with the separation between
the charges. This is exactly what we need for color confinement in 
QCD. 

Turning to QCD we immediately notice two important differences:
one conceptual and one technical. The first difference is that in QCD 
we want chromoelectric, not chromomagnetic flux tubes to form.
The vacuum medium must repel chromoelectric fields. This can only 
be achieved by condensation of the magnetic charges, rather than the 
electric ones, as in the Meissner effect. Thus, if a mechanism of this 
type ensures  color confinement, it must be a {\em dual Meissner 
effect}. The problem is that in QCD
the classical 't Hooft-Polyakov 
monopoles do not exist as physical objects (particles).

Second, if in the Ginzburg-Landau model we can work in the weak 
coupling regime, where semi-classical methods are perfectly 
applicable, QCD is a genuinely strongly coupled theory, and we do not 
expect any semi-classical approach
to be valid except, perhaps, in qualitative pictures intended for 
orientation. 

A possible solution of the first problem was indicated by 't Hooft 
\cite{thooft}. Even though QCD, unlike the Georgi-Glashow model,
does not have magnetic monopoles as physical objects, whose 
existence is a gauge-independent fact, it still may have solutions
that {\em in a certain gauge} look like monopoles. The presence of
the appropriate field configurations, thus, will depend on the choice 
of the gauge. Nevertheless, one may hope, that being found in some 
gauge, they may turn out to be important  for implementing  the
dual Meissner effect in QCD. 

The second problem -- the strong coupling regime in QCD -- cannot 
be eliminated in this way, of course. Therefore, to built a fully 
controllable theoretical description, one must try to implement the 
't Hooft-Mandelstam idea, the dual Meissner effect leading to color 
confinement, 
 beyond the semiclassical approximation. How one could do this in 
QCD, and whether it is possible at all, is still unclear. At the same 
time, a remarkable progress was 
achieved in  non-Abelian supersymmetric gauge theories, where in 
certain
instances something similar to the dual Meissner effect can be 
rigorously proven \cite{SEIWIT}. 

We will return the 't Hooft suggestion of the QCD ``monopole" 
condensation shortly,
and now continue our general discussion of the confining phase.
It was already mentioned that in pure gluodynamics (QCD with no 
quarks) the area law is believed to take place for the Wilson loop.
Clearly, we cannot make experiments in pure gluodynamics,
but numerous numerical simulations on the  lattices seem to reveal 
this type of behavior (within usual uncertainties  and other natural 
limitations -- finite volume, {\em etc.} -- inherent to any numerical 
analysis).  There is an invariant clear-cut distinction between the
confining and the Higgs phases in the case when all fields appearing 
in the Lagrangian belong to the adjoint representation. One can 
consider the Wilson loop 
$$
W(C) = \mbox{Tr}\, \exp\oint (igA_\mu dx_\mu )\, , \,\,\, A_\mu =
A_\mu^a T^a
$$
in the fundamental representation (i.e. the generator matrices $T^a$
refer to the fundamental representation; for $SU(3)$, for instance, 
they are $\lambda^a/2$ where $\lambda^a$ are the Gell-Mann 
matrices). In the confining phase, for large contours $C$
the Wilson loop $W(C) \sim \exp (-\sigma\cdot\mbox{(area)})$.
The area law reflects the formation of the flux tube of the 
chromoelectric field attached to the probe fundamental (very) heavy 
quarks. The color charge cannot be screened, and the flux tube can 
not end. It either starts and begins at the color charges or forms 
closed contours.

In the Higgs phase the color field originating at the point of the color 
charge is exponentially screened. No long chromoelectric  flux tubes 
exist; the potential between two separated probe charges saturates
at some constant value. Correspondingly, 
 the
Wilson loop for large contours behaves as
$W(C) \sim \exp (-\Lambda\cdot\mbox{(perimeter)})$.

If we will treat $v$ in Eq. (\ref{GGL}) as a free parameter, at large 
$v$ we are in the Higgs phase, with the perimeter law for the Wilson 
loop, while at $v\ra 0$ we are presumably in the confining phase,
with the area law. At some critical value of $v$,
$$
v_*\sim \Lambda\, ,
$$ 
a phase transition from the Higgs to confinement phase must take 
place. 

Now, if we introduce, additionally, some dynamical fields in the 
fundamental representation, say, quarks, the Wilson loop no more 
differentiates between the two regimes. Indeed, the field of the static
probe quarks can be screened now by dynamical (anti)quarks.
The potential at large distances saturates at a constant value,
and the perimeter law always takes place. 

As a matter of fact, if the Higgs field itself is in the fundamental 
representation of the color group, there is no distinction at all 
between the confinement phase and the Higgs phase. As the
vacuum expectation value (VEV) of the
Higgs field continuously changes from large values to smaller ones,
we continuously flow from the weak coupling regime to the strong 
coupling one.  The spectrum of all physical states, and all other 
measurable quantities, change smoothly \cite{BRFS}. One can argue 
that
that's the case in many different ways. Perhaps, the most 
straightforward line of reasoning is as follows.
Using the Higgs field in the fundamental representation one can built
gauge invariant interpolating operators for {\em all} possible 
physical states. By physical states I mean a part of the Hilbert space, 
with the gauge equivalent points eliminated. The  
K\"{a}ll\'{e}n-Lehmann spectral function corresponding to these 
operators, which carries complete information on the spectrum,
obviously depends smoothly on $\langle \chi^*\chi \rangle$.
When the latter parameter is large the Higgs description is more 
convenient, when it is small it is more convenient to think in terms
of the bound states. There is no boundary, however. We deal
with a single Higgs/confining phase \cite{BRFS}. 

To elucidate this point in more detail
let us consider a specific model. Namely, we will introduce in the 
Lagrangian (\ref{GGL}), in addition to the adjoint Higgs $\phi$,
a complex doublet field $\chi^i$, $=1,2$, with the vacuum expectation 
value $\eta$. We will assume that VEV of the field $\phi$ vanishes, 
and will study the $\eta$ dependence of the physical quantities.

This model has a global $SU(2) $ symmetry, associated with the 
possibility of rotating the doublet $\chi^i$ into the
conjugated doublet $\epsilon^{ij}\chi_i^\dagger$. The $SU(2)$ 
symmetry of the $\chi$ sector becomes explicit if we introduce a 
matrix field
\beq
X = \left( \begin{array}{cc}
\chi^1 & -\chi^{2\dagger}  \\
\chi^2 & \chi^{1\dagger} 
\end{array} \right)\,  ,
\label{eks}
\eeq
and rewrite the Lagrangian in terms of this matrix,
\beq
\Delta{\cal L}_\chi = \frac{1}{2} \,
\mbox{Tr}\, D_\mu X^\dagger D_\mu X - \frac{\tilde\lambda}{4}
\left( \frac{1}{2} \,
\mbox{Tr}\, X^\dagger X -\eta^2\right)^2\, .
\label{lagchi}
\eeq
All physical states form representations of the global $SU(2)$.
Consider, for instance, $SU(2)$ triplets produced from the vacuum by 
the operators
\beq
W_\mu^a = -\frac{i}{2} \, \mbox{Tr}\, (X^\dagger 
\stackrel{\leftrightarrow}{D}_\mu X\sigma^a )\, , \,\,\,
a= 1,2,3 \,  .
\label{wbos}
\eeq
The lowest-lying states produced by these operators
in the weak coupling regime (i.e. when
$\langle \chi^\dagger\chi\rangle \gg \Lambda^2$) coincide
with the conventional $W$ bosons of the Higgs picture, up to a 
normalization constant. The mass of the $W$ bosons is 
$\sim g\eta$. On the other hand, if 
$\langle \chi^\dagger\chi\rangle \lsim \Lambda^2$
it is more appropriate to think of the bound states of the $\chi$ 
quanta forming vector mesons,  triplet with respect to the global 
$SU(2)$ (``$\rho$ mesons").
Their mass is $\sim\Lambda$. Continuous evolution of $\eta$ results 
in the continuous evolution of the mass of the corresponding states.
It is easy to check that the complete set of the gauge invariant
operators one can build in this model spans the whole Hilbert space
of the physical states \footnote{The absence of the phase transition
and the existence of a unified Higgs/confinement phase
in the case when the Higgs field is in the fundamental representation 
blocks any attempts of modeling ``slightly unconfined" quarks
by using a mechanism of De Rujula {\em et al.} \cite{DRuj}.
This mechanism simply does not exist.}.

I hasten to add that introducing  matter fields in the fundamental 
representation we do not necessarily kill all phase transitions. For 
example in the $SU(2)$ model considered above one can put 
$\eta=0$ and study the phase transition with respect to the 
expectation value of the adjoint field $\phi$. It is quite obvious that
for large $v$ we deal with the Abelian Coulomb phase, while when 
$v$ is small, confinement presumably takes place. What we do kill is 
the Wilson loop as the order parameter.  One can differentiate 
between the phases by using other criteria, however.

To this end one  assigns some ``external" flavor quantum 
numbers, say,  to the quark fields. One of the possibilities is the 
electric 
charge \footnote{I mean here the genuine electric charge,
not to be confused with the ``charges" of the QCD  monopoles and 
dyons. The latter are understood as the charges with respect
to some $U(1)$ subgroup of the original gauge group, $SU(3)$ in the 
case of
QCD, cf. Sect. 1.3. To consider the true electromagnetic interaction
we add an extra $U(1)$. Say, extended QCD including 
electromagnetism
has the gauge group $SU(3)_c\times U(1)_{em}$.}.
The quarks in QCD are fractionally charged, all up quarks have 
charges $2/3$ while all down quarks charges $-1/3$.  The 
electromagnetic interaction is external with respect to QCD and has 
nothing to do with color confinement. It is used just as a marker of 
particular states.
Some other flavor markers will do the job as well, say the Gell-Mann 
vector
$SU(3)$ existing in QCD with the massless $u,d,s$ quarks, or baryon 
number, and so on. 

Assume that {\em two} scalar fields in the {\em adjoint} 
representation are 
added in QCD, as
 in Eq. (\ref{GGL}). This is enough to break  the QCD  gauge group 
$SU(3)$ completely. 
When the   expectation values of these scalar fields are large,
the theory is in the (weakly coupled) Higgs phase. The states with 
the 
fractional baryon numbers (and fractional electric charge;
to be referred  as {\em fractionals} below) 
exist in the spectrum, as asymptotic 
states. Moreover, since the gauge coupling is small, these states are 
essentially undressed, and light. The states with the integer baryon 
numbers composed
of the quark-antiquark pairs or triquarks also exist but they are not 
necessarily bound. Their energy is higher than those with the 
fractional baryon numbers.

As we move to smaller vacuum expectation values
of the scalar fields, the  mass of the 
fractionals grows, since the fields they induce
become stronger. Bound states of the quark-antiquark pairs or 
triquarks become energetically advantageous. At a certain point the 
fractionals become infinitely heavy, so they are 
not seen in the observable spectrum. Those states 
which have finite mass and are seen have integer baryon numbers.
Presumably, there is a phase transition from the Higgs to the 
confining phase, although the order parameter  is not obvious in the 
case at hand.

There exists a dynamical scenario in which some fractionals
will still be seen in the spectrum, the so called
{\em oblique confinement} \cite{thooft,CR}. So far we discussed the 
dual 
Meissner effect, with condensed monopoles. The monopole 
condensation forces the chromoelectric field to form tubes
and makes quarks confined. In Sect. 1.3 we got acquainted with the 
dyon configurations in the Georgi-Glashow model. They were
characterized by non-zero magnetic and electric charges, 
simultaneously. The existence of dyons is inevitable in any gauge 
theory with magnetic monopoles. As was shown by Witten \cite{Ed},
introducing a non-vanishing vacuum angle $\vartheta$,
\beq
{\cal L}_\vartheta = \frac{\vartheta g^2}{32\pi^2} G_{\mu\nu}^a
 G_{\mu\nu}^a
\eeq
necessarily generates an electric charge $q$ for a particle
with the magnetic charge $m$,
\beq
q = \frac{\vartheta g^2}{8\pi^2} m\, .
\eeq
It is the  condensation of dyons that sets up the phase
of  the oblique confinement.

Let us discuss the phenomenon in more detail. Figure 1$a$
represents the spectrum of possible electric/magnetic charges
in the $SU(2)$ theory with $\vartheta = 0$
\footnote{The electric charge here has nothing to do 
with the conventional electromagnetism. This is the charge with 
respect to a $U(1)$ singled out by the 't Hooft Abelian projection. See 
the next section for further details.}. Elementary states with
the electric charge are along the horizontal axis.
If the matter fields are in the adjoint representation
their quanta have charges $0, \pm 1$ (point $A$). Two quanta can
have charges $\pm 2$, and so on. 
We may want to introduce quarks in the fundamental ($SU(2)$ 
doublet)
representation; their charges are $\pm 1/2$. 

\begin{figure}
  \epsfxsize=7cm
  \centerline{\epsfbox{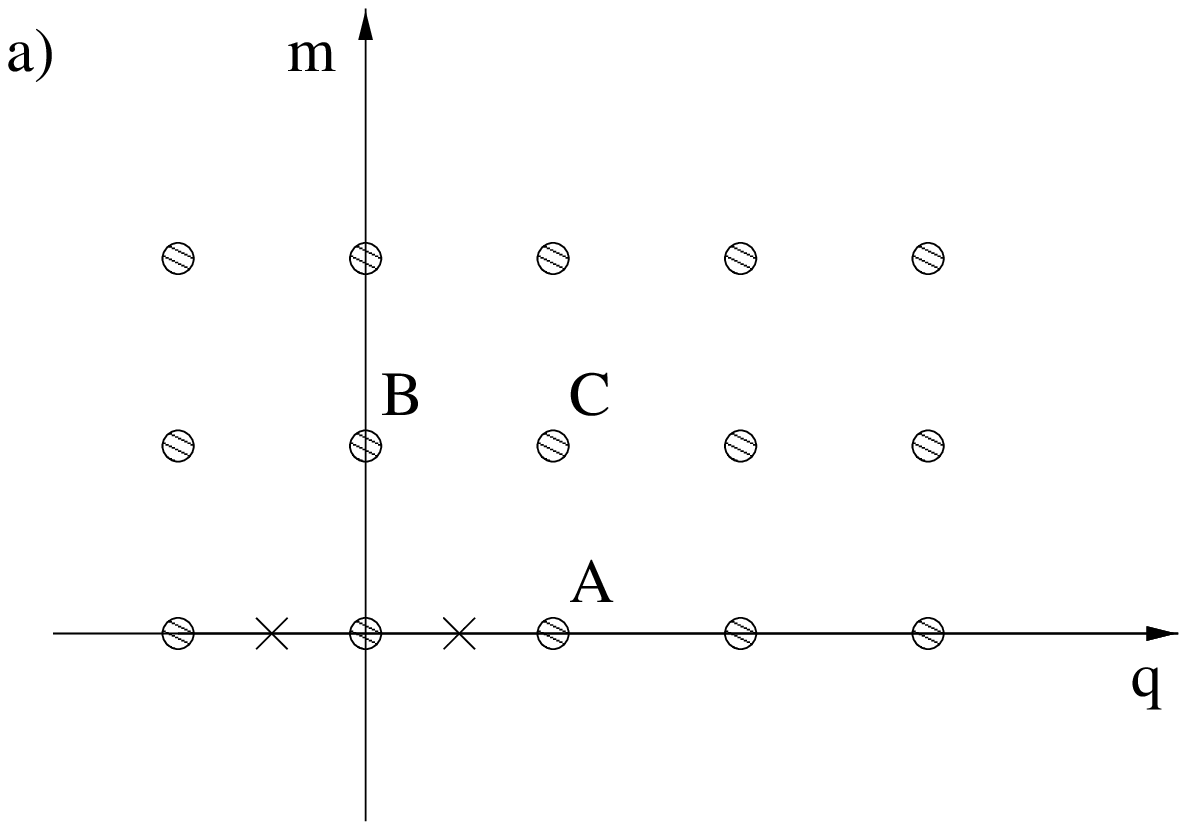} 
  \epsfxsize=7cm
  \epsfbox{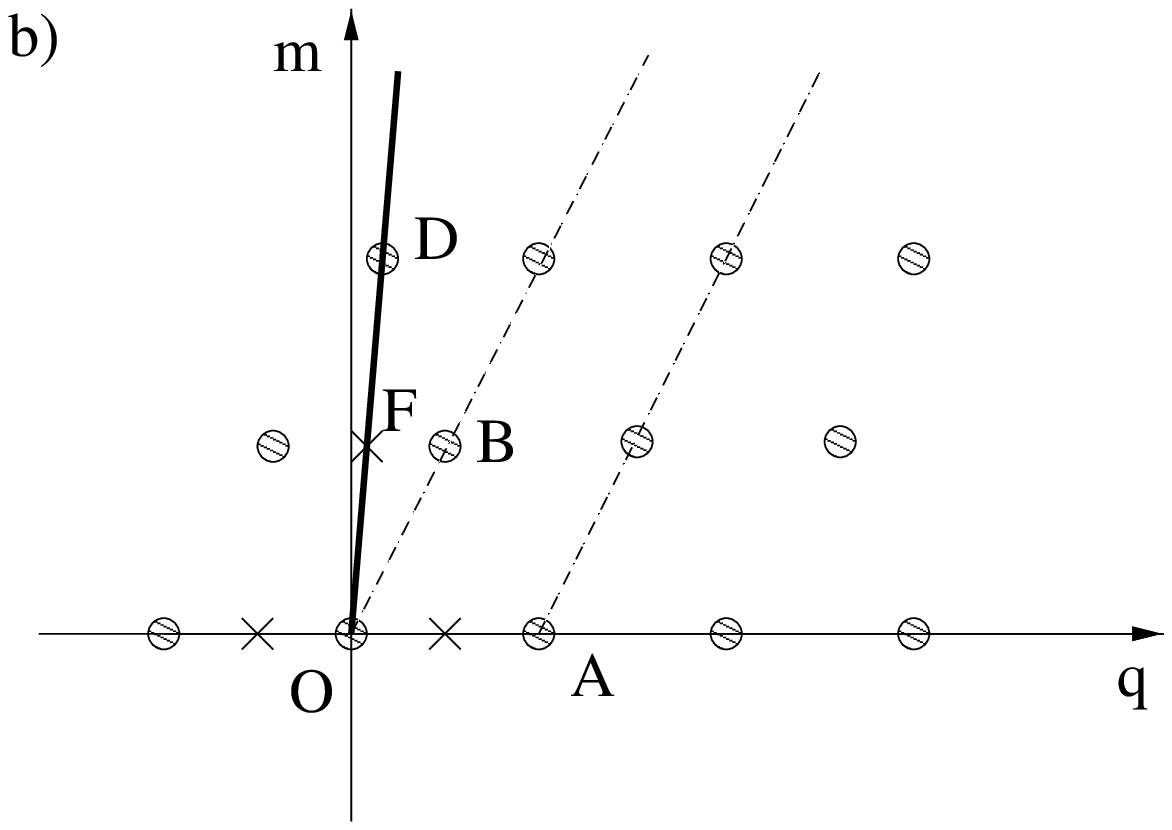}}
  \caption{
  The grids of the electric-magnetic charges in the $SU(2)$
gauge theory. The charges $\{q,m\}$ are measured with respect to 
the $U(1)$
subgroup of $SU(2)$, as in Sect. 1.5. The closed circles indicate 
the charges in the theory with the adjoint ($SU(2)$ triplet)
matter. The crosses indicate the electric charge of the
matter field in the fundamental ($SU(2)$ doublet) representation.
($a$) The rectangular grid corresponding to $\vartheta =0$. The 
point
denoted by $C$ is a bound state of the triplet matter quantum with
 the monopole.
Bound states of the doublet matter quanta with monopoles are 
possible too
(but not indicated). ($b$) The oblique grid
corresponding to $\vartheta =\pi +\varepsilon$, $0<\varepsilon \ll
\pi$. The state denoted by $D$ is a dyon with a small value of the
electric charge, which presumably condenses. The point
denoted by $F$ is an unconfined  bound state of the quark and dyon 
$B$.
Its external quantum numbers
are those of the quark.
}
\end{figure}

The magnetic monopole lies on the vertical axis (point $B$).
 It has $m=1,\,\, q=0$.
Other states on the vertical axis are anti-monopole, a pair of 
monopoles
and so on. All points which do not belong to the horizontal and 
vertical
axis are bound states of electric and magnetic quanta. Note that
the monopole condensation  automatically 
precludes from condensation all states
carrying the electric charge, since their mass squared
is positive (and infinite). The latter are confined. All states which
do not lie on the straight line connecting the origin with 
the point $B$
are confined by the flux tubes of the chromoelectric field 
attached to them. 

Increasing $\vartheta$ we deform the $\{q,m\}$ grid
in a continuous way. When $\vartheta$ is close to
$\pi$, but slightly larger than $\pi$, we arrive at the grid
shown on Fig. 1$b$. 
It is quite obvious that if $\vartheta\neq 0$ or $2\pi$
the grid of all possible values of $\{q,m\}$ is oblique. (That's where
the name {\em oblique confinement} comes from.) At
$\vartheta =\pi$ the objects with $m=1$ will have $q = \pm 1/2,
\pm 3/2$ and so on. It may well happen that the $m=1$ objects will 
not condense, since apart from the magnetic charge they have  
large (and opposite) electric charges. But among the $m=2$ 
monopoles there is one with the vanishing electric charge, and it is 
quite conceivable that energetically it is more expedient for these 
objects to condense. If $\vartheta$ is slightly larger
than $\pi$ the state suspect of condensing is that
marked by $D$ on Fig. 1$b$. 
All states that do not lie on the straight line connecting the
origin with $D$ 
 will be confined. This is a very peculiar 
confinement, 
however. Some of the states, carrying the external quantum
numbers of the fundamental quarks, exist in the observable 
spectrum. 
The quarks themselves (crosses on the horizontal axis)
do not lie on the $OD$ line and, hence, are confined. The bound state
of the quark and a dyon (the cross marked by the letter $F$)
does belong to this line, and is {\em not} confined. Since
the dyon has no baryon charge, or any other external quantum 
number,
the state $E$ has exactly the same baryon charge as the 
fundamental quark,
i.e. it is a fractional. 

One has to pay a price for 
having fractionals in the observable spectrum. As explained in 
\cite{thooft}, even though the quarks are described by the Fermi 
fields at the Lagrangian level, the states with the fractional 
(and non-vanishing) baryon charge to be  observed will all be 
bosons!

\subsection{QCD monopoles and Abelian projection}

Let us discuss how monopole-like configurations could emerge in 
QCD. 
Following  't Hooft we will impose an {\em incomplete} gauge  fixing 
condition in a special way. In the Georgi-Glashow model the Higgs 
mechanism breaks the original $SU(2)$ down to $U(1)$, which
paves the way to the emergence of the monopoles as physical 
objects. In QCD the gauge group remains unbroken, of course. Our 
task is to single out an $U(1)$ subgroup by imposing an appropriate  
gauge condition.  The monopole-like solution, obtained in this way 
certainly  cannot be interpreted as a physical  particle; it should be 
viewed rather as a first stage in a construction which, eventually, 
may
answer the question whether the dual Meissner effect takes place in 
QCD.

That the monopole-like solutions of the classical equations of motion 
are present in QCD can be seen in many different ways.
For instance, assume that (an incomplete) gauge condition is imposed 
in such a way that the time-dependent gauge transformations are 
forbidden. Whatever this gauge condition might be it still does not 
forbid the gauge transformations which depend on the spatial 
coordinates, $\vec x$. 
For simplicity we will additionally assume that the gauge group is 
$SU(2)$, not $SU(3)$ as in  genuine QCD.
Under this narrow class of gauge transformations $A_0^aA_0^a$
is obviously gauge invariant, since with respect to the 
spatial-dependent gauge transformations the time component of $A$
transforms homogeneously. If so, the model becomes identical to the 
BPS limit of the Georgi-Glashow model in the $A_0=0$ gauge,
which is known to possess the monopole solutions.

Let us elucidate the latter assertion in more detail. The Lagrangian
of the $SU(2)$ Yang-Mills theory in the above ``gauge", for the static 
field configurations, takes the form
\beq
{\cal L} = -\frac{1}{4}G_{ij}^a (\vec x) G_{ij}^a (\vec x)
+\frac{1}{2} D_iA_0^a(\vec x) D_iA_0^a(\vec x)\, ,
\label{LGGF}
\eeq
where
$$
D_iA_0^a = \partial_i A_0 +g\epsilon^{abc}A_i^bA_0^c
$$
is the covariant derivative. If we now rename $A_0^a \ra i\phi^a$,
the Lagrangian (\ref{LGGF}) will coincide with that of the
Georgi-Glashow model in the $A_0=0$ gauge (in the BPS
limit)
\footnote{Since $A_0^a \ra i\phi^a$ and is purely imaginary we deal 
here with an analytic continuation to the Euclidean space.}. The 
classical equations of motion are identical and should then 
have
identical solutions; in particular, the monopole solution (\ref{B8})
goes through. In the case at hand $v^2$ is the value of
$A_0^a A_0^a$ at the spatial infinity. Since the time-dependent gauge 
transformation are forbidden, this value is well-defined. 

I pause here to make a few explanatory remarks.
The 't Hooft-Polyakov monopole in the BPS limit satisfies the 
condition (\ref{B5}).
One immediately recognizes in this condition in our particular case 
the
self-duality equation $G_{\mu\nu} = 
(1/2)\epsilon_{\mu\nu\alpha\beta} G_{\alpha\beta}$ defining
the (multi)instanton solutions. Thus, heuristically it is clear
that the ``monopole" solution we have found must be
equivalent (up to a gauge transformation) to a chain of instantons. 
Since the ``monopole" mass 
is finite, the action is infinite; therefore, we deal here with an infinite 
chain of instantons. The fact that the equivalence does indeed take 
place was demonstrated in Ref. \cite{moninst} where it was 
established
that the monopole can be identified with a sequence of equally 
spaced instantons located at the time axis. The spacing
between the instanton centers is $8\pi^2 M_{\rm mon}^{-1}g^{-2}$, 
and their radii tend to infinity.
Some other chains were shown to be gauge-equivalent to dyons 
\cite{moninst1}.
The inverse transition, from instantons to monopoles,
was also studied. It was found that the instanton considered in
the Abelian projection (see below) coincides with a closed 
monopole loop centered at the instanton center \cite{moninst2}.
For a discussion of the  transition from the ``monopoles"
to the instanton chains and back see Ref. \cite{Simonov}.

Those readers who feel uneasy with the notion of monopoles in QCD
can think of the solution above as of a sequence of
instantons.
The instantons are, of course,  much more familiar to the QCD 
practitioners.

A related but somewhat more general line of reasoning
leading to the same monopole-like configurations was suggested by 't 
Hooft \cite{thooft}. His {\em Abelian projection} of QCD can be 
elucidated as follows. Let us consider some gauge non-invariant 
operator in the adjoint representation of the gauge group 
transforming homogeneously. In pure gluodynamics this may be,
say, $G_{12}^a$. If quarks are present one might
consider $\bar\psi_i\psi^j$ where $i$ and $j$ are color indices.
It may be more convenient, however, to introduce
(auxiliary) scalar fields in the adjoint representation, $\phi^a$, 
making them sufficiently heavy, so they do not affect the low-energy 
physics.
Generically, this operator will be denoted as
$X_i^j\equiv X^a (T^a)_i^j$ where the matrices  $T^a$ are
the generators of the gauge group. For simplicity we will assume that 
the gauge group is $SU(N)$; then, $(T^a)_i^j$ are $N$ by $N$
matrices generalizing the Gell-Mann matrices of QCD.
The gauge transformation acts on $X$ as
$X\ra UXU^{-1}$ where $U$ is arbitrary $x$ dependent matrix from
$SU(N)$. It is quite obvious that by choosing $U(x)$ in an
appropriate way it is always possible to diagonalize $X$.
In other words, the only components of $X^a$ surviving after the 
gauge transformation are those corresponding to the Cartan 
subalgebra of $SU(N)$: $a= 3, 8, 15$ and so on. This ``Abelization"
of the operator $X$ obviously explains why the gauge is referred to 
as the Abelian projection. 

The gauge condition above is an incomplete gauge,
since one can additionally  perform gauge transformations
corresponding to arbitrary rotations around the third, eighth and so 
on axis
without destroying the Abelian nature of $X$.
The diagonal form of the matrix $X$ is preserved under these 
rotations: all generators $T^3$, $T^8$, $T^{15}$, ... are diagonal. 
Thus, in the gauge at hand, $U(1)^{N-1}$ subgroup is singled out,  
$SU(N)$ gluodynamics
looks similar to QED, with $N-1$ different photons ($A^3, A^8,
A^{15}$, {\em etc.}) and $(1/2)N(N-1)$ ``matter fields"
(all off-diagonal $A$'s) charged with respect to the photon 
fields. The values of the charges are, generally speaking, different
for different photons. They depend on the group constants.

Intuitively it is clear that we are going to have the ``monopole" 
solutions. Formally, this can be seen as follows.
In the Abelian projection the operator $X$ is a diagonal matrix,
$X=$ diag$\{\lambda_1, \lambda_2, ... , \lambda_N\}$
where the eigenvalues $\lambda$ may be ordered,
$\lambda_1>\lambda_2>...>\lambda_N$. In some exceptional points in 
space, however, two out of $N$ eigenvalues can coincide, say
$\lambda_1$ and $\lambda_2$. This singles out the upper left 
$SU(2)$ 
corner of $SU(N)$ with a monopole sitting at the point where
$\lambda_1=\lambda_2$.

Indeed, close to this point we can focus on the upper left
two-by-two corner of the matrix $X$, assuming that the remaining 
part of 
the matrix is already 
diagonal, with non-coinciding eigenvalues. We have to diagonalize 
only the upper left corner.  Near the point ${\vec x}_0$ where
$\lambda_1=\lambda_2$ the two-by-two submatrix has the form 
$$
\lambda {\bf 1} + \varepsilon^a (\vec x) \sigma^a
$$
where {\bf 1} is the two-by-two unit matrix and $\sigma^a$ are the 
Pauli 
matrices. The condition $\lambda_1=\lambda_2$ at ${\vec x}_0$ 
means that
$\varepsilon^a = 0$ at $\vec x = {\vec x}_0$ ($a=1,2,3$). Since we 
have to ensure that
three (real) functions vanish, generically this can happen only on
manifold of dimension zero, i.e. in isolated points in space, as was 
mentioned. Moreover, near these points $\varepsilon^a (\vec x) $
has a hedgehog configuration, 
$\varepsilon^a (\vec x) \sim ( x - { x}_0)^a$, 
a characteristic feature of the
't Hooft-Polyakov monopole.

In the absence of a genuinely small parameter in QCD,
the semi-classical description sketched above cannot lead us too far
beyond a qualitative picture.
Definitely, there is no way one can study the monopole condensation,
a crucial element of the confinement mechanism-to-be,
in this approximation. One can try, however, to apply these ideas in 
the context of the lattice simulations, hoping to get precious insights
from numerical studies. Although work in this direction is far from 
completion, and many aspects remain unclear (see e.g.
Ref. \cite{monlatun}), some initial results are quite encouraging.
In particular, the abundance of the monopoles in the vacuum 
ensemble of  the lattice QCD was observed, and a connection with the 
chiral symmetry breaking conjectured \cite{monlat}. Approaches 
combining analytical methods with numerical analysis are under 
investigation (see e.g. \cite{monlat1}).

\vspace{0.2cm}

Summarizing, we have learned  that the gauge theories can be in the 
following phases: 

1) Coulomb (Abelian and non-Abelian; in the latter case 
the infrared limit is conformal, as will be discussed in detail in 
Sect. 3);

2) free (Landau);

3) Higgs (a possible version is a unified Higgs/confining phase);

4) confining (a possible version is oblique confinement).

Now that we are familiar with various
dynamical scenarios one expects to observe in different
gauge theories, we are ready to proceed to supersymmetric theories 
where, in some cases, it is possible to go far beyond qualitative 
ideas, towards exact solution,  using miracles of supersymmetry. 

\newpage

\section{Lecture 2. Basics  of Supersymmetric Gauge Theories}

\renewcommand{\theequation}{2.\arabic{equation}}
\setcounter{equation}{0}

The dream of every QCD practitioner is to find analytic solutions of 
two
most salient properties of QCD: color confinement and spontaneous 
breaking of 
the chiral symmetry.  In spite of two decades of vigorous efforts
very little progress is made in this direction. At the same time,
exciting developments took place, mostly in the last few years, 
in supersymmetric gauge theories -- close relatives of QCD. These 
developments can, eventually, lead 
to  a breakthrough in QCD. Even if this does not happen, they are 
very 
interesting on their own. It turns out that supersymmetry helps 
reveal  
 several 
intriguing and extremely elegant properties which  shed light on 
subtle
aspects of the gauge theories in general. In this section we will start 
our 
excursion in
supersymmetric gauge theories.
There is a long way to go, however, before we will be able to discuss 
a variety of 
fascinating  results obtained in this field recently. As a first step let 
me briefly review some basic 
elements 
of the formalism we will need below.

\subsection{Introducing supersymmetry}

Supersymmetry relates bosonic and fermionic degrees of freedom. A  
necessary condition for any theory to be supersymmetric is the 
balance
between the number of the bosonic and fermionic degrees of 
freedom,
having the same mass and the same ``external" quantum numbers, 
e.g.
color. Let us consider several simplest examples of practical 
importance.

A scalar complex field $\phi$ has two degrees of freedom (a particle
plus antiparticle). Correspondingly, its spinor superpartner is the
Weyl (two-component) spinor, which also has two degrees of 
freedom --
say, the left-handed particle and the right-handed antiparticle. 
Alternatively, 
instead of working with the complex fields, one can introduce real 
fields, with 
the same physical content: two real scalar fields $\phi_1$ and 
$\phi_2$ 
describing two ``neutral" spin-0 particles, plus  the Majorana (real 
four-component) spinor describing a ``neutral" spin-1/2 particle with 
two 
polarizations. (By neutral I mean that the corresponding antiparticles 
are 
identical to their particles). This family has a balanced number of the
degrees of freedom both in the massless and massive cases.  Below 
we will
see that in the superfield formalism it is described, in a concise form, 
by 
one {\em chiral superfield}. 

When we speak of the quark flavors in QCD we  count the Dirac 
spinors.
Each Dirac spinor is equivalent to two Weyl spinors. Therefore, in 
SQCD each 
flavor requires two chiral superfields. Sometimes, the superfields 
from this 
chiral pair are referred to as subflavors. Two subflavors comprise 
one flavor. 

Another important example is vector particles, gauge bosons (gluons 
in QCD,
$W$ bosons in the Higgs phase).  Each gauge boson carries two 
physical
degrees of freedom (two transverse polarizations). The appropriate 
superpartner is the Majorana spinor. Unlike the previous example 
the balance 
is achieved only for massless particles, since the massive vector 
boson 
has 
three, not two, physical degrees of freedom. The superpartner to the 
massless 
gauge boson is called gaugino. Notice that the mass still can be 
introduced 
through the (super)Higgs mechanism. We will discuss the Higgs 
mechanism in 
supersymmetric gauge theories later on. 

In  counting  the degrees of freedom above the external quantum 
numbers were left aside. Certainly, they should be the same for each 
member 
of the superfamily. For instance, if the gauge group is SU(2), the 
gauge bosons 
are ``color" triplets, and so are gauginos. In other words, the 
Majorana fields 
describing gauginos are provided by the ``color" index $a$ taking 
three 
different values, $a=1,2,3$. 

If we consider the free field theory with the balanced number of 
degrees of 
freedom, the vacuum energy vanishes. Indeed, the vacuum energy is 
the
sum of the zero-point oscillation frequencies for each mode of the 
theory,
\beq
E_{\rm vac} = \sum_{\vec k} \left\{ \omega^{\rm boson}_{\vec k}-
\omega^{\rm ferm}_{\vec k}\right\}\, .
\label{ffve}
\eeq
I remind that the modes are labeled by the three-momentum $\vec 
k$;
say, for massive particles 
$$
\omega_{\vec k}= \sqrt{m^2 +{\vec k}^2}\, .
$$
It is important that the boson and fermion terms enter with the 
opposite signs
and cancel each other, term by term. This observation, which can be 
considered as a precursor to supersymmetry, was made by Pauli in 
1950 
\cite{Pauli}!  If 
interactions
are introduced in such a way that supersymmetry remains 
unbroken, the 
vanishing of the vacuum energy is preserved in  dynamically 
nontrivial 
theories.

Balancing  the number of degrees of freedom is the necessary but
not sufficient condition for supersymmetry in  dynamically 
nontrivial 
theories, of course. All vertices must be supersymmetric too. This 
means that 
each line can be substituted by that of a superpartner. Let us 
consider, for 
instance, QED, the simplest gauge theory. We start from the 
electron-electron-photon coupling (Fig. 2$a$). Now,  as we already 
know,
in SQED the electron is accompanied by two selectrons (two, because
the electron is described by the four-component Dirac spinor rather 
than the 
Weyl spinor). Thus, supersymmetry requires 
the selectron-selectron-photon vertices,
 (Fig. 2$b$),
with the same coupling constant.
Moreover, the 
photon
can be substituted by its superpartner, photino, which generates the 
electron-selectron-photino vertex  (Fig. 2$c$), 
with the same coupling. In the old-fashioned language of the 
pre-SUSY 
era we would call this vertex the Yukawa coupling. In the 
supersymmetric 
language this is the gauge interaction since it generalizes the gauge 
interaction coupling of the photon to the electron. 

\begin{figure}
  \epsfxsize=12cm
  \epsfysize = 2cm
  \centerline{\epsfbox{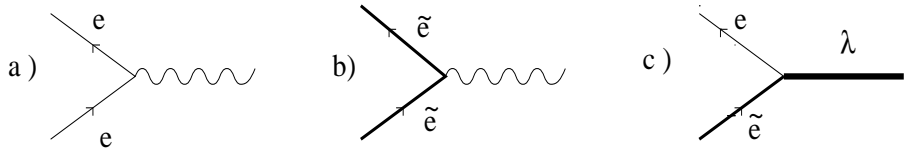}}
  \caption{ Interaction vertices in QED and its 
   supergeneralization, SQED.
   ($a$) $\bar e e\gamma$ vertex; ($b$) selectron coupling to photon;
   ($c$) electron-selectron-photino vertex.   
   All vertices have the same coupling constant. The quartic
   self-interaction of selectrons is also present, but not shown.}    
\end{figure}

With the above set of vertices one can show that the theory is 
supersymmetric at the  level of trilinear interactions, provided that 
the 
electrons and the 
selectrons are degenerate in mass, while the photon and photino 
fields are 
both massless. To make it fully supersymmetric one should also add 
some 
quartic terms, describing self-interactions of the selectron fields, as 
we will 
see shortly.

Now, the theory is dynamically nontrivial, the particles
-- bosons and fermions -- are not free and still 
$E_{vac} = 0$. This is the {\em first miracle} of supersymmetry.

The above pedestrian (or step-by-step) approach to 
supersymmetrizing the 
gauge 
theories is quite possible, in principle. Moreover, historically
the first supersymmetric model derived by Golfand and Likhtman, 
SQED,
was obtained in this way \cite{GL}. This is a painfully slow method, 
however,
which is totally out of use at the present stage of the theoretical 
development.
The modern efficient approach is based on the superfield formalism, 
introduced in 
1974 by Salam and Strathdee \cite{SS}  who replaced the 
conventional 
four-dimensional 
space by the superspace. 

\subsection{Superfield formalism: bird's eye view}

I will be unable to explain this formalism, even 
briefly. The reader is referred to the text-books and 
numerous excellent 
reviews, see the list of recommended literature at the
end.
Below some elements are listed mostly with the purpose of 
introducing 
relevant notations, to be used throughout the entire lecture course.
(Our notation, conventions and useful formulae are collected in 
Appendix.) 
 
If the conventional space-time is parametrized by the coordinate 
four-vector 
$x_\mu$,
the superspace is parametrized by $x_\mu$ and two Grassmann 
variables,
$\theta$ and $\bar\theta$. The Grassmann numbers obey all 
standard rules
of arithmetic except that they anticommute rather than commute 
with each 
other. In particular, the product of a Grassmann number with itself
is zero, for this reason.

With respect to the Lorentz properties, $\theta$ and $\bar \theta$ 
are 
spinors. As well known, the four-dimensional Lorenz group is 
equivalent to
$SU(2)\times SU(2)$ and, therefore, there exist  two types of spinors,
left-handed and right-handed, denoted by undotted and dotted 
indices, 
respectively;   $\theta_\alpha$  is the left-handed spinor while
$\bar\theta_{\dot\alpha}$ is the right-handed one ($\alpha , 
\dot\alpha = 
1,2$).  The indices of the
right-handed spinors are supplied by dots to emphasize the fact
that their transformation law does not 
coincide with that of the left-handed spinors.

The Lorentz scalars can be formed as a convolution of two dotted or
two undotted spinors, $\theta^\alpha\theta_\alpha$ or
$\bar\theta_{\dot{\alpha}}\bar\theta^{\dot{\alpha}}$,
with one lower and one upper index. Raising and lowering of indices
is realized by virtue of the antisymmetric (Levi-Civita) symbol,
$$
\bar\theta_{\dot{\alpha}} =\epsilon_{\dot{\alpha}\dot{\beta}}
\bar\theta^{\dot{\beta}}\, ,\,\,\,\,
\theta_\alpha =\epsilon_{\alpha\beta}\theta^\beta\, ,
$$
where
$$
\epsilon^{12} = \epsilon^{\dot{1}\dot{2}} = 1\, ;\,\,\,\,
\epsilon_{12} = \epsilon_{\dot{1}\dot{2}} = -1
$$
so that $\epsilon^{\alpha\beta}\epsilon_{\beta\gamma} 
=\delta^\alpha_\gamma$. When one  raises or lowers
the index of $\theta$ the $\epsilon$ symbol must be placed to
the left of $\theta$.

A shorthand notation when the indices
of the spinors are implicit is widely used, for instance,
$$
\theta^2 \equiv \theta^\alpha\theta_\alpha
$$
and
$$
\bar\theta^2\equiv \bar\theta_{\dot{\alpha}}\theta^{\dot{\alpha}}\, 
.
$$
Notice that in convoluting the undotted indices one writes
first the spinor with the upper index while for the dotted indices
the first spinor has the lower index.  The ordering is important since 
the 
elements
of the spinors are anticommuting Grassmann numbers.

It remains to be added that the vector quantities can be obtained
from two spinors -- one dotted and one undotted. Thus, 
$\theta_\alpha
\bar\theta_{\dot\alpha}$ transforms as a Lorentz vector.

Now, we can introduce the notion of supertranslations in the 
superspace
$\{ x, \theta, \bar\theta\}$. The generic supertransformation has the 
form
$$
\theta \ra \theta + \varepsilon \, ,\,\,\,\, 
\bar\theta \ra \bar\theta + \bar\varepsilon \, ,
$$
\beq
x_{\alpha\dot\beta} \ra x_{\alpha\dot\beta} +
2i\varepsilon_\alpha
\bar\theta_{\dot\beta} -
2i \theta_\alpha \bar\varepsilon_{\dot\beta} \, .
\eeq 
The supertranslations generalize conventional translations in the 
ordinary 
space. 

One can also consider the so called chiral and antichiral superspaces
(chiral realizations of the supergroup);
the first one does not explicitly contain $\bar \theta$ while the
second does not contain $\theta$. It is not difficult to see that a point
from the chiral superspace is parametrized by $\{ x_L, \theta\}$, and 
that
from the antichiral superspace is parametrized by $\{ x_R, 
\bar\theta\}$.
Here
\beq
(x_L)_{\alpha\dot\alpha} = x_{\alpha\dot\alpha} - 
2i\theta_\alpha\bar\theta_{\dot\alpha}\, ,\,\,\,
 (x_R)_{\alpha\dot\alpha}= x_{\alpha\dot\alpha} + 
2i\theta_\alpha\bar\theta_{\dot\alpha}\,  .
\eeq
Under this definition the supertransformations corresponding to the
shifts in $\theta$ and $\bar\theta$, respectively,
leave us inside the corresponding superspace. Indeed, if
$\theta \ra \theta + \varepsilon$ and $ \bar\theta \ra \bar\theta 
+ \bar\varepsilon$,
then 
\beq
(x_L)_{\alpha\dot\beta}\ra (x_L)_{\alpha\dot\beta} 
-4i \theta_{\alpha}\bar\varepsilon_{\dot\beta}
\, \,\, \mbox{and}\,\,\, 
(x_R)_{\alpha\dot\beta}\ra (x_R)_{\alpha\dot\beta} 
+4i\varepsilon_\alpha
\bar\theta_{\dot\beta} \, .
\label{AD1}
\eeq

Superfields provide a very concise description of supersymmetry 
representations. They are very natural generalizations of 
conventional fields. 
Say, the scalar field $\phi (x)$ in the $\lambda\phi^4$ theory is a 
function of 
$x$. Correspondingly, superfields are functions of $x$ and $\theta$'s.
For instance, the chiral superfield $\Phi (x_L, \theta )$ depends on 
$\theta$ and $x_L$ (and has no explicit $\bar\theta$ dependence).
If we Taylor-expand it in the powers of $\theta$ we get the 
following formula:
\beq
\Phi (x_L, \theta ) = \phi (x_L) +\sqrt{2}\psi (x_L)\theta  + \theta^2 
F(x) \, .
\label{chifi}
\eeq
There are no higher-order terms in the expansion since higher 
powers of 
$\theta$ vanish due to the Grassmannian nature of this parameter. 
For the 
same reason the argument of the last component of the chiral 
superfield, $F$, 
is set equal to $x$. The distinction between $x$ and $x_L$ is not 
important in 
this term. The last component of the chiral superfield is always called 
$F$.
$F$ terms of the chiral superfields are non-dynamical, they appear in 
the
Lagrangian without derivatives. We will see later that  $F$ terms 
play a 
distinguished role.

The lowest component of the chiral superfield is a complex scalar 
field
$\phi$, and the middle component is a Weyl spinor $\psi$. Each of 
these
fields describes two degrees of freedom, so the appropriate balance 
is
achieved automatically. Thus, we see that  superfield is a concise 
form of 
representing a set of components. The transformation law of the
components follows immediately from Eq. (\ref{AD1}), for instance,
$\delta\phi (x) =\sqrt{2}\psi (x) \varepsilon$, and  so on.

The antichiral superfields depend on $x_R$ and $\bar\theta$. The 
chiral and 
antichiral superfields  describe the matter sectors of the theories
to be studied below. The gauge field appears from the so called 
vector 
superfield $V$ which depends on both,
 $\theta$ and $\bar\theta$ and satisfies the condition 
$V=V^\dagger$.
The component expansion of the vector superfield has the form
$$
V(x,\theta,\bar\theta ) = C(x) +i\theta\chi (x)  - i\bar\theta\bar\chi 
(x) 
+\frac{i}{2}\theta^2 [M(x) + i N(x)] -\frac{i}{2}\bar\theta^2 [M(x) - i 
N(x)] -
$$
\beq
2\theta_\alpha \bar\theta_{\dot\beta} v^{\alpha\dot\beta} (x)
+2\left\{ i\theta^2\bar\theta_{\dot\beta}\left[ 
\bar\lambda^{\dot\beta}
-\frac{i}{4}\partial^{\alpha\dot\beta}\chi_\alpha \right] + 
\mbox{h.c.}\right\}
+ \theta^2\bar\theta^2\left[ D(x) -\frac{1}{4}\Box 
C(x)\right]\, .
\eeq
The components $C,D,M,N$ and $v^{\alpha\dot\beta}$ must be real to 
satisfy 
the condition $V=V^\dagger$. The vector field $v^{\alpha\dot\beta}$
gives its name to the entire superfield. 

The last component of the vector superfield, apart from a full 
derivative, is
called the `` $D$ term".  $D$ terms also play a special role. 

Let me say a few words about the gauge transformations.
For simplicity I will consider the case of the Abelian (U(1)) gauge 
group.
In the non-Abelian case the corresponding formulae become
more bulky, but the essence stays the same.

As well known , in nonsupersymmetric gauge theories the matter 
fields
transform under the gauge transformations  as
\beq
\phi (x)  \ra e^{i\alpha (x)}\phi (x) \, , \,\,\,  \phi (x) ^\dagger \ra 
e^{-i\alpha 
(x)} \phi (x) ^\dagger\, , 
\label{gau1}
\eeq
while the gauge field
\beq
v_\mu (x) \ra v_\mu (x) + \partial_\mu \alpha (x)\, ,
\label{gau2}
\eeq
where $\alpha (x) $ is an arbitrary function of $x$. Equations 
(\ref{gau1})
and (\ref{gau2}) prompt the supersymmetric version of the gauge 
transformations,
\beq
\Phi (x_L, \theta ) \ra e^{i\Lambda}\Phi (x_L, \theta )\, , \,\,\,
\bar \Phi (x_R, \bar\theta ) \ra e^{-i\bar\Lambda}\Phi (x_R, 
\bar\theta )\,
\label{sgau1}
\eeq
and
\beq
V\ra V  -i(\Lambda -\bar\Lambda)
\label{sgau2}
\eeq
where $\Lambda$ is an arbitrary chiral superfield, $\bar\Lambda$ 
is 
its
antichiral partner. $\bar\Phi e^V\Phi$ is then a gauge invariant 
combination playing the same role as ${\cal D}_\mu\phi^\dagger
{\cal D}_\mu\phi$ in non-supersymmetric theories. 
Let me parenthetically note that supersymmetrization of the gauge 
transformations, Eqs. (\ref{sgau1}), (\ref{sgau2}), was the path which 
led 
Wess 
and Zumino \cite{WessZ} to the 
discovery of the supersymmetric theories (independently of
Golfand and Likhtman).

In components
$$
C\ra C -i  (\phi - \phi^\dagger)\, ,\,\,\,
\chi \ra \chi - \sqrt{2}\psi \, , \,\,\, M+iN\ra M+iN -2F\,
$$
\beq
v_{\alpha\dot\beta}\ra v_{\alpha\dot\beta} +
\partial_{\alpha\dot\beta}
(\phi + \phi^\dagger )\,, \,\,\, \lambda\ra\lambda\,, \,\,\,
D\ra D\,  .
\eeq
We see that the $C, \chi , M$ and $N$ components of the vector 
superfield
can be gauged away. This is what is routinely done when the 
component 
formalism is used. This gauge bears the name of its inventors -- it is 
called the 
{\em Wess-Zumino} gauge. Imposing the 
Wess-Zumino gauge condition
in  supersymmetric theory one actually does not fix the gauge 
completely. 
The component Lagrangian one  arrives at in the Wess-Zumino gauge
 still possesses the gauge freedom with respect to 
 non-supersymmetric (old-fashioned) gauge transformations.

It remains to introduce spinorial derivatives. They will be denoted 
by
capital $D$ and $\bar D$,
\beq
D_\alpha = \frac{\partial}{\partial \theta^\alpha} - i 
\partial_{\alpha\dot\beta}
\bar\theta^{\dot\beta}\, , \,\,\,  \bar D_{\dot\alpha}
= -\frac{\partial}{\partial \bar\theta^{\dot\alpha}} + i 
\partial_{\beta\dot\alpha}
\theta^\beta\, .
\label{AD2}
\eeq
The relative signs in Eq. (\ref{AD2}) are fixed by the
requirements $D_\alpha (x_R)_{\beta\dot\gamma} = 0$
and $\bar D_{\dot\alpha} (x_L)_{\beta\dot\gamma} = 0$.

To make the spinorial derivatives distinct from the regular 
covariant
derivative the latter will be denoted by the script ${\cal D}$. 
The supergeneralization of the field strength tensor of the gauge field
has the form
$$
W_\alpha (x_L, \theta) = \frac{1}{8} \bar D^2 (e^{-V}D_\alpha e^V ) =
$$
\beq
i\lambda_\alpha (x_L) -\theta_\alpha D(x_L)
-i\theta^\beta G_{\alpha\beta} (x_L) +\theta^2 {\cal 
D}_{\alpha\dot\alpha}
\bar\lambda^{\dot\alpha} (x_L) \, ,
\label{doublew}
\eeq
where $G_{\alpha\beta}$ is the gauge field strength tensor in the 
spinorial 
form. 

This brief excursion in the formalism, however boring it might seem,
is  necessary for understanding physical results to be discussed 
below.  I will 
try to limit such excursions to absolute minimum,
but we will not be able to avoid them completely. Now, the stage is 
set, and 
we are ready to submerge in the intricacies of the
supersymmetric gauge dynamics.

\subsection{Simplest  supersymmetric models}

In this  section we will discuss some simple models. Our basic task is 
to reveal
 general features playing the key role in various unusual dynamical 
scenarios 
realized in supersymmetric gauge theories. One should keep in  mind 
that all theories with matter can be divided in two distinct classes:
chiral and non-chiral matter. The second class includes 
supersymmetric generalization of QCD, and all other models 
where each matter multiplet is accompanied by the corresponding 
conjugate representation. In other words, mass term is possible for 
all matter fields. Even if the massless limit is considered, the very 
possibility of adding the mass term is very important for dynamics.
In particular, dynamical SUSY breaking cannot happen in the
 non-chiral models. 

Models with chiral matter are those where the mass term is 
impossible. The matter sector in such models is severely constrained 
by the absence of the internal anomalies in the theory. The most 
well-known example of this type is the $SU(5)$ model with 
equal number of chiral quintets and (anti)decuplets.  Each quintet 
and anti-decuplet, together,  are called {\em generation}; when the 
number of generations is three
this is nothing but the most popular grand unified theory
of electroweak interactions. The chiral models are singled out by the 
fact that dynamical SUSY breaking is possible, in principle,  only in 
this class. In the present lecture course dynamical SUSY breaking
is not our prime concern. Rather, we will focus on various 
non-trivial dynamical regimes. Most of the regimes to be discussed 
below manifest themselves in the non-chiral models, which are
simpler. Therefore, the emphasis will be put on the  non-chiral 
models,  digression to the chiral models will be made occasionally.

\subsubsection{Supersymmetric gluodynamics}

To begin with we will consider supersymmetric generalization of 
pure 
gluodynamics -- i.e. the theory of gluons and gluinos. The Lagrangian
has the form \cite{Sergio}
\beq
{\cal L} = -\frac{1}{4g^2} G_{\mu\nu}^a G_{\mu\nu}^a 
+ \frac{\vartheta}{32\pi^2} G_{\mu\nu}^a \tilde G_{\mu\nu}^a 
+\frac{i}{2g^2}
\bar\lambda^a {\cal D}_\mu\gamma^\mu \lambda^a
\label{susyym}
\eeq
where $G_{\mu\nu}^a$ is the gluon field strength tensor, $\tilde 
G_{\mu\nu}^a$ is the dual tensor,  $g$ is the gauge coupling constant, 
$\vartheta$ is the 
vacuum angle,  and ${\cal D}_\mu$
is the covariant derivative. Moreover, $\lambda^a$
is the  gluino field, which can be described either by a 
four-component
Majorana (real) fields or two-component Weyl (complex) fields.

In terms of superfields
\beq
{\cal L} = \frac{1}{4g^2_0}\, \mbox{Tr}\,
\int d^2 \theta \, W^2 +\, \mbox{H.c.} \, ,
\label{susyg}
\eeq
where the superfield $W$ is a color matrix,
$$
W = W^a T^a\, ,
$$
$T^a$ are the generators of the gauge group (in the fundamental 
representation), Tr$T^aT^b = (1/2)\delta^{ab}$. It is very important 
that
the gauge constant $1/g^2_0$ in Eq. (\ref{susyg}) can be treated as
a complex parameter. The subscript 0 emphasizes the fact that
the gauge couplings in Eqs. (\ref{susyg}) and (\ref{susyym})
are different,
\beq
\frac{1}{g_0^2} = \frac{1}{g^2} - i\frac{\vartheta}{8\pi^2}\,  ,
\label{complg}
\eeq
its real part is the conventional gauge coupling while the imaginary 
part
is proportional to the vacuum angle. 
Thus, the gauge coupling becomes complexified in SUSY theories. This 
fact has 
far-reaching consequences. 

Equivalence between Eqs. (\ref{susyg}) and (\ref{susyym}) is clear 
from Eq. 
(\ref{doublew}).
The $F$ component of $W^2$ includes the kinetic term of the
gaugino field (or gluino, I will use these terms indiscriminately),
$$
\bar \lambda {\cal D}_\mu\gamma^\mu \lambda\, ,
$$
and that of the gauge field,
$$
G_{\mu\nu}^a G_{\mu\nu}^a + i G_{\mu\nu}^a\tilde 
G_{\mu\nu}^a\, .
$$

Superficially the model looks very similar to
conventional QCD; the only difference is that the quark fields 
belonging
to the fundamental representation of the gauge group in QCD are 
replaced
by the gluino field belonging to the adjoint representation in 
supersymmetric 
gluodynamics. Like QCD, supersymmetric gluodynamics is a strong 
coupling 
non-Abelian
theory. Therefore, it is usually believed that

\vspace{0.2cm}

$\bullet$ only colorless asymptotic state exist;

$\bullet$ the Wilson loop (in the {\em fundamental} representation)
is subject to the area law (confinement);

$\bullet$ a mass gap is dynamically generated; all particles in the 
spectrum 
are massive. 

\vspace{0.2cm}

I would like to stress the word ``believe" since  the above features 
are 
hypothetical. Although the theory does indeed look pretty similar to 
QCD,
supersymmetry brings in remarkable distinctions  -- some 
quantities turn out to be exactly calculable. Namely, we know that 
the gluino condensate develops,
\beq
\langle \lambda^{a \alpha} \lambda^a_\alpha \rangle
=\mbox{const.} \times \Lambda^3 e^{2\pi ik/N_c}
\label{gcond}
\eeq
where $N_c$ is the number of colors ($SU(N_c)$ gauge group is 
assumed and the vacuum angle $\vartheta$ is set equal to zero),
$\Lambda$ is the scale parameter of supersymmetric gluodynamics,
$k$ is an integer ($k=0,1,...,N_c-1$), and the constant in Eq. 
(\ref{gcond})
is  {\em exactly} calculable \cite{NSVZ1,SV4}. A discrete $Z_{2N_c}$
symmetry of the model, a remnant of the anomalous $U(1)$,  is 
spontaneously broken by the gluino 
condensate \footnote{I hasten to add that it was argued recently
\cite{Kovner} that supersymmetric gluodynamics actually
has two phases: one with the spontaneously broken $Z_{2N_c}$
invariance, and another, unconventional, phase where
the chiral $Z_{2N_c}$ symmetry is unbroken and the gluino 
condensate does not develop. Dynamics of the chirally symmetric 
phase is drastically different from what we got used to in QCD.
In particular, although no invariance is spontaneously broken,
massless particles appear, and no mass gap is generated. This 
development is too fresh, however, to be included in this lecture 
course.

The existence of the gluino condensate was anticipated
\cite{VENY}, from the analysis of the
so called  Veneziano-Yankielowicz effective
Lagrangian, even prior to the first dynamical calculation
\cite{NSVZ1}. The Veneziano-Yankielowicz Lagrangian,
 very useful for orientation, is not a genuinely Wilsonean
construction, and one must deal with it  extremely cautiously
in extracting consequences. For a recent discussion see Ref. 
\cite{Kovner}.}
down to $Z_2$. Correspondingly, there are $N_c$ degenerate vacua, 
counted by 
the integer parameter $k$. Supersymmetry is unbroken -- all vacua 
have the
vanishing energy density.

Moreover, the Gell-Mann--Low function of the model, governing the 
running
of the gauge coupling constant, is also exactly calculable \cite{NSVZ2},
\beq
\beta (\alpha ) = - \frac{3N_c\alpha^2}{2\pi} \, \frac{1}{1 - 
N_c\alpha /(2\pi)}
\, .
\label{beta}
\eeq
 By ``exactly" I mean that all orders of perturbation theory are 
known, and 
one can additionally show that in the case at hand there are no 
nonperturbative contributions. 

Equations (\ref{gcond}) and (\ref{beta}) historically were the first
examples of non-trivial (i.e. non-vanishing) quantities exactly 
calculated
in four-dimensional field theories in the strong coupling regime.
These examples, alone, show that the supersymmetric gauge 
dynamics
is full of hidden miracles. We will encounter many more examples in 
what 
follows. Eventually, after learning  more about supersymmetric 
theories, you 
will be able to understand how Eqs. (\ref{gcond}) and (\ref{beta})
are derived. But this will take some time. Here I would like only to 
add an
explanatory remark regarding the vacuum degeneracy in 
supersymmetric 
gluodynamics. At the classical level Lagrangian (\ref{susyym})
has a U(1) symmetry corresponding to the phase rotations of the
gluino fields,
\beq
\lambda\ra e^{i\alpha}\lambda
\,.
\label{phtr}
\eeq
The corresponding current is sometimes called the
 $R_0$ current; it is a  superpartner of the 
energy-momentum tensor and the supercurrent.  The $R_0$ current 
exists in 
any supersymmetric theory.  
Moreover, in conformally invariant   theories -- and supersymmetric 
gluodynamics is conformally invariant at the classical level -- it  is 
conserved 
\cite{SSF}.
In the spinor notation the $R_0$ current has the form
$J_{\alpha\dot\alpha}= 
\bar\lambda_{\dot\alpha}\lambda_{\alpha}$,
while in the Majorana notation the very same current takes the form
$J_\mu =\bar\lambda\gamma_\mu\gamma_5\lambda$. (Let me 
parenthetically note that the vector current of the Majorana gluino 
identically 
vanishes. The proof of this fact is left as an  exercise.)
The conservation of the axial current above is broken by the
triangle anomaly, 
$$
\partial^\mu J_\mu = \frac{N_c}{16\pi^2} G_{\mu\nu}^a
\tilde G_{\mu\nu}^a\, .
$$
So, there is no continuous $U(1)$ symmetry in the 
model.
By the same token, the conformal invariance is ruined by the 
anomaly in the 
trace of the energy-momentum tensor. 
As a matter of fact, the divergence of the 
$R_0$ current and the trace of the energy-momentum tensor can be 
combined 
in one superfield \cite{Grisa}. 

However, a remnant of the would-be symmetry remains, in the form 
of the 
discrete phase transformations of the type (\ref{phtr}) with
$\alpha = \pi k /N_c$. The gluino condensate further breaks this 
symmetry to 
$Z_2$ corresponding to $\lambda\ra -\lambda$. The number of the 
degenerate 
vacuum states, $N_c$, coincides with Witten's index for the $SU(N_c)$ 
theory \cite{Witten},
an invariant which counts the number of the boson zero energy 
states 
minus the  number of  the fermion zero energy states. If Witten's 
index is 
non-vanishing supersymmetry cannot be spontaneously broken, of 
course. 

An interesting aspect, related to the discrete degeneracy of the 
vacuum states, is the $\vartheta$ dependence. What happens with 
the vacua if $\vartheta\neq 0$? The question was answered in Ref.
\cite{SV4}. The $\vartheta$ dependence of the gluino condensate is
\beq
\langle\lambda\lambda\rangle_\vartheta
=\langle\lambda\lambda\rangle_{\vartheta 
=0}e^{\frac{i\vartheta}{N_c}}\, .
\label{AD3}
\eeq
This shows that the $N_c$ vacua are intertwined as far as the
$\vartheta$ evolution is concerned. When $\vartheta$ changes 
continuously from 0 to $2\pi$ the first vacuum becomes second, the 
second becomes third, and so on, in a cyclic way.

\subsubsection{SU(2) SQCD with one flavor}

As the next step on a long road leading us to understanding of
supersymmetric 
gauge dynamics  we will consider SUSY generalization of SU(2) QCD
with the matter sector consisting of one flavor. This model will
serve us as a reference point in all further constructions.

Since the gauge group is SU(2) we have three gluons and three
superpartners -- gluinos. 

As far as the matter sector is concerned, let us remember that 
one quark flavor in  QCD is described by a Dirac field, a doublet with 
respect to 
the
gauge group. One Dirac field is equivalent to two chiral fields:
a left-handed and a right-handed, both transforming according
to the fundamental representation of SU(2). Moreover, the 
right-handed
doublet is equivalent to the left-handed anti-doublet, which in turn
is equivalent to a doublet. The latter fact is specific to the SU(2) 
group,
whose all representations are (pseudo)real. Thus, the Dirac quark 
reduces
to two left-handed Weyl doublet fields.

Correspondingly, in SQCD each of them will acquire a scalar partner.
Thus, the matter sector will be built from two superfields,
$S_1$ and $S_2$. In what follows we will use the notation
$S_f^\alpha$ where $\alpha =1,2$ is the color index, and $f= 1,2 $
is a ``subflavor" index. Two subflavors comprise one flavor.
The chiral superfield has the usual form, see Eq. (\ref{chifi}).

In the superfield language the Lagrangian of the model can be 
represented in 
a very concise form
\beq
{\cal L} = \frac{1}{2g^2_0}\, \mbox{Tr}\,
\int d^2 \theta W^2 + 
\frac{1}{4} \int d^2\theta d^2\bar\theta
\bar S_f e^V S_f +
\left( \frac{m_0}{4}\int d^2\theta S^{\alpha f}S_{\alpha f}
+\mbox{ H. c.}\right)\, ,
\label{su2lagr}
\eeq
where the superfields $V$ and $W_\alpha$ are matrices in the color 
space,
for instance, $V\equiv V^a\tau^a/2$, with $\tau^a$ denoting the Pauli 
matrices. The subscript 0 indicates that the mass parameter and the 
gauge 
coupling constant are bare parameters, defined at the ultraviolet cut 
off. 
In what follows we will omit this subscript to ease the notation in 
several 
instances where
it is unimportant.

If we take into account the  rules of integration over the Grassmann 
numbers
we immediately see that
the integral over $d^2\theta$ singles out the $\theta^2$ component
of the chiral superfields $W^2$ and $S^2$, i.e. the   $F$ terms.  
Moreover, the
integral over $d^2\theta d^2\bar \theta$ singles out the
$\theta^2\bar \theta^2$ component of the real superfield $\bar 
Se^VS$, i.e. the 
 $D$ term.

Note that the SU(2) model under consideration, with {\em one flavor}
possesses a global SU(2) (``subflavor") invariance
allowing one to freely rotate the superfields $S_1 \leftrightarrow 
S_2$.
This symmetry holds even in the presence of the mass term, see Eq. 
(\ref{su2lagr}), and is specific for SU(2) gauge group, with
its pseudoreal representations. All indices corresponding to the
SU(2) groups (gauge, Lorentz and subflavor) can be lowered and 
raised by 
means of the $\varepsilon$ symbol, according to the general rules.

The Lagrangian presented in Eq. (\ref{su2lagr}) is not generic.
Renormalizable models with a richer matter sector usually allow 
for one more type of $F$ terms, namely
$$
\int d^2\theta S^3\, .
$$
These terms are called the Yukawa interactions, since one of the
vertices they include corresponds to a coupling of two spinors
to a scalar. Strictly speaking, they should be called
the super-Yukawa terms, since  spinor-spinor-scalar vertices
arise also in the (super)gauge parts of the Lagrangian. This jargon is 
widely 
spread,
however; eventually you will get used to it and learn how to avoid 
confusion.
The combination of the $F$ terms 
$S^2+ S^3$ is generically referred to as superpotential. The 
conventional 
potential of self-interaction of the scalar fields stemming from the 
given
superpotential is referred to as {\em scalar potential}.

It is instructive to pass from the superfield notations to components.
We will do this exercise now in some detail, putting emphasis on 
those 
features
which are instrumental in the solutions to be discussed below. Once 
the 
experience is accumulated the  need in the component notation
will subside.

Let us start from $W^2$. The corresponding $F$ term was already 
discussed in 
Sect. 2.3.1. There is one new important point, however. In Sect. 2.3.1 
we 
omitted the square of the $D$ term present in $W^2|_F$, see Eq. 
(\ref{doublew}),
\beq
\Delta{\cal L}^{(W)}_D =
\frac{1}{2g^2}D^a D^a \, .
\label{ld}
\eeq
If the matter sector of the theory is empty, this term is unimportant.
Indeed, the $D$ field enters with no derivatives, and, hence, can be 
eliminated 
from the Lagrangian by virtue of the equations of motion. With no 
matter
fields $D=0$. In the presence of the matter fields, however, 
eliminating $D$ we get a non-trivial term constructed from the scalar 
fields,
which is of a paramount importance. This point will be discussed 
later; here 
let me only note that the sign of $D^2$ in the Lagrangian, Eq. 
(\ref{ld}),
is unusual, positive.

The next term to be considered is $\int d^2\theta d^2\bar\theta\bar 
Se^V S$.
Calculation of the $D$ component of $\bar Se^V S$ is a more 
time-consuming 
exercise since we must take into account the fact that $S$ depends 
on $x_L$
while $\bar S$ depends on $x_R$; the both arguments differ from 
$x$.
Therefore, one has to expand in this difference. The factor $e^V$ 
sandwiched 
between $\bar S$ and $S$ covariantizes all derivatives. Needless to 
say
that the field $V$ is treated in the Wess-Zumino gauge.  It is not 
difficult to 
check that
$$
\frac{1}{4}
\int d^2\theta d^2\bar\theta\bar S_fe^V S_f
= \bar\psi_f i\not\!\!{\cal D} \psi_f
-\phi^\dagger_f{\cal D}^2 \phi_f +
$$
\beq
\left[ (\psi_f\lambda )\phi_f^\dagger +\mbox{H.c.}\right]
+ D^a \phi^\dagger_f T^a \phi_f
\label{barss}
\eeq
where $T^a$ are the matrices of the color generators. In the SU(2) 
theory 
$T^a=\tau^a/2$. Now we see why the 
$D^2$ term is so important in the presence of matter; $D^a$ does not 
vanish 
anymore. Moreover, using the equation of motion
we can express $D^a$ in terms of the squark fields, generating in this 
way
a quartic self-interaction of the scalar fields,
\beq
V_D =\frac{1}{2g^2} D^a D^a\, , \,\,\,\, D^a = -\frac{g^2}{2}
\left(\phi_1^\dagger\tau^a\phi_1 +
\phi_2^\dagger\tau^a\phi_2\right)\, .
\label{vd}
\eeq

In the old-fashioned language of the pre-SUSY era one would call the 
term
$(\psi_f\lambda )\phi_f^\dagger$ from Eq. (\ref{barss}) the Yukawa 
interaction. The SUSY practitioner would refer to this term as to the 
gauge 
coupling since it is merely a supersymmetric generalization of the
quark-quark-gluon coupling. I mention these terms here
because later on their analysis will help us establish the form of the 
conserved $R$ currents. 

\subsubsection{Vacuum  valleys}

Let us examine the $D$ potential $V_D$ more carefully, neglecting 
for the time 
being $F$  terms altogether. As well-known, the  energy of any state 
in any 
supersymmetric theory is positive-definite. The minimal energy 
state,
the vacuum, has energy exactly at zero. Thus, in determining the 
classical 
vacuum we must find all field configurations corresponding to 
vanishing 
energy. From Eq. (\ref{vd}) it is clear that in the Wess-Zumino
gauge the {\em classical space of vacua} (sometimes called
the moduli space of vacua) is defined by the $D${\em -flatness 
condition}
\beq
D^a = 0\,\,\, \mbox{for all}\,\, a  \, .
\label{dflat}
\eeq
More exactly, Eq. (\ref{dflat}) is called the Wess-Zumino gauge $D$ 
flatness 
condition. Since this gauge is always implied, if not stated to the
contrary, we will omit the reference to the Wess-Zumino gauge.
 
The $D$ potential $V_D$  represents
a quartic self-interaction of the scalar fields, of a very peculiar form.
Typically in the $\phi^4$ theory the potential has one -- at most 
several --
minima. In other words, the space of the vacuum fields 
corresponding to minimal energy, is a set of isolated points. The only 
example 
with a continuous manifold of points
of minimal energy which was well studied previously
is the spontaneous breaking of a global  continuous symmetry, say, 
U(1) 
(Sect. 1). In this case all points belonging to this vacuum manifold 
are 
physically equivalent. The $D$ potential (\ref{vd}) has a specific 
structure --
the minimal (zero) energy is achieved along entire directions 
corresponding to 
the solution of Eq. (\ref{dflat}). It is instructive to think of the 
potential 
as of a mountain ridge; the $D$ flat directions then present the flat 
bottom
of the valleys. Sometimes, for transparency,  I will call the $D$ flat  
directions 
{\em the vacuum valleys}. Their existence was first
noted in Ref. \cite{BDSF}. As we will see, different points belonging 
to  the 
bottom
of the valleys are physically inequivalent. This is a remarkable 
feature of the
supersymmetric gauge theories.
 
In the case of the SU(2) theory with one flavor  it is not difficult to 
find
the $D$ flat direction explicitly. Indeed, 
consider the scalar fields of the form
\beq
\phi_1 = v \left( \begin{array}{c}
1 \\  0
\end{array} \right)\, ,\,\,\, 
\phi_2 = v \left( \begin{array}{c}
0 \\  1
\end{array} \right)\, ,
\label{dflat2}
\eeq
 where $v$ is an arbitrary complex constant. It is obvious that
for any value of $v$ all $D^a$'s vanish. $D^1$ and $D^2$ vanish 
because
$\tau^{1,2}$ are off-diagonal matrices; $D^3$ vanishes after 
summation over
two subflavors. 

It is quite obvious that if  $v\neq 0$ the original  gauge symmetry 
SU(2)
is totally spontaneously broken.
Indeed, under the condition (\ref{dflat2}) all three gauge bosons
acquire masses $\sim gv$. Thus, we deal here with the 
supersymmetric
generalization of the Higgs phenomenon. Needless
to say that supersymmetry is not broken. It is instructive
to trace the reshuffling of the degrees of freedom before and after 
the
Higgs phenomenon. In the unbroken phase, corresponding to $v=0$,
we have three massless gauge bosons (6 degrees of freedom),
three massless gaugino (6 degrees of freedom), four matter
fermions (the Weyl fermions, 8 degrees of freedom), and
four matter scalars (complex scalars, 8 degrees of freedom).
In the broken phase three matter fermions combine with
the gauginos to form three massive Dirac fermions (12 degrees of 
freedom).
Moreover, three matter scalars combine with the gauge fields
to form three {\em massive} vector fields (9 degrees of freedom) 
plus three 
massive (real) scalars.
What remains massless? One complex scalar field,
corresponding to the motion along the  bottom of the valley, $v$,
and its fermion superpartner, one Weyl fermion. The
balance between the fermion and boson degrees of freedom is 
explicit.

A gauge invariant description of the system of the vacuum valleys 
was
suggested in Refs. \cite{ADS1,ADS2} (see also \cite{BDSF}). In these 
works it 
was noted that
the set of  proper coordinates parametrizing the space of the classical 
vacua is 
nothing else but the set of all independent (local) products of the 
chiral matter 
fields existing in the theory. Since this point is very important let me
stress once more that the  variables to be included in the set are
polynomials built 
from the fields of one and the same chirality {\em only}. These 
variables are 
clearly gauge invariant.

At the intuitive level this assertion is almost obvious. Indeed, if there
is a $D$ flat direction, the motion along the degenerate bottom of the
valley must be described by some effective (i.e. composite) chiral 
superfield, 
which is gauge invariant and has  no  superpotential.  The 
opposite is also true. If we are able to build
some chiral gauge invariant (i.e. colorless) superfield $\Phi$, as a 
local product
of the chiral matter superfields of the theory at hand, then the 
energy is 
guaranteed to vanish, since (in the absence of the $F$ terms)
all terms which might appear in the effective Lagrangian for $\phi$ 
necessarily  contain
derivatives. Here $\phi$ is the lowest component of the 
above
superfield $\Phi$. In 
other words, then, changing the value of $\phi$ we will be moving 
along the 
bottom of the valley. 

A formal proof of the fact that the classical vacua are fully described 
by the
set of local (gauge invariant) products of the chiral fields comprising 
the 
matter sector is given in the recent work \cite{Luty1} which 
combines 
and extends results scattered in the literature \cite{BDSF,ADS1,6,7}.

The approach based on the chiral polynomials is very convenient
for establishing the fact of the existence (non-existence)
of the moduli space of the classical vacua, and in counting the
dimensionality of this space. For instance, in the SU(2) model with 
one flavor there exists only one invariant,
$S^2\equiv S_{\alpha f} S_{\beta g} 
\epsilon^{\alpha\beta}\epsilon^{fg} = 
2v^2$.
Correspondingly, there is only one vacuum valley -- one-dimensional 
complex 
manifold. The remaining three (out of four) complex scalar fields are
 eaten up in the super-Higgs phenomenon by the vector fields,
which immediately tells us that a generic point from the bottom of 
the valley
corresponds to fully broken gauge symmetry.

In other cases we will have a richer structure of the moduli space of 
the 
classical vacua. In some instances no chiral invariants can be 
built at all. Then the $D$ flat directions are absent.

If the $D$ flat directions exist, and the gauge symmetry is 
spontaneously
broken, then the constraints of the type of 
Eq. (\ref{dflat2}) can be viewed as a gauge fixing condition. This is 
nothing 
else but the unitary gauge in SUSY. Those components of the matter 
superfields which are set equal to zero are actually eaten up by the
vector particles which acquire the longitudinal components through 
the
super-Higgs mechanism. 

Although constructing the set of the chiral invariants is helpful
in the studies of the general properties of the space of the  classical 
vacua,
sometimes it is still necessary to explicitly parametrize  the vacuum 
valleys,
just in the same way as it is done in Eq. (\ref{dflat2}).  As we have 
seen, this 
problem is trivially
solvable in the SU(2) model with one flavor. For higher groups and 
representations the general situation is much more complicated,
and the generic solution is not found.
Many useful tricks for finding explicit parametrization of the
vacuum valleys in particular examples were suggested in Refs. 
\cite{BDSF,ADS1,ADS2}. A few  simplest examples are considered 
below.
More complicated instances are considered in the literature.
For instance, a parametrization of the valleys in the
SU(5) model with two quintets and two antidecuplets was given in 
Ref. 
\cite{SVZ1} and  in the E(6) model with the 27-plet in Ref. 
\cite{Koga1}.
The  correspondence between the explicit parametrization of the $D$ 
flat
directions and the chiral polynomials was discussed recently more 
than once,
see e.g. \cite{Seib1} -- \cite{Gidd1}. I would like to single 
out Ref.
\cite{Gher1} where a catalog of the flat directions in the minimal 
supersymmetric standard model (MSSM) was obtained by analyzing 
all 
possible chiral polynomials and eliminating those of them which are 
redundant.

Once the existence of the $D$-flat directions is established at the 
classical level
one may be sure that a manifold of the degenerate vacua
will survive at the quantum level, provided no $F$ terms appear
in the action which might lift the degeneracy. Indeed, in this case the
 only impact of the quantum corrections
is providing an overall $Z$ factor in front of the kinetic  term, which 
certainly 
does not affect the vanishing of the $D$ terms.  The $F$ terms which 
could lift 
the degeneracy
must be either added in the action by hand (e.g. mass terms), or 
generated
nonperturbatively. A remarkable non-renormalization theorem
\cite{GRS} guarantees that no $F$ terms can be generated
perturbatively. We will return to the discussion of this
{\em second miracle} of SUSY in Sect. 2.4.

In this respect the 
supersymmetric theories are fundamentally different from
the non-supersymmetric ones. Say, in the good old $\phi^4$ theory 
with the 
Yukawa
interaction
$$
{\cal L} = |\partial_\mu\phi |^2 + \bar\psi 
i\partial_\mu\gamma^\mu\psi
+ (g\phi\bar\psi\psi +\mbox{H. c.})
$$
we could also assume that that the mass and self-interaction of the
scalar field vanish at the classical level. Then, classically, we will 
have a flat
direction -- any constant value of $\phi$ corresponds to the 
vanishing vacuum 
energy. However, this vacuum valley does not survive inclusion of 
the
quantum corrections. Already at the one-loop level both the mass 
term of the
scalar field, and its self-interaction, will be generated, and the 
continuous 
vacuum degeneracy will inevitably disappear.

\subsubsection{In  search of the valleys}

Although our excursion in the SU(2) model with one flavor is not yet 
complete,
the issue of the $D$ flat directions is so important in this range of 
problems
that
we pause here to do, with pedagogical purposes, a few  simple 
exercises. If you choose to skip this section in the first
reading it will be necessary to return to it later.

\vspace{0.2cm}

{\em $SU(N_c)$ model with $N_f$ flavors ($N_f < N_c$)} \cite{ADS1}

\vspace{0.1cm}

The matter sector includes $2N_f$ subflavors --
$N_f$ chiral fields in the fundamental representation
of $SU(N_c)$, $S^{\alpha f}$, and $N_f$ chiral fields in the 
antifundamental 
representation, $\tilde S_{\alpha f}$, where $\alpha = 1, ..., N_c$ and 
$f=1,... 
N_f$.
It is quite obvious that one can form $N_f^2$ chiral products of the
type
\beq
\tilde S_{\alpha}^fS^{\alpha g}\, ,\,\,\, f,g = 1,..., N_f\, .
\label{nf}
\eeq
All these chiral invariants are independent. Thus, the moduli space
of the classical vacua (the vacuum valley) is a complex manifold of 
dimensionality $N_f^2$,
parametrized by the coordinates (\ref{nf}). A generic point from the 
vacuum 
valley corresponds to spontaneous breaking of $SU(N_c)\ra SU(N_c-
N_f)$ 
(except for the case when $N_f=N_c-1$, when the original gauge 
group
is completely broken).
The number of the broken generators is $2N_cN_f -N_f^2$; hence,
the same amount of the complex scalar fields are eaten up
in  the super-Higgs mechanism. The original number of the 
complex scalar fields was $2N_cN_f$. The remaining $N_f^2$ degrees
of freedom are the moduli (\ref{nf}) corresponding to the motion 
along the bottom of the valley. 

In this particular problem it is not difficult to indicate a concrete 
parametrization of the vacuum field configurations. Indeed,
consider a set
\beq
S_1= \tilde S_1^\dagger = v_1 \left( \begin{array}{c}
1 \\  0 \\ 0 \\ ... \\ 0
\end{array} \right)\, ,
S_2 =\tilde S_2^\dagger = v_2 \left( \begin{array}{c}
0 \\  1 \\ 0 \\ ... \\ 0
\end{array} \right)\, , \,\,  ...\,\,  ,
S_{N_f} =\tilde S_{N_f}^\dagger = v_{N_f} \left( \begin{array}{c}
0 \\  0 \\... \\ 1 \\ ... \\ 0
\end{array} \right)\, 
\label{dflat3}
\eeq
where the unity in $S_{N_f}$ occupies the $N_f$-th line and 
$v_{1,2,..., N_f}$
are $N_f$ arbitrary complex numbers. It is rather  obvious that for 
this
particular set all $D$ terms vanish. In verifying this assertion it is 
convenient
to consider first those $D^a$'s which lie outside the Cartan subalgebra 
of
$SU(N_c)$. Since the corresponding $T^a$ matrices are off-diagonal 
each term 
in the sum
$\sum_f(S^\dagger T^a S+\tilde S^\dagger T^a\tilde S) $ 
vanishes individually. For the generators
from the Cartan subalgebra the fundamentals and anti-fundamentals 
cancel 
each other.

The point (\ref{dflat3}) is {\em not} a generic point from the bottom 
of the 
valley. This is clear from the fact that it is parametrized by only 
$N_f$ complex 
numbers.
To get a generic solution one observes that the theory is invariant
under the global $SU(N_f)\times SU(N_f)$ flavor rotations (the 
fundamentals
and antifundamentals can be rotated separately). On the other hand, 
the 
solution
(\ref{dflat3}) is not invariant. Therefore, we can apply 
a general $SU(N_f)\times SU(N_f)$  rotation to Eq. (\ref{dflat3}) 
without 
destroying the condition
$D^a=0$. It is quite obvious that the generators belonging to the 
Cartan 
subalgebra of $SU(N_f)\times SU(N_f)$  do not introduce new 
parameters.
The remaining rotations introduce $N_f^2 - N_f$ complex parameters,
to be added to $v_1$, ...,$v_{N_f}$, altogether $N_f^2$ parameters, as 
it was 
anticipated from counting the number of the chiral invariants.

\vspace{0.2cm}

{\em $SU(N_c)$ model with $N_f$ flavors ($N_f =  N_c +1$)}

\vspace{0.1cm}

The vacuum valley is parametrized by $N_f^2$ complex parameters,
although the number of the chiral invariants is larger,
$N_f^2 + 2N_f$. Not all chiral invariants are independent.
For further details see Sect. 3.1.

\vspace{0.2cm}

{\em $SU(5)$ model with one quintet and one (anti)decuplet}

\vspace{0.1cm}

This gauge model describes  Grand Unification,
with one generation of quarks and leptons. 
This is our first example of  non-chiral matter;
it is singled out historically -- the
 instanton-induced dynamical supersymmetry breaking was 
first found in this model \cite{SU5}.
The quintet field is
$V^\alpha$, the (anti)decuplet field is antisymmetric 
$X_{\alpha\beta}$.
It is quite obvious that there are no chiral invariants at all. Indeed, 
the only 
candidate, $VVX$, vanishes due to antisymmetricity of 
$X_{\alpha\beta}$.
This means that no $D$ flat directions exist. The same conclusion can 
be reached by explicitly parametrizing $V$ and $X$; inspecting then 
the $D$-flatness conditions one can conclude that they have no 
solutions, see e.g. Appendix A in Ref. \cite{Amati}.

\vspace{0.2cm}

{\em $SU(5)$ model with two quintets and two (anti)decuplets
and no superpotential}

\vspace{0.1cm}

This  model (with  a small tree-level superpotential term)  was the 
first 
example of the instanton-induced
supersymmetry breaking  in the weak coupling regime \cite{SU52}.
It presents another example of the anomaly-free chiral 
matter sector. Unlike the one-family model (one quintet and one 
antidecuplet) flat directions do exist (in the absence of 
superpotential). The system of the vacuum valleys in the two-family 
$SU(5)$  model  was analyzed in Ref.
\cite{SVZ1}. 
Generically, the gauge $SU(5)$
symmetry is completely broken, so that 24 out of 30 chiral matter 
superfields are eaten up in the super-Higgs mechanism. Therefore, 
the vacuum valley should be parametrized by six complex moduli.

Denote two quintets present in  the 
model as
$V_f^\alpha$ $(f=1,2)$, and two antidecuplets as 
$(X_{\bar g})_{\alpha\beta}$ where ${\bar g}=1,2$  and 
the matrices $X_{\bar g}$ are antisymmetric in color indices 
$\alpha ,\beta$. Indices $f$ and ${\bar g}$ reflect the 
$SU(2)_X \times SU(2)_V$
flavor symmetry of the model.
Six independent chiral 
invariants are
$$
M_{\bar g}=V_k X_{\bar g} V_l \epsilon^{kl}\,  ,
$$
$$
B_{{\bar g} f}=X_{\bar g} X_{\bar k} X_{\bar l} V_f 
\epsilon^{{\bar k}{\bar l}} \, ,
$$
where the gauge indices in the first line are convoluted in a 
straightforward manner $V^\alpha X_{\alpha \beta} V^\beta$, while 
in 
the second line one uses the
$\epsilon$ symbol, 
$$
X_{\bar g} X_{\bar k} X_{\bar l} V_f =
\epsilon^{\alpha\beta\gamma\delta\rho}
(X_{\bar g})_{\alpha\beta}(X_{\bar k})_{\gamma\delta}(X_{\bar
l})_{\rho\kappa}
(V_f)^\kappa\, .
$$
The choice of invariants above  implies
 that there are no moduli transforming as 
$\{4,\,2\}$ under the flavor group (such moduli vanish).

In this model the explicit parametrization of the valley is far from 
being obvious, to put it mildly. The most convenient strategy of the 
search is analyzing
the five-by-five matrix
\beq
D^{\alpha}_{\beta} = V^{\alpha}_f {\bar V}^f_{\beta} - 
2({\bar X}^{\bar g})^{\alpha\gamma}(X_{\bar g})_{\gamma\beta}
\label{55M}
\eeq
where ${\bar V}= V^\dagger$ and ${\bar X} = X^\dagger$.
If this 
matrix 
is proportional to the unit one, the vanishing of the $D$ terms is 
guaranteed.  (Similar strategy based on analyzing analogs of 
Eq. (\ref{55M}) is applicable in other cases as well). 

 A  solution of the $D$-flatness condition  which contains 7 real
parameters 
looks as follows:
$$
V_1 = \left(
\begin{array}{c}
a_1 \\ 0 \\ 0 \\a_4 \\ 0
\end{array}\right)\, , \,\,\,
V_2 = \left(
\begin{array}{c}
0 \\ a_2 \\ 0 \\ 0 \\ 0
\end{array}\right)\, , 
$$
\beq
X_1 =\frac{1}{\sqrt{2}}
\left(
\begin{array}{ccccc}
0 & 0 & 0 & s & 0 \\
0 & 0 & b & 0 & 0 \\
0 & -b & 0 & 0 & f\\
-s & 0 & 0 & 0 & 0 \\
0 & 0 & -f  & 0 & 0 
\end{array}\right)\, , \,\,\,
X_2 =\frac{1}{\sqrt{2}}
\left(
\begin{array}{ccccc}
0 & d & 0 & 0 & g \\
-d & 0 & 0 & 0 & 0 \\
0 & 0 & 0 & 0 & 0 \\
0 & 0 & 0 & 0 & h \\
-g & 0 & 0  & -h & 0 
\end{array}\right)\, , 
\label{MM6}
\eeq
where
$$
|a_1|^2 = r^2 \cos^2 \theta\, , \,\,\, 
|a_4|^2 =  r^2 \cos^2 \alpha \tan^2 \theta\, ,\,\,\,
|a_2|^2 = r^2 \tan^2\theta \frac{\cos^2\theta - 
cos^2\alpha}{\sin^2\theta + \cos^2\alpha }\, ,
$$
and
$$
|s|^2 = \frac{r^2 \cos^2 \alpha}{\cos^2 \theta {\sin^2\theta + 
\cos^2\alpha}}\, ,\,\,  |b|^2 = r^2 {\sin^2\theta + \cos^2\alpha}\, , 
\,\,
|f|^2 = r^2 \frac{\cos^2 \alpha}{\cos^2 \theta } {\frac{\cos^2\theta - 
cos^2\alpha}{\sin^2\theta + \cos^2\alpha }}\, ,
$$
\beq
|d|^2 = \frac{r^2}{\cos^2 \theta } {(cos^2\theta - cos^2\alpha)}\, ,
\,\,\,
|g|^2 = r \cos^2\alpha\, , \,\,\, |h|^2 = r^2\sin^2\theta\, .
\label{MM7}
\eeq

Thus, the absolute values  of matrix elements are parametrized by 
three real
parameters 
$r$, $\alpha$ and $\theta$. Additional 4 parameters appear {\em 
via} phases 
$\delta$
of 9 elements $a_i$, $s$, $b$, $f$, $d$, $g$, $h$. Three phases, out of
nine, are related to gauge rotations and are not observable in the 
gauge
singlet sector. Additionally,  there are two constraints,
$$
\delta_1 - \delta_4 = \delta_h - \delta_g\, ,
$$
$$
\delta_f - \delta_b = \delta_g - \delta_d\, ,
$$
which are readily  derived from vanishing of the  off-diagonal terms
in $D^{\alpha}_{\beta}$. 

Substituting the above expressions  it is easy to check that invariants
$$
B_{{\bar g_1}{\bar g_2}{\bar g_3} f}=X_{\{{\bar g_1}} X_{\bar g_2} 
X_{{\bar g_3}\}} V_f  
$$
symmetrized over $g_1$, $g_2$, $g_3$  (i.e. the $\{4,\,2\}$ 
representation
of 
$SU(2)_X \times SU(2)_V$) do 
vanish,  indeed. 

The most general valley parametrization depends on 12 real 
parameters,
while so  far we have only 7. The remaining five parameters
are  provided by $SU(2)_X \times SU(2)_V$ flavor rotations
of the configuration (\ref{MM6}). 

\vspace{0.1cm}

\subsubsection{Back to the SU(2) model -- dynamics of the flat 
direction}

After this rather lengthy digression into the general theory of the
vacuum valleys we return to our simplest toy model, SU(2) with one 
flavor.
The vacuum valley in  this case is parametrized by one complex 
number, $v$,
which can be chosen at will since for any value of $v$ the vacuum 
energy 
vanishes. One can quantize the theory near any value of $v$. 
If $v\neq 0$, the theory splits into two sectors -- one containing 
massive 
particles which form SU(2) triplets, and another sector which 
includes only 
one massless Weyl fermion and one massless complex scalar field.
These massless particles are singlets with respect to both SU(2) 
groups -- color 
and subflavor.

So far we totally disregarded the mass term in the action, assuming 
$m=0$.
If $m\neq 0$ the corresponding term in the superpotential lifts
the vacuum degeneracy, making the bottom of the valley non-flat.
Indeed, 
$$ 
F = mv\, ,
$$
the corresponding contribution in the scalar potential
 is 
$$
\Delta V = |mv|^2\, ,
$$
which makes the theory ``slide down" towards the origin of the 
valley.
Since the perturbative corrections do not renormalize the
$F$ terms, this type of behavior --
sliding down to the origin of the former valley --
is preserved to any finite order
in perturbation theory.

What happens if one switches on nonperturbative effects? 

The 
non-renormalization theorem \cite{GRS}, forbidding the occurrence of 
the $F$ 
terms,
does not apply to nonperturbative effects, which, thus, may or may 
not 
generate relevant $F$ terms. The possibility of getting a 
superpotential
can be almost completely investigated  by analyzing the general 
properties of 
the model at hand, with no explicit calculations.  Apart from the 
overall
numerical constant, the functional form of the superpotential, if it is 
generated, 
turns out to be fixed.

Let me elucidate this point in more detail. First, on what variables 
can the 
superpotential depend?
The vector fields are massive and are integrated over.  Thus, we are 
left with 
the matter fields only, and the only chiral invariant is
\beq
I = S^{\alpha f} S_{\alpha f}\, .
\eeq
The superpotential, if it exists, must have the form
\beq
{\cal L}_F = \int d^2\theta f (I(x_L, \theta )) + \mbox{H.c.}
\eeq
where $f$ is some function. Notice that the mass term has just this 
structure,
with $f(z) = z$. We will discuss the possible impact of the mass term 
later,
assuming at the beginning that $m=0$.

Now, our task is to find the function $f$ exploiting the symmetry 
properties of 
the model. At the classical level there exist two conserved currents.
One of them, the $R_0$ current, is the superpartner of the 
energy-momentum tensor and the supercurrent
\cite{FEZU}. The divergence of 
the 
$R_0$ current and the trace of the energy-momentum tensor can be 
combined 
in one superfield. The $R_0$ current exists in any supersymmetric 
theory, 
and, 
moreover, in conformally invariant   theories it  is conserved.
Indeed, since the trace of the energy-momentum tensor vanishes 
in conformally invariant theories the 
divergence of $R_0$  vanishes as well.  In our present model the 
$R_0$ 
current
 corresponds to the following rotations of the fields
\beq
\lambda_\alpha \ra e^{i\beta}\lambda_\alpha\, , \,\,\,
\psi_\alpha^f \ra e^{-(i/3)\beta}\psi_\alpha^f \, , \,\,\,
\phi_\alpha^f \ra e^{(2i/3)\beta}\phi_\alpha^f \, .
\label{rnot}
\eeq
If we denote this current by $R_\mu^0$, then
\beq
R_\mu^0 = \frac{1}{g^2}\, 
\bar\lambda\gamma_\mu\gamma_5\lambda
-\frac{1}{3}\sum_f \left( \bar\psi^f\gamma_\mu\gamma_5\psi^f
-2 \phi^{f\dagger} \stackrel{\leftrightarrow}D_\mu \phi^f\right)\, .
\label{rnotcur}
\eeq

The relative phase between $\psi$ and $\phi$ is established in the 
following 
way. Let us try to add the $S^3$ term to the superpotential. 
In the  model at hand it is actually forbidden by the color gauge 
invariance,
but we ignore this circumstance, since 
the form of the $R_0$ current is general, and in other models the 
$S^3$ term 
is perfectly  allowed. It violates neither supersymmetry, nor the 
conformal 
invariance (at the classical level), which is obvious from the fact
that its dimension is 3. The $S^3$ term  in the 
superpotential produces $\phi\psi^2$ term in the Lagrangian. In this 
way
we arrive at the relative phase between $\psi$ and $\phi$ indicated 
in Eq. 
(\ref{rnot}). The relative phase between $\lambda$ and $\psi$
is fixed by  requiring the term $\lambda\psi \phi^\dagger$ in the 
Lagrangian
(the supergeneralization of the gauge coupling) to be invariant under 
the 
$U(1)$ transformation at hand. In the superfield language the last
two transformations in Eq. 
(\ref{rnot}) can be concisely written as
\beq
S^f \ra \exp \left(\frac{2i}{3}\beta \right) S^f\, , \,\,\, 
\theta_\alpha
\ra \exp (i\beta )\theta_\alpha \, .
\eeq

The second (classically) conserved current is built from the matter 
fields,
\beq
J_\mu = \sum_f \left( \bar\psi^f\gamma_\mu\gamma_5\psi^f + 
 \phi^{f\dagger} \stackrel{\leftrightarrow}D_\mu \phi^f\right)\, .
\label{matcar}
\eeq
The corresponding $U(1)$ transformation is
\beq
\psi_\alpha^f \ra e^{i\gamma}\psi_\alpha^f \, , \,\,\,
\phi_\alpha^f \ra e^{i\gamma}\phi_\alpha^f \, ,
\label{matter}
\eeq
or, in the superfield notation, $S^f (x_L,\theta )\ra \exp (i\gamma 
)S^f 
(x_L,\theta )$.

The conservation of both axial currents is destroyed by the quantum 
anomalies. At one-loop level
\beq
\partial_\mu R^0_\mu = 
\frac{5}{3}\frac{\alpha}{4\pi}G_{\mu\nu}^a\tilde G_{\mu\nu}^a\, , 
\,\,\, 
\partial_\mu J_\mu = 
\frac{\alpha}{4\pi}G_{\mu\nu}^a\tilde G_{\mu\nu}^a\, .
\label{anomc}
\eeq
One can form, however, one linear combination of these two currents,
\beq
R_\mu = R_\mu^0 - \frac{5}{3}J_\mu
\label{rcurrent}
\eeq
which is anomaly free and is conserved even at the quantum level.
The occurrence of a strictly conserved axial current, the so called $R$ 
current,
is a characteristic feature of many supersymmetric models. In what
follows we will have multiple encounters with the $R$ currents in 
various 
models. The one presented in Eq. (\ref{rcurrent}) was given in Ref.
\cite{ADS1}.

Combining both transformations, Eqs. (\ref{rnot}) and (\ref{matter}),
in the appropriate proportion, see Eq. (\ref{rcurrent}), we conclude 
that
the SU(2) model under consideration is strictly invariant
under the following transformation
\beq
S^f \ra e^{-i\beta}S^f\, , \,\,\, \theta_\alpha \ra e^{i\beta} 
\theta_\alpha\, .
\label{rinv}
\eeq
This $R$ invariance leaves us with a unique possible choice for the
superpotential
\beq
{\cal L}_F = \mbox{const.} \times \int d^2\theta \frac{1}{I(x_L, 
\theta )} 
\ra \mbox{const.} \times\int d^2\theta \frac{\Lambda^5}{S^{\alpha f}
S_{\alpha f}}
\label{superpot}
\eeq
where $\Lambda$ is a scale parameter of the model, and the factor
$\Lambda^5$ has been written out on the basis of dimensional 
arguments.

Whether the superpotential is actually generated, depends on the
value of the numerical constant  above. In principle, it could have 
happened
that the constant vanished. However, since no general principle 
forbids the
$F$ term (\ref{superpot}), the vanishing seems highly improbable.
And indeed, the direct one-instanton calculation 
in the weak coupling regime (i.e. $v^2\gg\Lambda^2$)
shows 
\cite{ADS1,SVZ1}
that this term is generated.

The impact of the term (\ref{superpot}) is obvious. The 
corresponding extra 
contribution to the self-interaction energy of the scalar field is
\beq
\Delta V = |F|^2 = 4\frac{\Lambda^{10}}{|v|^6}
\label{scpo}
\eeq
where the numerical constant is included in the definition of 
$\Lambda$.
Thus, we see that the instanton-generated contribution ruins the 
indefinite 
equilibrium along the bottom of the valley, pushing the theory away 
from the
origin. As a matter of fact, in the absence of the mass term, $m=0$,
the theory does not have any stable vacuum at all since the
minimal (zero) energy is achieved only at $|v|\ra\infty$. We 
encounter here 
an example
of the {\em run-away vacuum} situation. 

Switching on  the mass term blocks the exits from the valley. Indeed,
now
\beq
F  = mv - 2\frac{\Lambda^{5}}{v^3}\, ,
\eeq
and the lowest energy state shifts to  a finite value of $v$.
 It is easy to see that now there are  two points at the 
bottom of 
the former valley where the energy vanishes, namely
\beq
v^2 = \sqrt{2}\frac{\Lambda^{5/2}}{m^{1/2}}\,\,\,\mbox{and}\,\,\, 
v^2 = - \sqrt{2}\frac{\Lambda^{5/2}}{m^{1/2}}\, .
\label{2sfv}
\eeq
In other words, the continuous vacuum degeneracy is lifted,
and only two-fold degeneracy survives; the theory has two vacuum 
states.
The number of the vacuum states could have been anticipated from
a general argument based on Witten's index \cite{Witten}. 

I pause here to make a few remarks. First, we observe that the 
supersymmetric version of QCD dynamically has very little in 
common with 
QCD. Indeed, the chiral limit of QCD, when all quark masses are set
equal to zero, is non-singular -- nothing spectacular happens in this 
limit 
except that the pions become strictly massless. At the same time, in 
the 
supersymmetric SU(2) model at hand the limit of the massless 
matter fields 
results in the run-away vacuum. This situation is quite general, and 
takes 
place in many
models, although not in all. 

Second, the analysis of the dynamics of the flat directions presented 
above
is somewhat simplified. Two subtle points deserve mentioning. 
The general form of the superpotential compatible with the 
symmetry
of the model was established in the massless limit.
In  this way we arrived at Eq. (\ref{superpot}).  The mass term was 
then
introduced to avoid the run-away vacuum. If $m\neq 0$ the $R$
current is not conserved any more, even at the classical level. To 
keep the 
invariance (\ref{rinv}) alive one must simultaneously rotate the
mass parameter,
\beq
m\ra  e^{4i\beta} m\, .
\eeq
Let us call this invariance, supplemented by the phase rotation of the
mass parameter, an {\em extended R symmetry}.

One could think of $m$ as of a vacuum expectation value
of some auxiliary chiral field, to be rotated in a concerted way in 
order
to maintain the $R$ invariance. (We will discuss this trick later on in 
more 
detail). It is clear then that multiplying Eq. (\ref{superpot}) by any 
function
of the dimensionless complex parameter $\sigma =mS^4/\Lambda^5$ 
is not 
forbidden by the extended symmetry. Extra arguments are needed to 
convince oneself that this additional function
actually does not appear. Let us assume it does. Then it should be 
expandable 
in the Laurent series of the type
$$
\sum_n\frac{C_n}{\sigma^n}\, .
$$
If negative  powers of $n$ were present then the function would 
grow
at large $|\sigma|$, a behavior one can immediately reject on  
physical 
grounds.
The masses of the heavy particles, which we integrate over to obtain 
the
superpotential, are proportional to $gv$; large values of $|v|$ imply 
heavier 
masses, which implies, in turn, that the impact on the 
superpotential should be weaker. Thus, all $C_n$'s with negative $n$
must vanish. Positive $n$ are not acceptable as well.
If positive  powers of $n$ were present then the function would blow 
off
at fixed $|v|$ and $m\ra 0$. At fixed $|v|$, however, no dynamically 
nontrivial 
singularity develops in the theory.  The only mechanism which could
provide  powers of $m$ in the denominator is 
a chain of instantons connected by one massless fermion line
depicted on Fig. 3. 
The corresponding contribution, however, is one-particle reducible
and should not be included in the effective Lagrangian.
This concludes our proof of the fact that Eq. (\ref{superpot}) is exact.

\begin{figure}
  \epsfxsize=12cm
  \centerline{\epsfbox{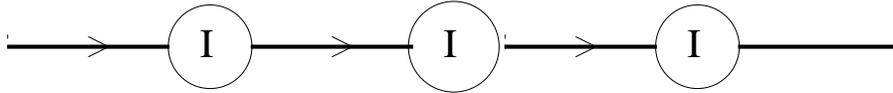}}
  \caption{ One-particle reducible graphs which might lead to
      $1/\sigma$ terms. This mechanism is not (and must not be) 
included
      in the superpotential. }
\end{figure}

The second subtle point is related to the discussion of the anomalies 
in the
$R_0$ and matter axial currents. The consideration presented above 
assumes
that both anomalies are one-loop. Actually, the anomaly in the $R_0$
current is multiloop \cite{SV2}. This fact slightly changes the form of 
the 
conserved $R$ current. The very fact of existence of the $R$ current 
remains intact. All expressions for the currents and charges 
presented above
refer to extreme ultraviolet where the gauge coupling (in 
asymptotically free theories) tends to zero. 
The  
final conclusion that the only superpotential
compatible with the symmetry of the model is that of Eq. 
(\ref{superpot})
is valid \cite{KSV}.

Thirdly, 
the consideration above (Eqs. (\ref{superpot}), (\ref{2sfv}))
strictly speaking, does not tell us what happens at the origin of the 
valley, $v^2 =0$, where all expressions become inapplicable.
Logically, it is possible to have an extra vacuum state characterized 
by $S^2=0$. This state would correspond to the strong coupling 
regime and will not be discussed here. The interested reader
is referred to Ref. \cite{Kovner}.

One last remark before concluding the section. Equation (\ref{scpo})
illustrates why different points from the vacuum valley are 
physically inequivalent. In the conventional situation of the 
pre-SUSY era, the spontaneous breaking of a global symmetry,
different vacua differ merely by a phase of $v$. Since physics 
depends on the ratio
$|\Lambda/v|$, this phase is irrelevant. In supersymmetric theories 
the vacuum valleys are typically non-compact manifolds. Different 
points are marked not only by the phase of $v$, but by its absolute 
value as well. The dimensionless ratio above is different in different 
vacua. In particular, if
$|\Lambda/v| \ll 1$ we are in the weak coupling regime;
if $|\Lambda/v| \sim 1$ we are in the strong coupling regime.

\subsection{Miracles of supersymmetry}

Two of many miraculous dynamical properties of SUSY have been 
already 
mentioned -- the vanishing of the vacuum energy and the 
non-renormalization theorem for $F$ terms. It is instructive to see 
how these 
features emerge in perturbation theory.

Let us start from the vacuum energy. Consider a typical two-loop 
(super)graph shown on Fig. 4. 

\begin{figure}   
  \epsfxsize=4cm
  \centerline{\epsfbox{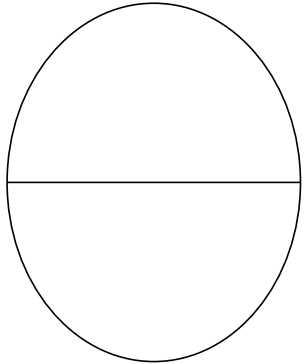}}
  \caption{A typical two-loop supergraph for the vacuum energy.}
\end{figure}

Each line on the graph represents  Green's function of some 
superfield. We do not even need to know what it is.  The crucial point 
is
that (if one works in the coordinate representation)  each interaction 
vertex
can be written as  an integral over $d^4x d^2\theta d^2\bar\theta$.
Assume that we substitute explicit expressions for Green's functions 
and
vertices in the integrand, and carry out the integration over the 
second vertex
keeping the first vertex fixed. As a result, we must arrive at an 
expression of 
the
form
\beq
\int d^4x d^2\theta d^2\bar\theta \times\mbox{ a function of}\, \, x\, 
,\theta\, 
,
\bar\theta \, .
\label{vacen}
\eeq
Since the superspace is homogeneous (there are no points
that are singled out, we can freely make translations, any point in 
the 
superspace is equivalent to any other  point) the function 
in Eq. (\ref{vacen}) can be only constant. If so, the result vanishes 
because of
the integration over the Grassmann variables $\theta$ and 
$\bar\theta$.

What remains to be demonstrated is that the one-loop vacuum 
graphs, not  representable in the form given on Fig. 4, 
also vanish. The one-loop 
(super)graph, however, is the same as for the free particles, and we 
know 
already that for free particles $E_{\rm 
vac}=0$, see Eq. (\ref{ffve}), thanks to the balance between the 
bosonic
and fermionic degrees of freedom. 

This concludes the proof of the fact that if the vacuum energy is
zero at the classical level it remains there to any finite order --
there is no renormalization. What changes if, instead of the vacuum 
energy,
we would consider renormalizations of the $F$ terms?

The proof presented above can be easily modified to include
this case as well. Technically, instead of the vacuum loops, we will 
consider 
now loop (super)graphs in a background field.

The basic idea is straightforward. In any supersymmetric theory 
there are
several -- at least four -- supercharge generators. 
In a generic background all supersymmetries are broken since the
background field is generically not invariant under  
supertransformations.
One can select such a background field, however, that leaves a part of 
the
 supertransformations as valid symmetries. For this specific 
background field 
some terms in the effective action will vanish, others will not. 
(Typically, $F$ 
terms do not vanish while $D$ terms do). The nonrenormalization 
theorems 
refer to those terms which do not vanish in the background field 
chosen.

Consider, for definiteness, the Wess-Zumino model \cite{WZM},
\beq
{\cal S}^{\rm WZ} = \frac{1}{4}\int d^4x d^2\theta d^2\bar \theta
\bar\phi \phi + \frac{1}{2}\left[ \int d^4 x d^2\theta\left( m\phi^2 + 
g\phi^3\right) +\mbox{H.c.}\right] \, .
\label{awz}
\eeq
An appropriate choice of the background field in this case is
\beq
\bar \phi_0 = 0\, , \,\,\, \phi_0 = C_1 +C_2^\alpha\theta_\alpha 
+C_3\theta^2\, 
,
\label{cbf}
\eeq
where $C_{1,2,3}$ are some constants and the subscript 0 marks
the background field. This choice assumes that $\phi$ and
$\bar\phi$ are treated as independent variables, not connected by 
the 
complex conjugation (i.e. we keep in mind a kind of analytic 
continuation).
The $x$ independent chiral field (\ref{cbf}) is invariant under the 
action of
$\bar Q_{\dot\alpha}$, i.e. under the transformations
$$
\delta\theta_\alpha = 0\, ,\,\,\, \delta\bar\theta_{\dot\alpha} =
\bar\varepsilon_{\dot\alpha} \, , \,\,\,
\delta x_{\alpha\dot\alpha} = - 2i 
\theta_\alpha\bar\varepsilon_{\dot\alpha} 
\, .
$$

Now, we proceed in the standard way -- decompose the superfields
$$
\phi = \phi_0 +\phi_{\rm qu}\, , \,\,\, \bar\phi = \bar\phi_0 
+\bar\phi_{\rm 
qu}\, ,
$$
where the subscript qu denotes the quantum part of the superfield, 
expand 
the action in $\phi_{\rm qu}, \bar\phi_{\rm qu}$, drop the linear 
terms and 
treat the remainder as the action for the quantum fields. We, then, 
integrate 
the quantum fields over, order by order, keeping the background 
field fixed.
The key element is the fact that in the problem we  get for the 
quantum fields
there still exists the exact symmetry under the transformations 
generated by
$\bar Q$.

This means that the boson-fermion degeneracy holds, just as in the 
``empty"
vacuum. All lines on the graphs of 
 Fig. 4 
have to be treated now as Green's 
functions in the 
background field (\ref{cbf}). After substituting these Green's 
functions and 
integrating over all vertices except the
first one we come to an expression of the type
$$
\int d^4 x d^2\bar\theta \times \,\, {\rm a} \,\,  \bar\theta
\,\,\mbox{independent 
function} = 0\, .
$$
The $\bar\theta$ independence follows from the fact that our 
superspace is
 homogeneous in the $\bar\theta$ direction even in the presence of 
the 
background field 
(\ref{cbf}). This completes the proof of the non-renormalization 
theorem for 
the 
$F$ terms.
Note that the kinetic term ($D$ terms) vanishes in the background
(\ref{cbf}), so nothing can be said about its renormalization (and it 
gets renormalized, of course). The above, somewhat non-standard, 
proof
of the Grisaru-Ro\v{c}ek-Siegel theorem was suggested in Ref. 
\cite{SV2}.

A word of caution is in order here. Our consideration tacitly assumes 
that
there are no massless fields which can cause infrared singularities.
Infrared singular contributions may lead to the so called {\em
holomorphic anomalies} \cite{SV3} invalidating the 
non-renormalization 
theorem. We will discuss the property of the holomorphy and the 
corresponding anomalies later, and now will illustrate how infrared 
singular
$D$ term renormalizations can effectively look as $F$ terms. Consider 
the
$D$ term of the form
$$
\int d^2\theta d^2\bar\theta \frac{D^2}{\Box} F(\phi )\, .$$
It can be rewritten as 
$$
 \int d^2\theta F(\phi )
$$
by using the property $[D^2 ,\bar D^2 ] \propto \Box$ and by 
integrating by 
parts in the superspace \cite{6}.  It is obvious that the $\Box^{-1}$
singularity can appear only due to massless poles. It was explicitly 
shown 
\cite{JJW}
that in the massless Wess-Zumino model such ``fake" $F$ terms 
appear
at the two-loop level. The origin of the two-loop and all higher order 
terms
in the Gell-Mann-Low function of supersymmetric gauge theories is 
the same 
-- they emerge as a ``fake" $F$ term which is actually an infrared 
singular
$D$ term \cite{SV2}. 

A recent discussion of the ``fake" $F$ terms is given in Ref. 
\cite{Lisa}.

\subsection{Holomorphy}

At least some of the miracles of supersymmetry can be traced back 
to
a remarkable property which goes under the name holomorphy.
Some parameters in SUSY Lagrangians,
usually associated with $F$ terms, are complex rather than real
numbers. Mass parameter in the superpotential is an obvious 
example.
Another example mentioned in Sect. 2.3.1 is the inverse
gauge coupling, $1/g_0^2$. Now, it is known for a long time, since the 
mid-eighties, that appropriately chosen quantities depend on these 
parameters analytically, with possible singularities in certain 
well-defined 
points. It is obvious that the statement that a function (analytically) 
depends 
on a complex variable is infinitely stronger than the statement that a 
function 
just depends on two real parameters. The power of holomorphy is 
such 
that one can obtain a variety of extremely non-trivial results ranging 
from 
non-renormalization theorems to exact $\beta$ functions, the first 
time ever 
in dynamically non-trivial four-dimensional theories. 

In this section we will outline  basic steps, keeping in mind that the 
corresponding technology will be of use more than once in what 
follows.
Let us consider, as an example, SU(2) SQCD with one flavor, Sect. 
2.3.5.
I have already mentioned that $m_0$, the complex mass parameter 
in
the action, can be viewed as a vacuum expectation value
(of the lowest component)  of an auxiliary chiral superfield, let us call 
it $M$.
It is important that $M$ is singlet with respect to the gauge group,
and, thus, say, fermions from $M$ do not contribute to the triangle 
anomalies.
One can think of the corresponding degrees of freedom as of very 
heavy
particles. Then the only role of $M$ is to develop $\langle M \rangle 
\neq 0$
which obviously does not violate SUSY and provides the mass term.

The theory is strictly invariant under the following phase 
transformations
\beq
W_\alpha \ra e^{i\beta}W_\alpha\, ,\,\,\, S_f\ra e^{-
i\beta}S_f\, ,\,\,\,
M\ra e^{4i\beta}M\, ,\,\,\, \theta_\alpha \ra 
e^{i\beta}\theta_\alpha\, 
.
\label{extri}
\eeq
This is an extended $R$ invariance -- extended, because it takes 
place in the
extended theory with the chiral superfield $M$ introduced by hand.

It is rather clear that  one chiral superfield can depend
only on the expectation value of another superfield of the same 
chirality --
otherwise transformation properties under SUSY would be broken. 
Thus,
the expectation value of $W^2$ can depend only on that of $M$; 
$\bar M$
cannot be involved in this relation. Equation (\ref{extri}) then tells 
us that
\beq
\langle \mbox{Tr}\, W^2 \rangle =\mbox{const.}  (M)^{1/2}  .
\eeq
In other words, the gluino condensate
\beq
\langle \lambda\lambda \rangle \propto \sqrt{m_0}\, ,
\label{llcond}
\eeq
and this relation is exact as far as the $m_0$ dependence is 
concerned.
It holds for small $|m_0|$ when the theory is weakly coupled, as well 
as
for large $|m_0|$, when we are in the strong coupling regime. 

A similar assertion is valid regarding the vacuum expectation value 
of 
$S^2$. Now, Eq. (\ref{extri}) tells us that
\beq
\langle S^2 \rangle =\mbox{const.}
 (M)^{-1/2}  \, ,
\eeq
implying, in turn, that
\beq
\langle \phi^2 \rangle \propto 1/\sqrt{m_0}\, .
\label{ppcond}
\eeq

It is worth emphasizing that the exact dependence of the 
condensates on the 
mass parameter established above refers to the bare mass 
parameter.
If we decided to eliminate the bare mass parameter $m_0$ in favor 
of the
physical mass of the Higgs field $m$, we would have to introduce
the corresponding $Z$ factor which depends on $m$ in a complicated
 non-holomorphic way.

Equations (\ref{llcond}) and (\ref{ppcond}), first derived in Ref. 
\cite{SV4} (a similar argument was also given in Ref. 
\cite{Amati}), 
lead to far reaching consequences. Indeed, since the functional 
dependence of 
the
condensates is fully established, we can calculate the relevant 
constant
at small $|m_0|$, when $\langle\phi^2\rangle $ is large, which 
ensures 
weak coupling regime.  I remind that the masses of the gauge bosons 
in this limit are
proportional
to 
$$
M_V^2 \propto |g^2\langle\phi^2\rangle |.
$$
The result will still be valid for large $|m_0|$,
in the strong coupling regime! This line of reasoning \cite{SV4}, 
based on 
holomorphy,  lies behind many advances achieved recently in SUSY 
gauge 
dynamics.

Let me parenthetically note that a nice consistency check to Eqs. 
(\ref{llcond}),  
(\ref{ppcond}) is provided by the so called Konishi anomaly 
\cite{Konishi1}.
In the model at hand the Konishi relation takes the form
\beq
\bar D^2\bar S^{\alpha f} e^VS^{\alpha f} = 4m_0 S_{\alpha f} 
S^{\alpha f} 
+\frac{1}{2\pi^2}\mbox{Tr}\, W^2 \, .
\label{konanom}
\eeq
This expression, or more exactly, the second term on the right hand 
side,  is 
nothing  but a supergeneralization of a the triangle anomaly in 
the 
divergence of the axial current of the matter fermions, cf.
Eq. (\ref{anomc}). Now, if SUSY is unbroken, the expectation value of
the left-hand side must vanish, since the left-hand side is a full 
superderivative. This fact implies that
\beq
4m_0 \langle S_{\alpha f} S^{\alpha f} \rangle = -
\frac{1}{2\pi^2}\langle \mbox{Tr}\, W^2 \rangle \,
\label{konia}
\eeq
which is consistent with Eqs. (\ref{llcond}),  (\ref{ppcond}). Another 
side 
remark: the exact  proportionality  of $\langle \phi^2 \rangle $   to 
$1/\sqrt{m_0}$ presents a somewhat different  proof of the fact that 
the 
instanton-generated superpotential (\ref{superpot}) is exact even in 
the 
presence
of the mass term, see Sect. 2.1.5.

One can take advantage of these observations in many ways. One
direction is finding  the exact $\beta$ function of the theory.
The idea is as follows. First we assume that $|m_0|$ is small and we 
are in the 
weak coupling regime. Then we are able to calculate the gluino 
condensate --
 in the weak coupling regime it is saturated by the one-instanton 
contribution. 
Since the functional dependence on $m_0$ is known we can
then proceed to the limit of large $|m_0|$, or small $\phi^2$.
Moreover, the vacuum expectation value of $\lambda\lambda $ is, in 
principle, a physically measurable quantity. The operator 
$\lambda\lambda $
has strictly vanishing anomalous dimension since it is the lowest 
component of 
the superfield $W^2$, and the upper component of the same 
superfield
contains the trace of the energy-momentum tensor. This means that
if $\langle \lambda\lambda\rangle $ is expressed in terms of the 
gauge 
coupling $g_0^2$ and the ultraviolet cut-off $M_0$,  when one 
changes the cut 
off, one should also change $g_0^2$ in a concerted way, to ensure
that $\langle \lambda\lambda\rangle $ stays intact.  In this way we
obtain a relation between the bare coupling constant and $M_0$, 
which is
equivalent to the knowledge of the $\beta$ function. 

More concretely, the one-instanton result for the gluino condensate
is \cite{NSVZ3}
\beq
\langle\lambda\lambda\rangle =
\mbox{const.}\times \frac{M_0^5e^{-8\pi^2/g_0^2}}{g_0^4 v_0^2}\, .
\label{instcond}
\eeq
This result is exact; only zero modes in the instanton background 
contribute in 
the calculation.
It is worth emphasizing that the expectation value of the scalar field 
appearing in Eq. (\ref{instcond}) refers to the bare field. The constant 
on the 
right-hand side is purely numerical; we will say more about this 
constant later 
on, but for the time being its value is inessential.

At the supersymmetric vacuum Eq. (\ref{konia}) must hold implying 
that
\beq
v_0^2 = \mbox{const.} \langle\lambda\lambda\rangle m_0^{-1} \, .
\label{vlm}
\eeq
Combining Eqs. (\ref{vlm}) and (\ref{instcond}) we conclude that
\beq
\langle\lambda\lambda\rangle =
\mbox{const.}M_0^{5/2}m_0^{1/2}e^{-4\pi^2/g_0^2}\frac{1}{g_0^2}\, .
\label{condeg}
\eeq

When analyzing the response of $g_0^2$ with respect to the 
variations of 
$M_0$ one should keep in mind that $m_0$ also depends on $M_0$, 
implicitly. Indeed, the physical (low-energy) values of the 
parameters are
kept fixed. This means that we fix the renormalized value of the 
mass,
$m = m_0Z^{-1}$, where $Z$ is the $Z$ factor renormalizing
the kinetic term of the matter fields, 
$$
(\bar S^{\alpha f} e^VS^{\alpha f})_0 \ra Z (\bar S^{\alpha f} 
e^VS^{\alpha 
f})_0\, 
.
$$
In this way we arrive at the conclusion that the combination
$
M_0^{5/2}e^{-4\pi^2/g_0^2}Z^{1/2}g_0^{-2 }
$
is invariant. Differentiating it with respect to $\ln M_0$ we find the
$\beta$ function,
\beq
\beta (\alpha) = -\frac{\alpha^2}{2\pi}\, \frac{5+\gamma }{1 -\alpha 
/\pi}\, ,
\label{su2beta}
\eeq
where the $\beta$ function is defined as
$$
\beta (\alpha_0) = {d\alpha_0}/{d\ln M_0}\, ,\,\,\, \alpha_0 
=g_0^2/4\pi \, ,
$$
and $\gamma$ is the anomalous dimension of the matter fields,
$$
\gamma =d\,\ln Z /d\, \ln M_0\, .
$$
Note that due to the SU(2) subflavor symmetry of the model at hand
both matter fields, $S_1$ and $S_2$, have one and the same 
anomalous 
dimension.

Equation (\ref{su2beta}) is a particular case of the general $\beta$ 
function,
sometimes referred to as NSVZ $\beta$ function,
\beq
\beta (\alpha ) =
-\frac{\alpha^2 }{2\pi}\left(1-\frac{T(G) \alpha}{2\pi}\right)^{-1}
\left[ 3T(G) -\sum_i T(R_i)(1-\gamma_i)\right]\, ,
\label{NSVZbeta}
\eeq
which can be derived in a similar manner \cite{NSVZ2}. Here $T(G)$ 
and 
$T(R)$ are the so called Dynkin indices defined as follows. Assume 
that
the gauge group is $G$, and we have a field belonging to the 
representation 
$R$ of the gauge group. If $T^a$ is the generator matrix of the group 
$G$ in 
the representation $R$ then
$$
\mbox{Tr} \, (T^aT^b) = T(R) \delta^{ab}\, .
$$
More exactly, $T(R)$ is one half of the Dynkin index. Moreover, 
$T(G)$ is
$T(R)$ for the adjoint representation. Note that for the fundamental 
representation of the unitary groups $T(R) =1/2$. The sum in Eq. 
(\ref{NSVZbeta}) runs over all subflavors. 

As is clear from its derivation, the NSVZ $\beta$ function
implies the Pauli-Villars regularization. It can also be derived
purely perturbatively, with no reference to instantons, using only
holomorphy properties of the gauge coupling \cite{VZS}. The relation 
of
this $\beta$ function to that defined in other, more conventional
regularization schemes is investigated in Ref. \cite{JJN}.

In some theories the NSVZ $\beta$ function is exact -- there are
no corrections, either perturbative or nonperturbative, as is the case
in the SU(2) model with one flavor. In other models it is exact only 
perturbatively -- nonperturbative corrections do modify it. 
The most important example \cite{Nathan} of the latter kind is $N=2$ 
supersymmetry.

The idea of using the analytic properties of chiral quantities for 
obtaining 
exact results
was  adapted for the case of  superpotentials in 
Ref. \cite{seibhol}. As a matter of fact, I have already discussed some
elements of the procedure  suggested in Ref. \cite{seibhol}, in Sect. 
2.3.5, in 
analyzing possible mass dependence 
of nonperturbatively generated superpotential in the SU(2) model. 
Let me 
summarize here
 basic stages of the procedure in the general form, and give a few 
additional  examples. Simultaneously, as a byproduct,
we will obtain a different proof of some of the non-renormalization 
theorems
considered in Sect. 2.4. What is remarkable is that, unlike the proof 
presented
in Sect. 2.4, the one given below will be   valid both perturbatively 
and 
nonperturbatively.

Thus, our task is establishing possible renormalizations of the 
superpotential 
${\cal W}$ in a given model. All (complex) coupling constants of the 
model 
appearing as coefficients in front of $F$ terms --
denote them generically by $h_i$ --  are treated as vacuum 
expectation values 
of some (auxiliary) chiral superfields. The set of $\{ h_i \}$ may 
include the 
mass parameters and/or Yukawa constants. Let us assume that if all
$h_i = 0$, the model considered possesses a non-anomalous global 
symmetry
group ${\cal G}$; however, the couplings  $h_i \neq 0$ break this 
symmetry.
Since $h_i$'s are treated now as auxiliary chiral superfields one can 
always
define transformations of these superfields in such a way as to 
restore
the global symmetry ${\cal G}$. Then the calculated superpotential
depending on the dynamical chiral superfields and on the auxiliary 
ones, 
should be invariant under this extended ${\cal G}$. This constraint 
becomes
informative if we take into account the fact that the calculated 
superpotential
${\cal W}$ must be a holomorphic function of all chiral superfields.
Thus, ${\cal W}$ can depend on $h_i$ but cannot depend on $h_i^*$. 
A few 
additional rules apply. The effective superpotential may depend
on the dynamically generated scale of the model $\Lambda$. It is 
clear
that negative powers of $\Lambda$ are forbidden since the result 
should be 
smooth in the limit when the interaction is switched off. Moreover, if 
we 
ensure that the theory is in the weak coupling regime, the possible  
powers of
$\Lambda$ are only those associated with one, two, three and so on 
instantons, i.e. $3T(G) -\sum_f T(R_f)$,  $2(3T(G) -\sum_f T(R_f))$, 
and so on, 
since the instantons are the only source of the nonperturbative 
parameter
$\Lambda$ in the weak coupling regime. Finally, one more condition 
comes 
from analyzing the limit of the small bare couplings $h_i$. This limit 
can be 
often treated perturbatively. Sometimes additional massless fields 
appear
in the limit $h_i=0$, which are absent for $h_i\neq 0$. When these 
fields
are integrated out and not included in the effective action, ${\cal W}$
may develop a singularity at $h_i=0$.

To illustrate the power and elegance of this approach \cite{seibhol} 
let us turn again to the Wess-Zumino model, Eq. (\ref{awz}). If the 
bare mass
and the coupling constant vanish, $m=g=0$, then the model has two
U(1) global invariances -- one associated with the rotations of the 
matter field,
$\phi \ra e^{i\alpha }\phi $,
and another one is the $R_0$ symmetry, $\phi \ra e^{2i\beta /3 
}\phi \, , \,\,\, 
\theta \ra e^{i\beta}\theta$.  To maintain both invariances with 
$m$ and
$g$ switched on we demand that
$$
m\ra m e^{-2i\alpha}\, , \,\,\, g\ra g e^{-3i\alpha}\,
$$
and 
$$
m\ra me^{2i\beta /3 }\, ,\,\,\, g \ra g 
$$
under $U(1)_\phi$ and $U(1)_R$, respectively.
The most general renormalized superpotential compatible with these
symmetries obviously has the form
\beq
{\cal W} = m\phi^2 f\left(\frac{g\phi }{m}\right)\, ,
\label{natwz}
\eeq
where $f$ is an arbitrary function. Let us expand it in the power 
series and 
consider the coefficient in front of $\phi^n$.  From Eq. (\ref{natwz})
it is clear that the corresponding terms has the form
$$
g^{n-2} m^{3-n} \phi^n\, .
$$
The balance of the powers of the coupling constant and $m$ is such 
that
this contribution could only be associated with the 1-particle 
reducible
tree graphs, which should not be included in the effective action.
Therefore, we conclude that there is no renormalization of the 
superpotential,
${\cal W} = {\cal W}_0$.

An example of a more sophisticated situation is provided by the
$SU(N)$ theory with $N_f$ flavors and the tree level superpotential 
\cite{ADS3}
\beq
{\cal W}_0 = h M \sum_f \tilde Q_f Q_f +h' M^3 \,
\label{ads3}
\eeq
where $Q_f$ is a chiral superfield belonging to the fundamental
representation of $SU(N)$ while $\tilde Q_f$ is a chiral field in the 
anti-fundamental representation. The color indices are summed over; 
the 
additional chiral field 
$M$ is color-singlet.

Now, if $h$ and $h'$ are set equal to zero, $M$ obviously decouples,
and the global symmetries of the model are those of the massless 
SQCD
plus one extra global invariance associated with the rotations of the
$M$ field. Massless SQCD is invariant under
$$
SU(N_f)_L\times SU(N_f)_R\times U(1)_V\times U(1)_R\, .
$$
The conserved $R$ charge is established from consideration of the 
anomaly
relations analogous to Eq. (\ref{anomc}). Indeed, it is not difficult to 
obtain 
that
\beq
\partial_\mu R^0_\mu = 
\left(  N-\frac{N_f}{3}\right) 
\frac{\alpha}{4\pi}G_{\mu\nu}^a\tilde G_{\mu\nu}^a\, , \,\,\, 
\partial_\mu J_\mu = N_f
\frac{\alpha}{4\pi}G_{\mu\nu}^a\tilde G_{\mu\nu}^a\, ,
\label{anomqcd}
\eeq
where the $R_0$ and $J$ currents are defined in parallel  to those
in the SU(2) model, see Eqs. (\ref{rnotcur}) and (\ref{matcar}). This 
means 
that the conserved $R$ current has the form
\beq
R_\mu = R^0_\mu - \frac{N-(N_f/3)}{N_f} J_\mu\, .
\label{CRC}
\eeq
Using this expression it is not difficult to calculate that the $R$ 
charge of the 
matter field is $(N_f-N)/N_f$. The $R$ charge of the $M$ field can be 
set equal 
to zero.

Now, the superpotential (\ref{ads3}) explicitly breaks both, $U(1)_R$
and $U(1)_M$.  To restore the symmetry we must ascribe to $h$ and 
$h'$
the following charges
$$
h\ra \left( -1, 2\frac{N}{N_f}\right)\, , \,\,\, 
h'\ra  \left( -3, 2 \right)\, , 
$$
where the first charge is with respect to $U(1)_M$
while the second charge is with respect to $U(1)_R$. If $h'=0$ the 
model
has a rich system of the vacuum valleys. Let us assume that we 
choose the
one for which the expectation value of $Q$ fields vanishes, but the
expectation value of $M\neq 0$. Moreover, we will assume that
$h\langle M\rangle$ is large. In this ``corner" of the valley the 
matter fields
are heavy, and can be integrated over. At very low energies the only
surviving (massless) field is $M$.  Our task is to find the effective 
Lagrangian 
for the $M$ field. By inspecting the above charge assignments
one easily establishes that the most general form of the effective 
superpotential compatible with all the assignments is
$$
{\cal W} = h' M^3 f (x)\, , \,\,\, x = \frac{\Lambda^{3 -(N_f/N)} 
(hM)^{N_f/N}}{h'M^3} \, ,
$$
where $\Lambda$ is a dynamically generated scale of the (strongly
coupled) SU(N) gauge theory. If $\Lambda \ra 0$ there should be no 
singularities. This implies that $f(x)$ is expandable in positive 
powers
of $x$. However, the behavior of the effective superpotential
at $h'\ra 0$ should also be smooth. These two requirements
fix the function $f$ up to a constant,
$$
f(x) = 1 +\alpha x \, , \,\,\, \alpha = \mbox{const.}\, ,
$$
and
\beq
{\cal W} = h' M^3  + \alpha \, \Lambda^{3 -(N_f/N)} (hM)^{N_f/N} \, .
\label{esp2}
\eeq
The first term is the same as in the bare superpotential, the second 
term is
generated nonperturbatively \cite{ADS4}. Note that the 
non-analytical
behavior at $h=0$ is due to the fact that at $h=0$ there are massless
matter fields, and we integrated them over assuming that they are 
massive.
The superpotential (\ref{esp2}) grows with $M$. This
is natural since the interaction becomes stronger as $M$ increases. 
The 
superpotential (\ref{esp2}) leads to a supersymmetric minimum at 
$M\neq 0$.

Concluding this section I would like to return to one subtle and very 
important point in this range of questions, holomorphic anomalies. 
Consider Eq. (\ref{condeg}) for the gluino 
condensate.
So far we have studied the analytic  dependence of this quantity on 
$m_0$.
At the same time, however, $1/g_0^2$ is also a coefficient of the $F$ 
term,
which can be viewed as an expectation value of an auxiliary
chiral superfield, dilaton/axion. One is tempted to conclude then that
the dependence of $\langle \lambda \lambda\rangle$ on $1/g_0^2$ 
must be
holomorphic, and its functional form must follow from consideration 
of the 
invariances of the theory. Is this the case?

The answer is yes and no. Let us examine the transformation 
properties
of the SU(2) theory with respect to the matter U(1) rotations, Eq. 
(\ref{matter}), supplemented by the rotation of the mass parameter
$m_0\ra m_0e^{-2i\gamma}$. At the classical level the theory is 
invariant.
The invariance is broken, however, by the triangle anomaly. In  
order to 
restore the invariance we must simultaneously shift the vacuum 
angle
$\vartheta$ (not to be confused with the supercoordinates 
$\theta_\alpha$!),
\beq
\vartheta\ra\vartheta - \Delta\mbox{Arg}(m_0) =\vartheta + 
2\gamma \, .
\label{thetas}
\eeq
Taking into account the fact that $\vartheta =$ Im$(-8\pi^2/g_0^2)$ 
we 
conclude 
that if Eq. (\ref{condeg}) contained no pre-exponential factor $g_0^{-
2}$
everything would be perfect -- $\lambda\lambda$ would be a 
holomorphic 
function of  $g_0^{-2}$, precisely the one needed for invariance of
$\lambda\lambda$. The pre-exponential factor $g_0^{-2}$ spoils the 
perfect
picture. As a matter of fact, one can see that in the pre-exponential it 
is
Re($g_0^{-2}$) which enters, and the holomorphy in $g_0^{-2}$ is 
absent.
The reason is the holomorphic anomaly associated with infrared 
effects.
In Refs. \cite{SV2,SV3} it was first noted that all formal theorems 
regarding 
the holomorphic dependences on the gauge coupling constant
are valid only provided we define the gauge coupling through the 
Wilsonian 
action which, by definition, contains no infrared contributions. What 
usually 
one deals with (and refers to as  the action) is actually the generator 
of 
the 
one-particle irreducible vertices. In the absence of the infrared 
singularities 
these two notions coincide; generally speaking, they are different, 
however.
In particular, the gauge coupling constant in the Wilsonian action,
$g_W^{-2}$ is related to $g_0^{-2}$ by the following expression
\beq
\frac{8\pi^2}{g_W^2} =\frac{8\pi^2}{g_0^2} - T(G) \ln \mbox{Re} 
g_0^{-2} 
\label{wils}
\eeq
(in pure gluodynamics, without matter).
The gluino condensate is holomorphic with respect to  $g_W^{-2}$.

The holomorphic anomaly in the gauge coupling due to massless 
matter fields 
was also observed
in Ref. \cite{DKL} in the stringy context, see also \cite{LM}.

\subsection{Supersymmetric instanton calculus}

As  was mentioned more than once, the instanton calculations,
combined with specific features of supersymmetry, were 
instrumental 
in establishing various exact results in supersymmetric gluodynamics 
and other theories. We will continue to exploit them in further 
applications. Needless to say that I will be unable to
present supersymmetric instanton calculus to the degree needed for 
practical uses. The interested reader is referred to Ref. \cite{instvol}.
Here I will limit myself to a few fragmentary remarks.

Technically, the most remarkable feature making the instanton 
calculations in supersymmetric theories by far more manageable 
than
in non-supersymmetric ones is a residual supersymmetry in the 
instanton background field. It is clear that picking up a particular
external field we typically break (spontaneously) supersymmetry:
SUSY generators applied to this field act non-trivially. However,
the self-dual (or anti-self-dual) Yang-Mills field, analytically 
continued to the Euclidean space, to which the instanton belongs, 
preserves a half of supersymmetry. Depending on the sign of the 
duality relation either $Q_\alpha$ or $\bar Q_{\dot\beta}$ act 
trivially, i.e. annihilate the background field \cite{Zum,DADI}.

The fact that a part of supersymmetry remains unbroken in the
instanton background leads to far-reaching consequences.
Indeed, the spectrum of fluctuations around this background
remains degenerate for bosons and fermions from one and the same 
superfield, and the form of the modes is in one-to-one 
correspondence \cite{DADI}, for all modes except the zero modes. An 
immediate consequence is vanishing of the one-loop quantum 
correction in the instanton background. Unsurprisingly, a more 
careful study shows \cite{NSVZ1} that all higher quantum corrections 
vanish as well.

Thus, the result of any instanton calculation is essentially determined 
by the zero modes alone. The problem reduces to quantum 
mechanics of the zero modes. The structure of the zero modes is 
governed by a set of relevant symmetries of the theory
under consideration \cite{NSVZ3}. Therefore, all quantities that are 
saturated by instantons 
reflect the most general and profound geometrical properties of the 
theory.
One of the examples, the gluino condensate, was already considered 
above. In Sect. 3 we will discuss another example --
the instanton-induced modification of the quantum moduli space
in SQCD with $N_f=N_c$.

Historically the first application of instantons 
in  supersymmetric  gluodynamics was the calculation of the gluino 
condensate \cite{NSVZ1}
in the strong coupling regime. I mention this result here because
although it is 15 years old, there is an intriguing  mystery associated 
with it.

Let us
consider for definiteness the
$SU(2)$ gluodynamics. In this case there are four gluino zero modes 
in the
instanton field and hence, there is no direct instanton contribution to 
the 
gluino condensate $\langle \lambda\lambda \rangle$. At the same 
time the instanton does contribute to the 
correlation 
function
\begin{equation}
\langle \lambda^a_\alpha (x)\lambda^{a\alpha}(x)
,\lambda^b_\beta (0)\lambda^{b\beta}(0) \rangle\, ,
\label{I1}
\end{equation} 
Here $a,b =1,2,3$ are 
the color
indices and $\alpha ,\beta = 1,2$ are the spinor ones. An explicit  
instanton calculation  shows that the correlation 
function (\ref{I1}) is equal to a non-vanishing constant.

At first sight this result might seem supersymmetry-breaking since
the instanton does not generate any boson analog of Eq. (\ref{I1}).
Surprising though it is, supersymmetry does not forbid
(\ref{I1}) provided that this two-point function is actually an
$x$ independent constant. 
For purposes which will become clear shortly let us sketch here the 
proof of the above assertion. 

Three elements are of importance: (i) the supercharge 
${\bar Q}^{\dot\beta}$ acting on the vacuum state annihilates it; 
(ii) ${\bar Q}^{\dot\beta}$ commutes with $\lambda\lambda$; 
(iii) the derivative $\partial_{\alpha\dot\beta}(\lambda\lambda)$ is 
representable as the anticommutator of ${\bar Q}^{\dot\beta}$ and 
$\lambda^\beta G_{\beta\alpha}$. (The spinor notations are used.)
The second and the third point follow from the fact that 
$\lambda\lambda$ is the lowest component of the chiral superfield 
$W^2$, while $\lambda^\beta G_{\beta\alpha}$ is its middle 
component.

Now, we differentiate Eq. (\ref{I1}),  substitute 
$\partial_{\alpha\dot\beta}(\lambda\lambda)$ by $\{
{\bar Q}^{\dot\beta}, \lambda^\beta G_{\beta\alpha}\}$ and obtain 
zero.  Thus,  supersymmetry requires the $x$ derivative of (\ref{I1}) 
to 
vanish \cite{NSVZ1}. 
This is exactly what happens if the correlator (\ref{I1}) is  a constant.
		
If so, one can compute the result at short
distances where it is presumably saturated by small-size instantons, 
and, 
then, the very same constant is predicted at large distances, 
$x\rightarrow
\infty$. On the other hand, due to the cluster decomposition property
which must be valid in any reasonable theory the correlation 
function (\ref{I1})
at $x\rightarrow
\infty$
reduces to $\langle\lambda\lambda \rangle^2$. Extracting the 
square root we 
arrive at a
(double-valued) prediction for the gluino condensate. 

(The same line of reasoning is applicable in other similar problems, 
not only for the gluino condensate. The correlation function of the 
lowest components of any number of superfields of one and the same 
chirality, if non-vanishing, must be constant. By analyzing the 
instanton zero modes it is rather easy to catalog all such correlation 
functions, in which the instanton contribution does not vanish.
Thus, for $SU(N)$ gluodynamics one ends up with the $N$-point
function of $\lambda\lambda$. Inclusion of the matter fields,
clearly, enriches the list of the instanton-induced ``constant" 
correlators, but not too strongly \cite{ROVE}. The general strategy 
remains the same as above in all cases.)

Many questions immediately come to one's mind in 
connection with
this argument. First, if the gluino condensate is non-vanishing and 
shows 
up in a roundabout instanton calculation through (\ref{I1}) why it is 
not 
seen
in the direct instanton calculation of $\langle\lambda\lambda 
\rangle$?
Second, the constancy of the two-point function (\ref{I1}) required 
by SUSY
is ensured in the concrete calculation by the fact that the instanton 
size
$\rho$ turns out to be of order of $x$. The larger the value of $x$ the 
larger $\rho$
saturates the instanton contribution. For small $x$ this is alright.
At the same time at $x\rightarrow\infty$ we do not expect any 
coherent 
fields with the size of order $x$ to survive in the vacuum; 
such coherent fields would 
contradict our current ideas of the infrared-strong confining theories 
like
SUSY gluodynamics. If there are no large-size coherent fields in the 
vacuum 
how can one guarantee the $x$ independence of (\ref{I1}) at all 
distances?

A tentative  answer to the first question might be found in the 
hypothesis put forward by 
Amati {\em et al}. 
\cite{Amati}. It was assumed that, instead of providing us with the 
expectation value of 
$\lambda\lambda $ in the given vacuum, instantons in the strong 
coupling regime  yield
an average value of $\langle\lambda\lambda \rangle$ in all possible 
vacuum 
states. If  there exist two vacua, with the opposite
signs of $\langle\lambda\lambda \rangle$,
the conjecture of Amati {\em et al.} would explain
why  instantons in the strong coupling regime  do not generate 
$\lambda\lambda$ directly.

When we do the instanton calculation in the weak coupling regime
(the Higgs phase) the averaging over distinct vacua
does not take place. In the weak coupling regime, we have a marker: 
a large
classical expectation value  of the Higgs field tells us in what 
particular  vacuum 
we 
do our instanton calculation. In the strong coupling regime, such a
marker is absent, so that the recipe of Amati {\em et 
al.} seems plausible. 

This is not the end of the story, however. One  of the 
instanton computations which was done  in the mid-eighties 
\cite{NSVZ3}  remained a puzzle defying theoretical 
understanding for years. 
The result for $\langle\lambda\lambda\rangle$
obtained in the strong coupling regime 
(i.e. by following the program outlined after Eq. (\ref{I1})) does {\em 
not} match
$\langle\lambda\lambda\rangle$ calculated in 
an indirect  way, as we did in Sect. 2.5 -- extending the theory by 
adding one flavor, doing the calculation in the weakly coupled Higgs 
phase, and then returning back to SUSY gluodynamics by exploiting 
the holomorphy of the condensate in the mass parameter.
In Ref. \cite{NSVZ3} it was shown that
\beq
\langle\lambda\lambda\rangle_{\rm scr}^2 =
\frac{4}{5} \langle\lambda\lambda\rangle_{\rm wcr}^2\, ,
\label{DISCR}
\eeq
where the subscripts scr and wcr mark the strong and weak coupling 
regime calculations.

The hypothesis of Amati {\em et al.}, by itself,  does not explain the 
discrepancy 
 (\ref{DISCR}). If there are only two vacua characterized by 
$\langle\lambda\lambda \rangle =\pm \Lambda^3$, 
the gluino condensate is not affected by 
the averaging over these two vacuum states,
since the contributions of these two vacua to Eq. (\ref{I1}) are equal. 
If, however, there 
exist an {\em extra} 
 zero-energy state with $\langle\lambda\lambda \rangle = 0$ 
 involved in the averaging, the final result in the strong 
coupling regime
is naturally different from that obtained in the weak coupling regime
in the {\em given } vacuum. Moreover, the value of the condensate 
calculated in the strong coupling approach should be smaller, 
consistently
with Eq. (\ref{DISCR}). At the moment there seems to be no other 
way out of the dilemma 
\cite{Kovner}. The conclusion of the existence of the extra 
vacuum with $\langle\lambda\lambda \rangle = 0$ 
is quite
radical,
and, perhaps, requires further verification, in particular, in
connection with the Witten index counting.  What is beyond any 
doubt, however is that the combination of instanton calculus with 
holomorphy and other specific features of  supersymmetry
provides us with the most powerful tool we have ever had in 
four-dimensional field theories. 

It remains to be added that the interest to technical aspects
of supersymmetric instanton calculus \cite{NSVZ3}  was 
revived recently in connection with the Seiberg-Witten solution of 
the $N=2$ theory. The solution was obtained \cite{SEIWIT} from  
indirect arguments, and it was tempting to verify it 
by direct instanton calculations \cite{Khoze}.  Such calculations 
require extension of supersymmetric instanton calculus
to $N=2$, which was carried out, in a very elegant way, in Ref.
\cite{Yung}, see also \cite{Khoze}. 

\vspace{0.2cm}

\begin{center}
*******

\end{center}
\vspace{0.2cm}

Concluding this section let me briefly summarize the main lessons.

First, the most remarkable feature of the structure of SUSY gauge 
theories 
with matter is the existence of the vacuum valleys -- classically flat  
directions along which the energy vanishes. This degeneracy may or 
may not 
be lifted dynamically, at the quantum level. SU(2) model with one 
flavor is
an example of the theory where the continuous degeneracy  is lifted, 
and the 
quantum vacuum has only discrete (two-fold) degeneracy. If this 
does not 
happen, the classically flat directions give rise to {\em quantum 
moduli space
of supersymmetric vacua}. This feature is the key element of  the 
recent 
developments pioneered by Seiberg. 

Second, holomorphic dependences of various chiral quantities 
enforced by supersymmetry lie behind numerous miracles occurring 
in SUSY 
gauge
theories -- from specific non-renormalization theorems to the exact 
$\beta$ 
functions. This is also an important element of dynamical scenarios to 
be 
discussed below.

Now the stage is set and we are ready for more adventures and 
surprises
in supersymmetric dynamics.

\newpage

\section{Lecture 3. Dynamical Scenarios in SUSY Gauge Theories -- 
Pandora's 
Box?}

\renewcommand{\theequation}{3.\arabic{equation}}
\setcounter{equation}{0}

In the first two sections I summarized what was known (or assumed) 
about 
the intricacies of the gauge dynamics in the eighties. In the following 
two 
sections we will discuss the discoveries and exciting results of the 
recent 
years.
I should say that the current stage of development was open by 
Seiberg,
and many ideas and insights to be discussed today I learned from 
him or 
extracted
from works of his collaborators. 

Remarkable facets of the gauge dynamics will open to us. First of all, 
we will encounter non-conventional patterns of the 
chiral symmetry breaking. The chiral symmetry breaking is one of 
the most
important phenomena of which very little was known, beyond
some empiric facts referring to QCD. In the eighties, when our 
knowledge of the gauge dynamics was less mature than it is now,
it was believed that the massless fermion condensation obeys
 the so called maximum attraction channel (MAC) hypothesis
\cite{MAC}. In short,  one was supposed to consider the one-gluon 
exchange between fermions, find a channel with such quantum 
numbers that the attraction was maximal, and then assume  the 
condensation of the fermion pairs in this particular channel. 
The concrete quantum numbers of the fermion condensates imply 
 a very specific pattern of the chiral symmetry breaking.

In SQCD we will find patterns {\em contradicting the MAC 
hypothesis}.
This means that the chiral condensates are not governed by the 
one-gluon exchange, even qualitatively. The basic tool for exploring 
the 
 chiral condensates is the 't Hooft matching condition.
It was exploited  for this purpose previously many times, in the 
context 
most relevant to   us  in Ref. \cite{antiMAC}. Combining 
supersymmetry (the fact of the existence of the vacuum manifold) 
with the matching condition drastically enhances  the method. 

The second remarkable finding is the observation of conformally
invariant theories in four dimensions in the strong coupling regime.
The crucial instrument in revealing such theories
is  Seiberg's ``electric-magnetic" duality in the infrared domain,
connecting with each other two distinct gauge theories --
one of them is strongly coupled while the other is weakly coupled.
One can view the gluons and quarks of the weakly coupled theory
as bound states of the gluons and quarks of its dual partner. 
If so, composite gauge bosons can exist!
The arguments in favor of the ``electric-magnetic" duality
are again based on the 't Hooft matching condition (combined with 
supersymmetry) and some additional indirect consistency checks. 

\subsection{$SU(N_c)$ QCD with $N_f$ flavors -- preliminaries}

The SU(2) model considered previously is somewhat special
since all representations of SU(2) are (pseudo)real. For this reason
the flavor sector of this model possesses an enlarged symmetry.
Thus, for one flavor we observe the flavor SU(2) symmetry,
which is absent if the gauge group is, say, SU(3). Now we will 
consider
a more generic situation. If not stated to the contrary the gauge 
group
is assumed to be $SU(N_c)$ with $N_c >2$. 
Peculiarities 
of the orthogonal, symplectic or exceptional groups will be 
briefly discussed in Sect. 4. In accordance with Witten's index,
if the matter sector  consists of non-chiral matter
allowing (at least, in principle) for a mass term for all matter fields,
supersymmetry is unbroken.

To describe $N_f$ flavors one has to introduce $2N_f$ chiral 
superfields,
$Q^i$ in the representation $N_c$ and $\tilde Q_j$ in the 
representation
$\bar N_c$. To distinguish between the fundamental and 
anti-fundamental 
representations the flavor indices used are superscripts and 
subscripts, 
respectively. 

The Lagrangian is very similar to that of the $SU(2)$ model,
\beq
{\cal L} = \frac{1}{2g^2_0}\, \mbox{Tr}\,
\int d^2 \theta W^2 + 
\frac{1}{4} \int d^2\theta d^2\bar\theta
\left( Q^\dagger e^V Q + \tilde Q^\dagger e^{-V} \tilde Q\right)  +
\left( \frac{1}{2}\, \int d^2\theta \, {\cal W}(Q, \tilde Q )
+\mbox{ H. c.}\right)\, ,
\label{suNlagr}
\eeq
 where ${\cal W}(Q, \tilde Q )$ is a superpotential which may or may 
not be 
present. Then the scalar potential has the form
\beq
V = \frac{1}{2g^2} D^aD^a + \sum_Q \left\vert \frac{\partial{\cal 
W}(Q, \tilde Q 
)}{\partial Q}\right\vert^2 + 
\sum_{\tilde Q} \left\vert \frac{\partial{\cal W}(Q, \tilde Q 
)}{\partial \tilde 
Q} 
\right\vert^2\, .
\label{scapot}
\eeq
An example of possible superpotential is a generalized mass term,
$$
{\cal W} = m^j_i\tilde Q_j Q^i
$$
where $m^j_i$ is a mass matrix. Most often we will work under 
conditions of 
vanishing superpotential, ${\cal W} =0$.

It is convenient to introduce two $N_f\times N_c$ matrices of the 
form
\beq
q =\left\{ Q^1, Q^2, ..., Q^{N_f}\right\}\, ,\,\,\,  \tilde q= \left\{\tilde 
Q_1, \tilde 
Q_2,..., \tilde Q_{N_f}\right\}\, .
\label{qmatr}
\eeq
The rows of these matrices correspond to different values of the 
color 
index.
Thus, in the first row the color index is 1, in the second row 2, {\em 
etc.},
$N_c$ rows altogether. Both matrices can be globally rotated in the 
color and 
flavor spaces. Let us assume first that $N_f < N_c$. Then, by applying 
these 
rotations one can always reduce the matrix $q$ to the form
\beq
q= \left( \begin{array}{cccc}
a_1 & 0 & ... & 0 \\
0 & a_2 & ...&0 \\
... & ... & ... & ...\\
0 & 0 & ... & a_{N_f} \\
0 & 0 & ... & 0 \\
... & ... & ... & ...
\end{array} \right)\, .
\eeq
If $\tilde q = q$ we are at the bottom of the vacuum valley -- the 
corresponding energy vanishes. The gauge invariant description is 
provided
by the composite chiral superfield,
\beq
M^i_j = \tilde Q_j Q^i \, .
\eeq
The points belonging to the bottom of the valley are parametrized by 
the 
expectation value of $M^i_j$. Generically, if we are away from the 
origin
$SU(N_c)$ gauge group is broken down to $SU(N_c-N_f)$. The first 
group has
$N_c^2 -1$ generators, the second one has $ (N_c-N_f)^2 - 1$ 
generators.
Thus, the number of the chiral fields eaten up in the super-Higgs 
mechanism 
is
$2N_cN_f - N_f^2$. Originally we started from $2N_cN_f$ chiral 
superfields;
$N_f^2$ remain massless -- exactly the number of degrees of 
freedom 
in $M^i_j$. There are exceptional points. When det $M = 0$ the 
unbroken 
gauge subgroup is larger than $SU(N_c-N_f)$, and, correspondingly, 
we have
 more than $N_f^2$  massless particles. At the origin of the vacuum 
valley
the original gauge group $SU(N_c)$ remains unbroken.
 
The situation changes if the number of flavors is equal to or larger 
than
the number of colors. Indeed, if $N_f>N_c$ the generic form of the 
matrix $q$, 
after an appropriate rotation in the flavor and color space, is
\beq
q= \left( \begin{array}{ccccccc}
a_1 & 0 & ... & 0 & 0 & ... & 0\\
0 & a_2 & ... & 0 & 0 & ... & 0\\
... & ... & ... & ... & ... & ... & ... \\
0 & 0 & ... & a_{N_c} & 0 & ... & 0\\
\end{array} \right)\, .
\eeq
The condition defining the bottom of the valley (the vanishing of the
energy) is 
\beq
|a_i|^2 - |\tilde a_i|^2 = \mbox{ constant independent of}\,\,\, i
\eeq
where $i = 1,2, ... , N_c$. At a generic point from the bottom of the 
valley
the gauge group is completely broken. The gauge invariant chiral 
variables
parametrizing the bottom of the valley (the moduli space) now are
\beq
M^i_j = \tilde Q_j Q^i \, , \,\,\, B = Q^{[i_1} ...  Q^{i_{N_c}]}
\, , \,\,\, \tilde B = \tilde Q_{[j_1} ...  \tilde Q_{j_{N_c}]}\, ,
\label{MBB}
\eeq
where the color indices in $B$ and $\tilde B$  (they are not written 
out 
explicitly) are contracted with the help of the $\varepsilon$ symbol;
the flavor indices $i_1, ... , i_{N_c}$ and $j_1, ... , j_{N_c}$ then come 
out 
automatically antisymmetric, and  the square brackets in Eq. 
(\ref{MBB})
remind us of this antisymmetrization. {\em A priori}, the number of 
the 
variables $B$ and $\tilde B$ is $C_{N_f}^{N_c}$ each, where  
$C_{N_f}^{N_c}$
are the combinatorial coefficients,
$$
C_{N_f}^{N_c} = \frac{N_f !}{N_c ! (N_f-N_c)!}\, ,
$$
 since one can pick up $N_c$ flavors out of the total set of $N_f$ in 
various
ways.  If we try to calculate now the number of moduli, assuming 
that all 
those indicated in Eq. (\ref{MBB}) are independent, we will see that
this number does not match the number of the massless degrees of 
freedom.
Let us consider two examples, $N_f = N_c$
and $N_f = N_c + 1$.

\vspace{0.3cm}

${\bf N_f=N_c}$

\vspace{0.2cm}

The original number of the chiral supefields  is $2N_fN_c$; since the 
gauge 
symmetry is completely broken the number of the ``eaten" 
superfields is
$N_c^2 -1$; the number of the massless degrees of freedom is, thus,
$N_f ^2 +1 = N_c^2 +1$. The number of moduli in Eq. (\ref{MBB}) is
$N_f^2 +2$. One chiral variable is, thus, redundant.

\vspace{0.3cm}

${\bf N_f=N_c+1}$

\vspace{0.2cm}

The  number of the chiral supefields  is $2N_fN_c$; since the gauge 
symmetry 
is completely broken the number of the ``eaten" superfields is
$N_c^2 -1$; the number of the massless degrees of freedom is $N_f ^2 
$. The 
number of moduli in Eq. (\ref{MBB}) is
$N_f^2 +2N_f$, i.e. $2N_f$ chiral variables are, thus, redundant.

\vspace{0.2cm}

In the first case, $N_f=N_c$, the constraint eliminating the redundant
chiral variable is
\beq
\mbox{det} \{ M\} = B\tilde B \, ,
\label{nfnc}
\eeq
while in the second example, $N_f=N_c+1$, by using merely
the definitions of the moduli in Eq. (\ref{MBB}), it is
not difficult to obtain
\beq
B_i \{ M\} ^i_j = \{ M\} ^i_j \tilde B^j = 0\, ,
\eeq
and
\beq
\mbox{minor}\{ M\} ^j_i = B_i\tilde B^j\, ,
\label{minor}
\eeq
where the left-hand side of the last equation is the minor
of the matrix $\{ M\} $ (i.e. $(-1)^{i+j}\times$ determinant of the 
matrix obtained from 
$M$
by omitting the $i$-th row and the $j$-th column. 
Note that det $\{ M\} $ vanishes in this case. For brevity
we will sometimes write Eq. (\ref{minor}) in a somewhat sloppy 
form
\beq
\mbox{det}\, M  \left( M^{-1}\right)_i^j = B_i\tilde B^j\, .
\label{nf+1}
\eeq
At the classical level one could, in principle, eliminate the redundant 
chiral 
variables
using Eqs. (\ref{nfnc}) or (\ref{nf+1}). One should not hurry with this 
elimination, however, since at the quantum level the classical moduli 
fields
are replaced by the vacuum expectation values of $M^j_i $ and $B$, 
$\tilde B$,
and although generically the total number of the massless degrees of
freedom does not change, the quantum version of constraints 
(\ref{nfnc}) and 
(\ref{nf+1}) may (and will be) different. Moreover, at some specific 
points
from  the valley the number of the massless degrees of freedom 
may increase, as we will see shortly. 

SQCD with $N_f$ flavors and no tree-level superpotential
has the following  global symmetries free from the internal  
anomalies: 
\beq
SU(N_f)_L\times SU(N_f)_R \times U(1)_B \times U(1)_R
\label{AD5}
\eeq
where the conserved $R$ current is introduced in Eq. (\ref{CRC}), and
 the quantum numbers of the matter multiplets with
respect to these symmetries are collected in Table 1. 

\begin{table}
\begin{center}
\begin{tabular}{|c|c|c|c|c|}\hline
 $ ~~$~~~  & ~~ $SU(N_f)_L$ ~~ & ~~$SU(N_f)_R$ ~~ &~~ $U(1)_B$ ~~ &
~~$U(1)_R$~~ \\\hline 
$Q$ & $ N_f$  & 1 & 1 &  $ (N_f - N_c)/N_f$ \\ 
$\tilde{Q}$ & 1 & $\bar{N}_f$ & $-1 $&$ (N_f -  N_c)/N_f$\\
\hline
\end{tabular}
\end{center}
\caption{The quantum numbers of the matter fields
with respect to the global symmetries (\ref{AD5}) in SQCD with $N_f$ 
flavors. The $R$ charges indicated refer to the lowest components of 
the superfields.}
\end{table}

(For discussion of the subtleties in the $R$ current definition
see Ref. \cite{KSV}. These subtleties, being conceptually important,
are irrelevant for our consideration).

The $SU(N_f)_L\times SU(N_f)_R \times U(1)_B $ transformations   
act
only on the matter fields in an obvious way, and do not affect the 
superspace coordinate $\theta$. As for the extra global symmetry
$U(1)_R$ it is defined in such a way that it acts nontrivially on the
supercoordinate $\theta$ and, therefore, acts differently on the  
spinor
and the scalar or vector components of superfields. The $R$ charges  
in
Table 1 are given for the lowest component of the chiral
superfields. If the $R$ charge of the boson component of the given
superfield is $r$ then the $R$ charge of the fermion component
is, obviously, $r-1$.  A part of the above global symmetries
is spontaneously broken by the vacuum expectation values
of $M^i_j$ and/or $B,\tilde B$.

Unlike the $N_c=2$, $N_f=1$ model discussed in Sect. 2 instantons 
do not 
lift the classical degeneracy, and the bottom of the valley remains 
flat.
The easiest way to see this is to consider a generic point
from the bottom of the valley, far away from the origin,
where the theory is in the weak coupling regime, 
and try to write the most general superpotential, compatible with all 
exact 
symmetries (it must be symmetric even under those symmetries 
which may
turn out to be spontaneously broken) \cite{ADS1,ADS2}.   The 
symmetry 
under $SU(N_f)_L\times SU(N_f)_R \times U(1)_B$ is guaranteed
if we assume that the superpotential ${\cal W}$ depends on 
det$M$. What about the $R$ symmetry?

For $N_f=N_c$ the $R$ charge of the matter superfield vanishes, as is 
clear 
from Table 1. Since the superpotential must have  the $R$ charge 2,
it is obvious that it cannot be generated. For $N_f>N_c$ the $R$ 
charge
of the matter fields does not vanish, and, in principle, one
could have written 
$$
{\cal W} \propto \left(
\frac{\Lambda^{3N_c - N_f}}{\mbox{det}\, M}\right)^{\frac{1}{N_c-
N_f}}\, ,
$$
an expression which has the right dimension (three) and the correct 
$R$ charge
(two). However, the dimension of $\Lambda$ does not match the 
instanton
expression which can produce only $\Lambda^{3N_c - N_f}$ (and in 
the weak
coupling regime the instanton is the only relevant nonperturbative 
contribution). What is even more important, for $N_f>N_c$ 
the determinant of $M$ vanishes identically. This fact alone
shows that no superpotential can be generated, and the flat direction
remains flat \cite{DDS,ADS1,ADS2}. 

The argument above demonstrates again the power of holomorphy.
In non-supersymmetric theories one could built a large number of 
invariants
involving $Q, \tilde Q, Q^\dagger $ and $\tilde Q^\dagger$. In  SUSY 
theories,
as far as the $F$ terms are concerned, one is allowed to use only $Q$ 
and 
$ \tilde Q $ which constraints the possibilities to the extent nothing is 
left.

In summary, for $N_f \geq N_c$ the vacuum degeneracy is not 
lifted.
At the origin of the space of moduli, where $B=\tilde B = 0$ and
$M$ has fewer than $N_c-1$ non-zero eigenvalues,
the gauge symmetry is not fully broken. At this points, the classical 
moduli
space is singular. Far away from the origin, when the
expectation values of the squark  fields are
large, the distinction between the classical and quantum moduli 
space should 
be unimportant. In the vicinity of the origin, however, this 
distinction may be 
crucial. Our next task is to investigate this distinction. Needless to say
that just the vicinity of the origin is the domain of most interesting 
dynamics. Since the Higgs fields are in the fundamental 
representation, we are always in the Higgs/confining phase. 
Far away from the origin the theory is in the weak coupling regime
and is fully controllable by the well understood methods
of the weak coupling. In the vicinity of the origin the theory is
in the strong coupling regime. The issues to be investigated are the 
patterns of the spontaneous breaking of the global symmetries
and the occurrence of the composite massless degrees of freedom at 
large distances. Here each non-trivial theoretical result 
or
assertion is a precious asset, a miraculous achievement.

\subsection{The quantum moduli space}

Relations (\ref{nfnc}) and (\ref{nf+1}) are constraints on the classical 
composite 
fields. Since in the quantum theory the vacuum valley is 
parametrized
by the {\em expectation values} of the fields, which may get a 
contribution 
from quantum fluctuations, these relations may alter. In other words, 
the 
quantum moduli space need not exactly coincide with the classical 
one.
Only in the limit when the vacuum expectation values of the fields
parametrizing the vacuum valley become large, much larger than
the scale parameter of the underlying theory, we must be able to 
return
to the classical description.

To see that the quantum moduli space does indeed differ from the 
classical 
one we will consider here, following Ref. \cite{Seib1}, the same two 
examples, 
$N_f = N_c$ and $N_f = N_c+1$. The general strategy used in these 
explorations is the same as was discussed in detail in Sect. 2, in 
connection 
with $SU(2)$ SQCD \cite{SV4}: in order to analyze the theory along 
the 
classically flat
directions one adds the appropriately chosen mass terms (sometimes,
other superpotential terms as well), solves the theory in the weak 
coupling 
regime, and then analytically continues to the limit where the 
classical
superpotential vanishes.

\vspace{0.3cm}

${\bf N_f = N_c}$ 

\vspace{0.1cm}

Introduce a mass term for all quark flavors, or, more generically,
the quark mass matrix
\beq
m_i^j Q^i\tilde Q_j \, ,
\label{mt}
\eeq
(it can always be diagonalized, of course). If we additionally assume
that the mass terms for $N_f -1$ flavors are small and the mass term
for one flavor is large then we find ourselves
in a situation where an effective low-energy theory is that of $N_c 
-1$
flavors. From Sect. 2 we know already that in this case
the $SU(N_c)$ symmetry is totally broken spontaneously, the theory 
is in the
weak coupling phase, instantons generate a superpotential,
and this superpotential, being combined with Eq.
(\ref{mt}), leads to \cite{ADS1,ADS2,NSVZ3,Amati}
$$
M^{i }_{j} = \langle Q^i \tilde Q_{j}\rangle = \Lambda^{2} ({\rm det}\,  
m)^{1 
\over N_c}\left({1 \over m}\right)^{i }_{j}\, ,
$$
\beq
B = \langle Q^{i_1} ... Q^{i_{N_c}} \rangle \epsilon_{{i_1} ... {i_{N_c}}} = 
0\, ,\,\,\,
\tilde B = \langle \tilde Q_{j_1} ... \tilde Q_{j_{N_c}} \rangle 
\epsilon^{{j_1} ... 
{j_{N_c}}} = 0\, .
\label{qmod}
\eeq

Although this result was obtained under a very specific assumption
on the values of the mass terms, holomorphy tells us that it is exact. 
In 
particular, one can tend $m_i^j\ra 0$, thus returning
to the original massless theory. 
Equation (\ref{qmod}) obviously implies that
\beq
{\rm det}\, M - B\tilde B = \Lambda^{2N_c} \, .
\label{qmodnf}
\eeq
It is instructive to check that this relation stays valid even if 
$B\neq 0$, $\tilde B \neq 0$. To this end one must introduce, 
additionally,
a superpotential $\beta B +\tilde\beta\tilde B$ where $\beta$ and
$\tilde\beta$ are some constants, and redo the instanton calculations.
If $\beta\neq 0$ and
$\tilde\beta\neq 0$, the instanton-induced superpotential changes,
non-vanishing values of $B$ and $\tilde B$ are generated,
the vacuum expectation values $M^{i }_{j}$ change as well,
but the relation (\ref{qmodnf}) stays intact. 

Far from the origin, where the semiclassical analysis is applicable,
the quantum moduli space (\ref{qmodnf}) is close to the classical 
one.
A remarkable  phenomenon happens near the origin \cite{Seib1}. In 
the 
classical
theory where the gluons were massless near the origin,
the classical moduli space was singular. Quantum effects eliminated
the  massless modes by creating a mass gap \footnote{The latter 
statement is 
not quite correct. Massless moduli fields still persist. What is 
important,
however, is that the gluons acquire a dynamical ``mass".}. 
Correspondingly,
the singular points with $B=\tilde B = 0$  and  vanishing 
eigenvalues of $M$ are eliminated from the moduli space.

In the weak coupling regime dynamics is rather trivial and boring.
Let us consider the most interesting domain of the vacuum valley, 
near the origin, 
in more detail, ``in a microscope". There are several points
that are special, they are characterized by 
 an enhanced global
symmetry. For instance, if 
\beq
B=\tilde B = 0\,\,\, \mbox{ and}\,\,\,  M^i_j=\Lambda^2\delta^i_j \, ,
\label{bbvac}
\eeq
the original global $SU(N_f)\times SU(N_f)$ symmetry is 
spontaneously 
broken
down to the diagonal $SU(N_f)$, while the $U(1)_R$ remains 
unbroken (the
$R$ charges of $Q$ and $\tilde Q$ vanish, see Table 1). 
We are in the vicinity of the origin, where all moduli are either 
of order of $\Lambda^2$ or vanish. Hence, the fundamental gauge 
dynamics
of the quark (squark) matter is strongly coupled. We are in the 
strong 
coupling regime. 

The spontaneous 
breaking of the global symmetry implies the existence of the 
massless 
Goldstone
mesons which, through supersymmetry, entails, in turn,
the occurrence of the massless (composite) fermions. These fermions
reside in the superfields $M$, $B$, and $\tilde B$. Their quantum 
numbers with respect to the unbroken symmetries are indicated 
in Table 2.

\begin{table}
\begin{center}
\begin{tabular}{|c|c|c|c|}\hline
 $ ~~$~~~  & ~~ $SU(N_f)$ ~~ & ~~ $U(1)_B$ ~~ &
~~$U(1)_R$~~ \\
\hline 
$\psi_M$ &$N_f^2-1$&0&$-1$\\
$\psi_B$& 1&$N_f$&$-1$\\
$\psi_{\tilde B}$& 1&$-N_f$&$-1$\\
\hline
$\psi_Q$ & $ N_f$   & 1 &  $ -1$ \\ 
$\psi_{\tilde{Q}}$ &  $\bar{N}_f$ &$ -1$ &$ -1$\\
$\lambda$ & 1&0&1\\
\hline
\end{tabular}
\end{center}
\caption{The quantum numbers of the composite massless  fermions 
with respect to the unbroken global symmetries in SQCD with 
$N_f=N_c$}
\end{table}

For convenience Table 2 summarizes also the quantum numbers of
the fundamental fermions --
quarks and gluino.
A remark is in order concerning the multiplet of the massless
fermions $\psi_M$. Since $M$ is an $N_f\times N_f$ matrix, naively 
one might 
think that the number of these fermions is $N_f^2$. Actually we must 
not 
forget that we are interested in small fluctuations of 
the moduli fields $M$, $B$ and $\tilde B$
around the expectation values (\ref{bbvac}) subject to the constraint
(\ref{qmodnf}). It is easy to see that this constraint implies that
the matrix of fluctuations $M^i_j -\Lambda^2\delta^i_j$ is traceless,
i.e. the fluctuations form the adjoint ($(N_f^2 -1)$-dimensional)
representation of the diagonal $SU(N_f)$.

Massless composite fermions in the gauge theories are subject to a 
very 
powerful 
constraint known as the {\em 't Hooft consistency condition} 
\cite{thooft1}. As 
was
first noted in \cite{dolgov}, the  triangle anomalies of the $AVV$ 
type in the 
gauge theories with the 
fermion
matter  imply the existence 
of  infrared singularities in the matrix elements of  the axial
currents. (Here  $A$ and $V$ stand for the axial 
and vector currents, respectively). These singularities  are 
unambiguously 
fixed 
by the short-distance (fundamental) structure of the theory even if 
the theory 
at hand
is in the strong coupling regime and cannot be solved in the infrared. 
The  massless composite fermions in the theory, if present,    must  
arrange 
themselves  in 
such a
way as to match  these singularities. If they cannot, the 
corresponding
symmetry is spontaneously broken, and the missing
infrared singularity is provided by the Goldstone-boson poles 
coupled to the 
corresponding
 broken generators.  This device -- the 't Hooft consistency condition, 
or 
anomaly matching -- is widely used in the strongly 
coupled gauge theories: from QCD to technicolor, to supersymmetric 
models; 
it   
allows  one to check various conjectures about the massless 
composite states. 
(For 
a pedagogical 
review see e.g. \cite{ShifPR}.)  

In our case we infer the existence of the massless fermions
from the fact that a set of moduli exists, plus supersymmetry. 
Why do we need to check the matching of the $AVV$ triangles? If 
we know for sure
the pattern of the symmetry breaking -- which symmetry
is spontaneously broken and which is realized linearly --
the matching of the $AVV$ triangles for the unbroken currents must 
be automatic. 
The condensates   indicated in Eq. (\ref{bbvac}) suggest that the axial
$SU(N_f)$ is spontaneously broken 
while the $R$ current and the baryon current are unbroken. Suggest, 
but do not prove!
For in the strong coupling regime other (non-chiral) condensates 
might develop too. For instance, on general grounds one cannot 
exclude
the condensate of the type $\langle M^i_jB^\dagger\rangle $ which 
will 
spontaneously break the baryon charge conservation. Since this 
superfield is non-chiral the holomorphy consideration is  
inapplicable. 
If the anomalous triangles with the baryon current do match,
it will be a  strong argument showing that
no additional condensates develop, and the pattern of the 
spontaneous symmetry breaking can be read off from Eq. 
(\ref{bbvac}).  Certainly, this is not a complete rigorous proof,
 but, rather, a very strong indication.

What is
extremely  unusual in the pattern implied by Eq. 
(\ref{bbvac})
is the survival of an unbroken axial current (the axial component of 
the $R$ 
current). 
We must verify that this scheme of the
symmetry breaking is compatible with the spectrum of 
the massless composite fermions residing  in the superfields $M$, 
$B$, and 
$\tilde 
B$. 

The 't Hooft consistency 
conditions, to be  analyzed in the general case,  refer   to the so called 
{\em external} anomalies of the $AVV$ type. More exactly, one  
considers 
those axial 
currents, 
corresponding 
to global
symmetries of the theory at hand, which are non-anomalous inside 
the theory 
{\em per se},  but
acquire anomalies in  weak external backgrounds. For instance, in  
QCD
with several flavors the singlet axial current is internally anomalous
-- its divergence is proportional to $G\tilde G$ where $G$ is the gluon
field strength tensor. Thus, it should not be included in the set of the 
 't Hooft consistency 
conditions to be checked. The non-singlet currents are 
non-anomalous in 
QCD itself, but become anomalous if one includes the photon field, 
external 
with
respect to QCD. These currents must be checked. The anomaly in the 
singlet 
current does not lead to  the
statement of the infrared singularities in the current while the 
anomaly in the non-singlet currents does. Those 
symmetries
that are internally anomalous, are non-symmetries. 

In our case we first list all those symmetries which are
supposedly realized linearly, i.e. unbroken. 
After listing all relevant  currents  we then saturate
the corresponding triangles. The diagonal $SU(N_f)$ symmetry
which remains unbroken is induced by the vector current, not axial.
The same is true with regards to $U(1)_B$. The conserved 
(unbroken) $R$ 
current has the axial component. Therefore,  the list we must 
consider 
includes the
following triangles
$$
U(1)_R^3\, , \,\,\,  U(1)_R SU(N_f)^2\, , \,\,\,  U(1)_RU(1)_B^2\, .
$$
One more triangle is of a special nature. One can consider the 
gravitational 
field as external, and study the divergence of the $R$ current in this 
background. This divergence is also anomalous,
\begin{equation}
\partial^\mu (\sqrt{-
g}R_{\mu})\,=\,\frac{1}{192\pi^2}\left[\mbox{const.}\right] 
\epsilon_{\mu\nu\lambda\delta} R^{\mu\nu\sigma\rho}
R^{\lambda\delta}_{\sigma\rho} \; ,
\label{Rgrav}
\end{equation} 
where $g$ is the metric and $R^{\mu\nu\sigma\rho}$ is the 
curvature
of the gravitational background. The constant in the square brackets
depends on the particle content of the theory, and must be matched
at the fundamental  and composite fermion level. This gravitational 
anomaly in the $R$ current is routinely referred to as $U(1)_R$. 
Thus, 
altogether
we have to analyze four triangles. Let us start, for instance, from
$U(1)_R SU(N_f)^2$. The relevant quantum numbers of the
fundamental-level fermions ($\psi_Q$,  $\psi_{\tilde Q}$ and
$\lambda$), and the composite fermions ($\psi_M$,
$\psi_B$ and $\psi_{\tilde B}$) are collected in Table 2. At the 
fundamental
level we have to take into account only $\psi_Q$ and  $\psi_{\tilde 
Q}$
since only these fields have both $U(1)_R$ charge and transform 
non-trivially
with respect to $SU(N_f)$. The corresponding triangle is proportional
to $-N_fT_{\rm fund} - N_f T_{\rm anti-fund} \equiv - N_f$. The 
factor $N_f$
appears since we have $N_f$ fundamentals
and $N_f$ anti-fundamentals.
 Here $T$ is (one half of) the Dynkin index defined as follows. 
Assume we 
have the matrices of the generators of the
group $G$ in the representation $R$. Then
$$
\mbox{Tr}\, (T^a T^b )_R = T_R\, \delta^{ab}\, .
$$
For the fundamental representation $T=1/2$ while for the adjoint
representation of $SU(N)$ the index $T = N$. Now, let  us calculate 
the
same triangle at the level of the composite fermions. From Table 2 it 
is
obvious that we have to consider only $\psi_M$, and the 
corresponding
contribution is $-T_{\rm adjoint} = - N_f$. The match is perfect.

 The balance in the  $U(1)_R^3$ triangle looks as follows.
At the fundamental level we include $\psi_Q$,  $\psi_{\tilde Q}$ and
$\lambda$ and get $2N_f^2(-1)^3 + (N_f^2 -1) = -N_f^2 -1$.
At the composite fermion level we include $\psi_M$,
$\psi_B$ and $\psi_{\tilde B}$ and get $(N_f^2 -1)(-1)^3 -2 =
-N_f^2 -1$.

By the same token one can check that $U(1)_B^2U(1)_R$ triangle
gives $-2N_f^2$ both at the fundamental and composite levels. The 
$U(1)_R$
case requires a special comment.  The coupling of all fermions to 
gravity
is universal. Therefore, the  coefficient
in Eq. (\ref{Rgrav}) merely counts the number of the
fermion degrees of freedom 
weighed with their $R$ charges. At the fundamental level
we, obviously, have $- N_f^2- N_f^2 +(N_f^2-1) = - N_f^2 -1$,
while
at the composite level the coefficient is $-(N_f^2-1) -1-1 = - N_f^2 
-1$.
Again, the match is perfect.

Thus, the massless fermion content of the theory
is consistent with the regime implied by Eq. (\ref{bbvac}) --
spontaneous breaking of the chiral $SU(N_f)_R\times
SU(N_f)_L$ down to vector 
$SU(N_f)$. The baryon and the $U(1)_R$ currents remain unbroken. 
This regime is rather similar to what we have in ordinary QCD.
The  unconventional aspect, as was stressed above,
is the presence of the conserved unbroken $R$ current which has the 
axial
component.

This does not mean, however, that all points
from  the vacuum valley  
are so reminiscent of  QCD. Other points are characterized by
 different
dynamical regimes, with drastic distinctions in the  most salient 
features
of the emerging picture. To illustrate this  statement
let us consider, instead of Eq. (\ref{bbvac}), another point
\beq
B = -\tilde B =\Lambda^{N_f}\, ,\,\,\,  M^i_j = 0 \, .
\label{anpo}
\eeq
This point is characterized by a fully unbroken  chiral 
$SU(N_f)\times 
SU(N_f)$
symmetry,  in addition to the  unbroken  $R$ symmetry. The only 
broken
generator is that of $U(1)_B$. 

This regime is exceptionally unusual from the point of view
of the QCD practitioner.
As a matter of fact, the emerging picture is directly opposite to
what we got used to in QCD: the axial $SU(N_f)$ generators remain 
unbroken
while the vector baryon charge generator is spontaneously broken.

As is well-known, spontaneous breaking of the vector symmetries is
forbidden in QCD \cite{VW}. The no-go theorem of Ref. \cite{VW}
is based only on very general features of QCD -- namely, on the
vector nature of the quark-gluon vertex. Where does the
 no-go theorem fails in  SQCD?

The answer is quite obvious. The spontaneous breaking of
the  baryon charge generator in SQCD,  apparently defying the no-go 
theorem of Ref. \cite{VW}, is due to the fact that
in SQCD we have scalar quarks (and the quark-squark-gluino 
interaction)
which invalidates the starting assumptions of the theorem.

Moreover,  in QCD general arguments, based on the 't Hooft 
consistency 
condition and $N_c$ counting, strongly disfavor the possibility
of the linearly realized axial $SU(N_f)$  \cite{ColW}. Although I 
do
not say here that the consideration of Ref. \cite{ColW} proves the 
axial 
$SU(N_f)$  to be spontaneously broken in QCD, the living
space left for this option  is extremely narrow. The linear realization 
is
not ruled out   at all only because the argument of Ref. \cite{ColW} 
is based on an assumption regarding the $N_c$ dependence 
(discussed below) 
which is
absolutely natural but still was not derived from the first principles. 
Certain subtleties which I cannot explain now due to time limitations
might, in principle, invalidate this assumption. Leaving aside these -- 
quite 
unnatural -- subtleties one can say that the linear realization of
the axial $SU(N_f)$ is impossible in QCD. 

At the same time, this is exactly what happens in SQCD in the regime
specified by Eq. (\ref{anpo}). Again, the scalar quarks are to blame 
for the
failure of the argument presented in  Ref. \cite{ColW}.  In QCD it is 
difficult to 
imagine how massless baryons could saturate  anomalous triangles
since the baryons are composed of $N_c$ quarks; the corresponding
contribution naturally tends to be suppressed as $\exp (-N_c)$ at
large $N_c$. In SQCD there exist fermion states built from one
quark and one (anti)squark whose contribution to the triangle is not 
exponentially 
suppressed. 

After this introductory remark it is time to check that the 't Hooft
consistency conditions are indeed saturated. The triangles to be 
analyzed are
$$
SU(N_f)^3\, , \,\,\,   SU(N_f)^2U(1)_R\, , \,\,\,  U(1)_R^3\, , \,\,\,
\mbox{and}\,\,\, U(1)_R\, .
$$
The $SU(N_f)$ symmetry is either $SU(N_f)_R$ or $SU(N_f)_L$,
but the  triangles are the same for both. It is necessary to take
into account the fact that the fluctuations around the expectation 
values
(\ref{anpo}) subject to the constraint (\ref{qmodnf}) are slightly 
different 
than
those indicated in Table 2. Namely, the matrix of fluctuations
$M^i_j $ need not be traceless any longer; correspondingly, there are 
$N_f^2$
 fermions in  this matrix  transforming as $(N_f ,\, \bar N_f )$
representation of $SU(N_f)_R\times SU(N_f)_L$. At the same time the
fluctuations of $B$ and $\tilde B$ are not independent now,
so that $B-\Lambda^{N_f}= \tilde B +\Lambda^{N_f}$. One should 
count only 
one of them. The $U(1)_R$ quantum numbers remain intact, of 
course.

With this information in hands,  matching of the triangles becomes
a straightforward exercise. For instance, the $SU(N_f)^3$ triangle 
obviously 
yields
$N_f D_{\rm fund}$ both at the quark and composite levels. Here 
$D_{\rm fund}$ is the cubic Casimir operator for the fundamental 
representation defined as follows
$$
{\rm Tr }\, (T^a\{ T^b ,T^c\}) = d^{abc} D\, ;
$$
the matrices of the generators $T^a$ are taken in the given 
representation, 
and  the braces denote
the anticommutator; $d^{abc}$ stand for the $d$ symbols. 
The $SU(N_f)^2U(1)_R$ triangle  yields
$-N_f T_{\rm fund}$ both at the quark and composite levels. Both 
triangles,
$SU(N_f)^3$ and $SU(N_f)^2U(1)_R$,
are saturated by $\psi_M$. Passing to $U(1)_R^3$ we must add
the gluino contribution at the fundamental level and that of 
$\psi_B$ at the composite level. At both levels the coefficient of the 
$U(1)_R^3$
triangle is $-N_f^2 -1$. Finally, $U(1)_R$ counts the number of 
degrees
of freedom weighed with the corresponding $R$ charges. The 
corresponding coefficient again turns out to be the same, $-N_f^2 -1$.

Summarizing, the massless composite fermions residing in the
moduli superfields $M, B, \tilde B$ saturate all anomalies induced by 
the 
symmetries
that are supposed to be realized linearly. The conjecture of the
unbroken $SU(N_f)_R\times SU(N_f)_L$ and spontaneously broken
$U(1)_B$ at the point (\ref{anpo}) goes through.
As a matter of fact, some of these anomaly matching conditions were 
observed
long ago, in Refs. \cite{ADS1,ADS2}. 

Sometimes it is convenient to mimic the constraint
(\ref{qmodnf}) by introducing a Lagrange multiplier superfield
$X$ with the superpotential
\beq
{\cal W} = X \left( {\rm det }\, M -B\tilde B -\Lambda^{2N_f}\right)\, 
.
\label{mimic}
\eeq

We could treat, in a similar fashion, any point belonging to the 
quantum moduli space (\ref{qmodnf}). For instance, we could travel 
from (\ref{bbvac}) to (\ref{anpo}) observing how the regime 
continuously changes from the broken axial $SU(N_f)$ to the broken 
baryon number. 

Concluding this part we remind that the case of the gauge group 
$SU(2)$ is 
exceptional.
Indeed, in this case, the matter sector consisting
of 2 fundamentals and 2 anti-fundamentals has $SU(4)$ global flavor
symmetry, rather than $SU(2)\times SU(2)$. This is because all
representations of $SU(2)$ are (pseudo)real, and fundamentals
can be transformed into anti-fundamentals and {\em vice versa} by 
applying
the $\epsilon^{\alpha\beta}$ symbol. This peculiarity was already 
discussed
in detail in Sect. 2. 
Under the circumstances the pattern of the global symmetry 
breaking is somewhat 
different, and the saturation of the anomaly triangles must be 
checked anew.
Although this is a relatively simple exercise, we will not do it here. 
The 
interested 
reader is referred to \cite{Seib1}.

\vspace{0.3cm}

${\bf N_f = N_c+1}$ 

\vspace{0.1cm}

The 
general strategy is the same as in the previous case. We introduce 
the mass
term (\ref{mt}) assuming that two eigenvalues of the mass matrix 
are large
while others are small. Then two heavy flavors can be integrated 
over, leaving 
us with the theory with $N_f = N_c -1$, which can be analyzed in the 
weak
coupling regime.  A superpotential is generated on the vacuum 
valley.
Using this superpotential it is not difficult to get 
the vacuum expectation values of the moduli fields $M, B_i , \tilde B^j 
$.
They turn out to be   constrained by the following relation
\cite{Seib1}:
\beq
{\rm det}\, M \left(\frac{1}{M}\right)_i^j -B_i\tilde B^j =
\Lambda^{2N_c -1} m_i^j \, .
\label{c+1}
\eeq

Note that the vanishing of the determinant, det$M=0$,
which at the classical level automatically follows from the
definition of $M^i_j$, is gone for the quantum VEV's
$\langle Q^i\tilde Q_j\rangle$. This is most readily seen
if the mass matrix $m^j_i \ra m\delta^j_i$. In this case
 $\langle Q^i\tilde Q_j\rangle \sim (\Lambda^{2N_c-
1}m)^{1/N_c}\delta_j^i$,
$i,j = 1, ..., N_f$. 

In the massless limit $m_i^j\ra 0$  the quantum constraint 
(\ref{c+1}) 
coincides
with the classical one (\ref{minor}), or (\ref{nf+1}). Thus, the 
quantum and 
classical moduli spaces are identical. 
Every point from the vacuum valley can be reached by
adding appropriate perturbations to the Lagrangian (i.e.
mass terms and/or $\beta B +\tilde\beta \tilde B$).

The only point which deserves special 
investigation is the origin, which, unlike the situation $N_f=N_c$,  
remains 
singular. This  is a signature of  massless fields.
Classically we have  massless gluons and massless moduli fields. 
In the strong coupling regime we expect the gluons to acquire a 
dynamical
mass gap. The classical moduli subject to the constraint 
(\ref{nf+1})  need not be the only composite
massless states, however. Other composite
massless states may form too. We will see shortly that
they actually appear. At the origin, when $M=B=\tilde B =0$,
{\em all} global symmetries of the Lagrangian are
presumably  unbroken.
In particular, the axial $SU(N_f)$ is realized linearly. Although we 
have 
already learned , from the previous example, that such a regime
seems to be  attainable in SQCD (in sharp contradistinction with QCD), 
the case 
$N_f=N_c 
+1$
is even more remarkable -- we want {\em all} global 
symmetries
to be realized linearly. (For $N_f = N_c$, in the vacuum where
the axial $SU(N_f)$ symmetry was unbroken, the baryon charge 
generators 
were 
spontaneously broken.)

At the origin (and near the origin) the theory is in the strong 
coupling regime. Let us examine the 
behavior 
of the theory in this domain more carefully. When the expectation 
values of 
all moduli fields vanish, the global symmetry $SU(N_f) \times
SU(N_f) \times U(1)_B \times U(1)_R$ is unbroken provided no other 
(non-chiral) condensates develop. Is this solution self-consistent?

To answer this question we will try to match  all corresponding 
anomalous $AVV$ triangles; in  this 
case 
we have seven triangles, 
$$
SU(N_f)^3\, ,\,\,\,  SU(N_f)^2U(1)_B\, ,\,\,\,  SU(N_f)^2U(1)_R\, ,
$$
\beq
U(1)_RU(1)_B^2 \, ,\,\,\, U(1)_R^3\, , \,\,\, U(1)_R^2U(1)_B
\, , \,\,\,  U(1)_R\, .
\label{avvtr}
\eeq
They must be matched by the composite massless baryons residing 
in $M$, 
$B_i$
and $\tilde B^j$. As we will see shortly, to achieve the matching
we will need to consider all components of $M$, $B_i$
and $\tilde B^j$ as independent, ignoring the constraint
\beq
{\rm det}\, M \left(\frac{1}{M}\right)_i^j -B_i\tilde B^j = 0
\label{paramvv}
\eeq
defining the vacuum valley both at the classical and the quantum 
levels.
In other words, we will have to deal with a larger number of  
massless fields
than one could infer from the parametrization (\ref{paramvv})
of the vacuum valley. The constraint (\ref{paramvv}) on the vacuum 
valley 
will reappear due to the fact that the expanded set of the 
massless fields
gets a superpotential ${\cal W}(M,B,\tilde B )$. The requirement of 
the 
vanishing of the $F$ term
will return us Eq. (\ref{paramvv}).

Thus, our first task is to verify the matching. The quantum numbers
of the fundamental quarks and the composite massless fermions
can be inferred from Table 1. For convenience we collect them in 
Table 3.

\vspace{0.2cm}

\begin{table}
\begin{center}
\begin{tabular}{|c|c|c|c|c|}\hline
 $ ~~$~~~  & ~~ $SU(N_f)_L$ ~~ & ~~$SU(N_f)_R$ ~~ &~~ $U(1)_B$ ~~ &
~~$U(1)_R$~~ \\
\hline 
$\psi_M$ & $ N_f$  &$\bar{N}_f$ &0&$ \frac{2}{N_f}-1$\\
$\psi_B$ &$\bar{N}_f$& 1 &$N_f-1$&$-\frac{1}{N_f}$\\
$\psi_{\tilde B}$ &1 &${N}_f$&$-N_f+1$&$-\frac{1}{N_f}$\\
\hline
$\psi_Q$ & $ N_f$  & 1 & 1 &  $ \frac{1}{N_f}-1$ \\ 
$\psi_{\tilde{Q}}$ & 1 & $\bar{N}_f$ & $-1$ &$ \frac{1}{N_f}-1$\\
$\lambda$ & 1&1&0&1\\
\hline
\end{tabular}
\end{center}
\caption{The quantum numbers of the massless fermions
in SQCD with $N_f=N_c+1$}
\end{table}

Since we already  have a considerable experience in matching the 
$AVV$ 
triangles, I will not discuss all triangles from Eq. (\ref{avvtr}). As an 
exercise 
let us do just one of them, namely $U(1)_R^3$.
In this case, at the fundamental level we have the $\psi_Q, 
\psi_{\tilde Q}$
and $\lambda$ triangles which yield
$$
2N_cN_f \left(\frac{1}{N_f}-1\right)^3 +N_c^2 - 1\, .
$$
At the composite level the $U(1)_R^3$ anomalous triangle
is contributed by $\psi_M$ ($N_f^2$ degrees of freedom), $\psi_{B}$ 
and 
$\psi_{\tilde B}$ (each has $N_f$ degrees of freedom). Thus, we get
$$
N_f^2 \left(\frac{2}{N_f}-1\right)^3 - 
2N_f\left(\frac{1}{N_f}\right)^3\, .
$$
Both expressions reduce to
$$
-N_f^2 +6N_f -12 +\frac{8}{N_f}-\frac{2}{N_f^2}\, .
$$
Other triangles match too, in a miraculous way. Namely,
$$
SU(N_f)^3 \ra  (N_f-1) D_{\rm fund}\, ,
$$
$$
SU(N_f)^2U(1)_R \ra  -\frac{(N_f -1)^2}{N_f} T_{\rm fund}\, ,
$$
$$
U(1)_B^2U(1)_R \ra - 2(N_f-1)^2\, , 
$$
$$
SU(N_f)^2U(1)_B \ra  (N_f -1) T_{\rm fund}\, ,
$$
$$
U(1)_R^2U(1)_B \ra 0 \, ,
$$
$$
U(1)_R \ra -N_f^2+2N_f -2 \, .
$$
The matching discussed above, was  observed many years ago in Ref.
\cite{Amati} where the spectrum of the composite  massless particles
corresponding to the unconstrained $M,B$ and $\tilde B$
was conjectured.

Thus, the above spectrum of the composite  massless particles 
appearing at 
the origin of the
vacuum valley in the $N_f=N_c+1$ theory is self-consistent. We 
know, 
however,
that the vacuum valley in the model at hand is characterized by Eq.
(\ref{paramvv}). The situation seems rather puzzling.
How the constraint (\ref{paramvv}) might appear?

The answer to this question was given by Seiberg \cite{Seib1}.
If the massless fields, residing in the unconstrained $M,B$ and 
$\tilde B$,
acquire a superpotential, then the vacuum values of the moduli fields
are obtained through the condition of vanishing $F$ terms. The 
``right" 
superpotential will lead to Eq. (\ref{paramvv}) automatically.

So, what is the right superpotential? If it is generated, several 
requirements
are to be met. First, it must be invariant under all global symmetries
of the model, including the $R$ symmetry. Second, the vacuum 
valleys
obtained from this superpotential must correspond to Eq. 
(\ref{paramvv}).
Third, away from the origin the only massless fields must be those
compatible with the constraint (\ref{paramvv}).

All these requirements are satisfied by the following superpotential
\cite{Seib1}:
\beq
{\cal W} =\frac{1}{\Lambda^{2N_f-3}}
\left( B_iM^i_j\tilde B^j -{\rm det}\, M\right) \, .
\label{spot}
\eeq
It is obvious that the condition of vanishing of the $F$ terms
corresponding to $M_j^i$ identically coincides with Eq. 
(\ref{paramvv});
moreover, vanishing of the $F$ terms
corresponding to $B$ and $\tilde B$ yields to remaining constraints,
$B_i M_j^i = \tilde B^j M_j^i = 0$. Once we move away
from the origin,  the moduli $M_j^i$ grow,
the fields $B_i, \tilde B^j$ acquire  masses and can be integrated out.
This eliminates $2N_f$ degrees of freedom. This is exactly the
amount of the redundant degrees of freedom, see Sect. 3.1.  The 
emerging
 low-energy theory for the remaining degrees
of freedom  $M_j^i$ has no superpotential. When 
$M_j^i\gg\Lambda^2$, the fields $B_i, \tilde B^j$ are very heavy,
and the low-energy description based on Eq. (\ref{spot}) is 
no longer legitimate. It is interesting to trace the fate of the
baryons $B_i, \tilde B^j$ in the process of this evolution from small
to large values of $M_j^i$.
This question has not been addressed in the literature so far.

Let us pause here to summarize the features of the dynamical regime 
taking place in  the $N_f = N_c +1$ model. The space of vacua
is the same at the classical and quantum levels, the origin being 
singular
due to the existence of the massless degrees of freedom. 
Since we have Higgs fields in the fundamental representation 
the theory is in the Higgs/confining phase; at  the origin 
and
near the origin the theory is strongly coupled and ``confines" in the 
sense that physics is
adequately described in terms of gauge invariant composites and
their interactions. We think that at  the origin all global symmetries 
of the 
Lagrangian
are unbroken. The number of the massless degrees
of freedom here is larger than the dimensionality of the space of 
vacua.
To get the right description of the space of vacua one needs a 
superpotential,
and such a superpotential is generated dynamically. It is a 
holomorphic
function of the massless composites. The vacuum valley for this
superpotential coincides with the quantum moduli space of the 
original 
theory.
As we move along the vacuum valley away from the origin there is 
no phase
transition -- the theory smoothly goes into the weak coupling
Higgs phase.  The ``extra" massless fields become massive,
and irrelevant for the description of the vacuum valley.

The dynamical regime with the above properties got a special
name --
 now it is referred to as $s$-{\em confinement}.

Seiberg's example of the $s$-confining theory was the first, but not 
the last.
Other theories with the similar behavior were found, see e.g. 
\cite{s1} -- \cite{s7}. The set of the $s$-confining models 
includes even
such exotic one as the gauge group $G_2$ (this is an exceptional 
group),
with five fundamentals \cite{s6,s7}. As a matter of fact, it is not 
difficult to 
work out a general  strategy allowing one to carry out a systematic
search of all $s$-confining theories. This was done in Ref.  \cite{s8}. 
Without submerging into excessive technical detail
let me outline just one basic point of the procedure
suggested in \cite{s8}. 

A necessary condition of the $s$-confinement is generation of a
superpotential at the origin of the moduli space, a holomorphic
function of relevant moduli fields. Generically, the  form of this 
superpotential, dictated by the $R$ symmetry plus the dimensional 
arguments  is
\beq
{\cal W} \propto \left[ \prod_\ell S_\ell^{2T(R_\ell 
)}\Lambda^{3T(G)-
\sum_\ell T(R_\ell )}\right]^{1/(\sum_\ell T(R_\ell ) -T(G))}\, .
\label{s-sup-pot}
\eeq
The product (sum) runs over all matter fields present in the theory.
For instance, in the case of SQCD for each flavor we have to include 
two
subflavors.  I remind that $T(R)$ is (one half of) the Dynkin index.
Particular combinations of the superfields in the
product are not specified; they depend on the particular 
representations
of the matter fields with respect to the gauge group. What is 
important
is only the fact that they all are homogeneous functions of $S_\ell$'s,
of order $2T(R_\ell )$. Note that the combination appearing in Eq.
(\ref{s-sup-pot}) is the only one which has correct properties under
renormalization, i.e. compatible with the NSVZ $\beta$ function.

Now, if we want the origin to be analytic (and this is a feature of the
$s$-confinement, by definition), we must ensure that
\beq
\sum_\ell T(R_\ell ) -T(G) = 1
\label{sconfc}
\eeq
(more generically, 1/integer). This severely limits the choice
of possible representations since the Dynkin indices are integers.
For instance, if the matter sector is vector-like,  there exist only two 
options:  
(i)  Seiberg's model, $SU(N_c)$ color with $N_c+1$ flavors (i.e. 
$N_c+1$   
fundamentals and $N_c+1$
anti-fundamentals; (ii) $SU(N_c)$ color with one antisymmetric 
tensor plus its 
adjoint plus three flavors. 

Not to make a false impression I hasten to 
add that 
some models that satisfy Eq. (\ref{sconfc}), are not $s$-confining.
A few simple requirements to be met, which comprise a sufficient 
condition
for $s$-confinement, are summarized in Ref. \cite{s8}, which gives 
also
a full list of the $s$-confining theories.

\subsection{Conformal window. Duality}

Our excursion towards larger values of $N_f$ must be temporarily
interrupted here -- the methods we used so far fail at $N_f=N_c+2$.
One can show that the quantum moduli space coincides with the 
classical 
one, just as in the case $N_f=N_c+1$. However, at the origin of the 
moduli 
space, description of the large-distance behavior of the theory in 
terms of the 
massless
fields residing in $M, B$ and $\tilde B$ does not go through. These 
degrees
of freedom are irrelevant for this purpose; the dynamical regime of 
the theory 
in the infrared is different.
To see that $M, B$ and $\tilde B$ do not fit suffice it to try to 
saturate the 't Hooft triangles corresponding to the unbroken global 
symmetries,  in the same vein as we did previously for $N_f=N_c+1$.
There is no matching! As we will see shortly, the dynamical regime
does indeed change in passing from $N_f=N_c+1$ to $N_f=N_c+2$.
The correlation functions of the $N_f=N_c+2$ theory at large 
distances
are those of a free theory, like in massless electrodynamics.
But the number of free degrees of freedom (``photons" and 
``photinos") is different from
from what one might expect naively. Namely, we will have three 
``photons" and three ``photinos" in the case at hand, in addition
to $(N_c+2)\times 2$ free ``fermion" fields. These photons and 
photinos, in a sense, may be considered as the bound states of the 
original gluons, gluinos, quarks and squarks. 

To elucidate this, rather surprising, picture
we will have  to make a jump in our travel along the $N_f$ axis, 
leave the domain of $N_f$ close to 
$N_c$ for a 
while,
and turn to much larger values of $N_f$. The critical points on  the
$N_f$ axis are  $3N_c$ and $3N_c/2$. That's where a
{\em conformal window} starts and ends. We will return to the 
$N_f=N_c+2$ and $N_f=N_c+1$
theories 
later on.

At first, let me remind a few well-known facts from ordinary 
non-supersymmetric QCD. 
The Gell-Mann-Low function in QCD has the form \cite{PRD}
$$
\beta_{\rm QCD} (\alpha_s ) = -\beta_0 \frac{\alpha_s ^2}{2\pi}
- \beta_1 \frac{\alpha_s ^3}{4\pi^2} - ...\, ,
$$
\beq
\beta_0 = 11 -\frac{2}{3} N_f\, , \,\,\, \beta_1 = 51 -\frac{19}{3} 
N_f\,  .
\label{betaQCD}
\eeq
At small $\alpha_s$ it is negative since the first term always 
dominates. This is the celebrated asymptotic freedom. With the scale 
$\mu$ decreasing the running gauge coupling constant grows,
and the second term becomes important. Generically the second term 
takes over the first one at $\alpha_s/\pi \sim 1$, when all terms in 
the $\alpha_s$ expansion are equally important, i.e. in the strong 
coupling regime. Assume, however, that for some reasons the first 
coefficient $\beta_0$ is abnormally small, and this smallness does 
not 
propagate to higher orders. Then the second term catches up with 
the first one when $\alpha_s/\pi \ll 1$, we are in the weak coupling 
regime, and higher order terms are inessential. Inspection of Eq. 
(\ref{betaQCD}) shows that this happens when $N_f$ is close
to $33/2$, say 16 or 15 ($N_f$ has to be less than $33/2$ to ensure 
asymptotic freedom). For these values of $N_f$ the second coefficient
$\beta_1$ turns out to be negative! This means that the $\beta$ 
function develops a zero in the weak coupling regime, at
\beq
\frac{\alpha_{s}^{*}}{2\pi} = \frac{\beta_0}{-\beta_1} \ll 1\, .
\eeq 
(Say, if $N_f = 15$ the critical value is at 1/44.)
This zero is nothing but the infrared fixed point of the theory. 
At large distances $\alpha_s \ra \alpha_{s}^{*}$,
and $\beta (\alpha_{s}^{*}) = 0$, implying that the trace of the 
energy-momentum vanishes, and the theory is in the conformal 
regime. There are no localized particle-like states in the spectrum of 
the theory; rather we deal with massless unconfined interacting 
quarks and gluons; all correlation functions at large distances exhibit 
a power-like behavior. In particular, the potential between two 
heavy static quarks at large distances $R$ will behave as
$\sim \alpha_{s}^{*}/R$. The situation is not drastically different 
from 
conventional QED. The corresponding dynamical regime is, 
thus, a non-Abelian Coulomb phase. As long as $\alpha_{s}^{*}$
is small, the interaction of the massless quarks and gluons
in the theory is weak at all distances, short and large, and is 
amenable to the standard perturbative treatment (renormalization 
group, etc.). 
QCD becomes a fully calculable theory. 

There is nothing remarkable in the observation that, for a certain
choice of $N_f$, quantum chromodynamics becomes
conformal and weakly coupled in the infrared limit. 
Belavin and Migdal  played with this model over 20 years ago
\cite{MB}. They were quite excited explaining how great it
would be if $N_f$ in our world  were close to 16, and the theory 
would be in the infrared  conformal
regime, with calculable anomalous dimensions. Later on this idea
was discussed also by Banks and Zaks \cite{BZ}. Alas, we do not live 
in the world with $N_f\approx 16$...

What is much more remarkable is the existence of the  infrared 
conformal regime in SQCD for large couplings, $\alpha_*/\pi\sim 1$.
This fact, as many others in the given range of questions, was 
established by Seiberg \cite{Sei-conf}. The discovery of the strong 
coupling conformal regime \cite{Sei-conf} is based on the so-called
{\em electric-magnetic duality}. Although the term suggests the 
presence of electromagnetism and the same kind of duality under 
the substitution $\vec E \leftrightarrow \vec B$ 
one sees in the Maxwell theory, actually both elements,
``electric-magnetic" and ``duality" in the given context are nothing 
but remote analogies,
as we will see shortly. 

Analysis starts from consideration of SQCD with $N_f$
slightly smaller than $3N_c$. More exactly, if
\beq
\varepsilon \equiv 1-\frac{N_f}{3N_c}\, 
\eeq
we assume that
$N_c\ra\infty$, and $0<\varepsilon \ll 1$. It is assumed also that
we are at the origin of the moduli space -- no fields develop VEV's. 
By examining the NSVZ 
$\beta$ function, Eq. (\ref{NSVZbeta}), it is easy to see that in this 
limit the
first coefficient of the $\beta$ function is abnormally small,
and the second coefficient is positive and is of a normal order of 
magnitude,
\beq
\beta_0 = 3N_c\varepsilon\, , \,\,\, \beta_1 = -3N_c^2 + {\cal 
O}(\varepsilon )\, .
\eeq
To get $\beta_1$ I used the fact that
\beq
\gamma (\alpha ) = - \frac{N_c^2-1}{2N_c}\,\frac{\alpha}{\pi} +
{\cal O}(\alpha ^2 )
\label{anomdim}
\eeq
in the model considered (for a pedagogical review see e.g. the last 
paper in Ref. 
\cite{VZS}). 
There is a complete parallel with the conformal QCD, with 15 or 16 
flavors, discussed above. The numerator of the NSVZ $\beta$ 
function
vanishes at 
\beq
\gamma (\alpha_* ) \equiv \gamma_* 
= 1-3\frac{N_c}{N_f}\ra -\varepsilon\,
\,\,\, {\mbox or} \,\,\,   \frac{N_c\alpha_*}{2\pi} =\varepsilon\, .
\eeq
The vanishing of the $\beta$ function marks the onset of the
conformal regime in the infrared domain; the fact that
$\alpha_*$ is small means that the theory is weakly coupled
in the infrared (it is weakly coupled in the ultraviolet too since it is 
asymptotically free).
 
Here comes the breakthrough observation of Seiberg. 
Compelling arguments can be presented indicating  
that the original 
theory with the  $SU(N_c)$ gauge group (let us call this theory 
``electric"),  and another theory, with  the $SU(N_f-N_c)$ 
gauge group, the same number of flavors  $N_f$ as above, and a 
specific Yukawa interaction (let us call this theory ``magnetic"),
 flow to one and the same limit in the
infrared asymptotics. The corresponding 
Gell-Mann-Low functions of both theories
vanish  at their corresponding critical values of the coupling 
constants. Both theories are in the non-Abelian Coulomb (conformal) 
phases.
By inspecting Eq. (\ref{NSVZbeta}) it is easy to see that, when
$\alpha_*$ in the electric theory approaches zero (i.e. 
$\varepsilon\ra 0$), in the magnetic theory $\gamma (\alpha_*)$
approaches $-1$, i.e. the theory becomes strongly coupled. 
The opposite is also true. When the magnetic theory becomes weakly 
coupled, i.e.
\beq
N_f\ra \frac{3}{2} N_c\,\,\,\mbox{from above}\, ,\,\,\,
(\alpha_*)_{\rm magn} \ra 0\, ,
\eeq
in the electric theory $(\gamma_*)_{\rm electr}\ra -1$, and the 
electric theory is strongly coupled in the infrared. This reciprocity 
relation is, probably, the reason why the correspondence between 
the two theories is referred to as the electric-magnetic duality.
It is worth emphasizing that the correspondence takes place only in 
the infrared limit. By no means the above two theories are totally 
equivalent to each other; their ultraviolet behavior is completely 
different. If $N_f<(3N_c)/2$ the magnetic theory looses asymptotic 
freedom. Thus, the {\em conformal window}, where both theories
are asymptotically free in the ultraviolet and conformally invariant 
in the infrared extends in the interval
 $3N_c/2 \leq N_f \leq 3N_c$. 

The fact that the conformal window cannot stretch below $3N_c/2$
is seen from consideration of the electric theory {\em per se},
with no reference to the magnetic theory. Indeed, the total
(normal + anomalous) dimension of the matter field $Q\tilde Q$ in
the infrared limit is equal to 
$$
 d= 2 +\gamma_* = \frac{3(N_f-N_c)}{N_f}\, .
$$
No physical field can have dimension less than unity; this is 
forbidden by the K\"{a}ll\'{e}n-Lehmann spectral representation. 
If $d=1$ the field is free. The dimension $d$ reaches unity
exactly at $3N_c/2$. Decreasing $N_f$ further and assuming that
the conformal regime $\beta (\alpha_*)=0$ is still preserved would 
violate
the requirement $d\geq 1$.

Let us describe the electric  and magnetic theories in more detail.
I will continue to denote the quark fields of the first theory as $Q$,
while those of the magnetic theory will be denoted by $ {\cal Q} $.
Both have $N_f$ flavors, i.e. $2N_f$ 
chiral superfields in the matter sector.
The same number of flavors is necessary to ensure
that global symmetries of the both theories are identical.

 The magnetic theory, 
additionally, has $N_f^2$ colorless ``meson" superfields
$ {\cal M}^i_j$ whose quantum numbers are such as if they were 
built from 
a quark and an antiquark. The meson superfields are coupled to the 
quark ones of the magnetic theory through a superpotential
\beq
{\cal W} = f {\cal M}^i_j {\cal Q}_i\tilde {\cal Q}^j \, .
\label{msp}
\eeq
The quantum numbers of the fields belonging to the matter sectors
of the magnetic and electric theories are summarized in Tables 4 and 
5.

\begin{table}
\begin{center}
\begin{tabular}{|c|c|c|c|c|}\hline
 $ ~~$~~~  & ~~ $SU(N_f)_L$ ~~ & ~~$SU(N_f)_R$ ~~ &~~ $U(1)_B$ ~~ &
~~$U(1)_R$~~ \\\hline 
$\psi_Q$ & $ N_f$  & 0 & 1 &  $  - N_c/N_f$ \\ 
$\psi_{\tilde{Q}}$ & 0 & $\bar{N}_f$ &$ -1 $&$ -  N_c/N_f$\\
\hline
\end{tabular}
\end{center}
\caption{The quantum numbers of the massless fields of the 
``electric" theory from the dual pair}
\end{table}

\begin{table}
\begin{center}
\begin{tabular}{|c|c|c|c|c|}\hline
 $ ~~$~~~  & ~~ $SU(N_f)_L$ ~~ & ~~$SU(N_f)_R$ ~~ &~~ $U(1)_B$ ~~ &
~~$U(1)_R$~~ \\ \hline 
$\psi_{\cal Q} $ & $ \bar{N}_f$  & $0$ & $N_c/(N_f-N_c)$ &  $ (N_c-
N_f)/N_f$
 \\ 
$\psi_{\tilde{\cal Q}}$ & $0$ & $N_f$ & $- N_c/(N_f-N_c)$ &$ (N_c-
N_f)/N_f$
\\ 
$\chi =\psi_{\cal M}$ &  $N_f$ & $\bar{N}_f$ & $0$ & $(N_f - 
2N_c)/N_f$ \\
\hline
\end{tabular}
\end{center}
\caption{The quantum numbers of the massless fields of the 
``magnetic" theory from the dual pair}
\end{table}

The quantum numbers of the meson superfield are fixed by the 
superpotential (\ref{msp}).
Note a very peculiar relation between the baryon charges of the 
quarks in the electric and magnetic theories. This relation shows that
the quarks of the magnetic theory cannot be expressed, in any 
polynomial way, through quarks of the electric theory. 
The connection of one to another is presumably extremely non-local 
and complicated. The explicit connection between the operators
in  the dual pairs is known only for a handful of operators
which have a symmetry nature \cite{Kut}. 

We can now proceed to the arguments establishing the equivalence 
of these two theories in the infrared limit. The main tool we have at 
our 
disposal for establishing the equivalence is again the 't Hooft 
matching, the 
same line of reasoning as was used above in verifying various 
dynamical regimes in $N_f=N_c$ and $N_f = N_c +1$ models. Since 
we are at the origin of the moduli space, all global symmetries are 
unbroken, and one has to check six highly non-trivial matching 
conditions corresponding to various
triangles with the $SU(N_f)$, $U(1)_R$ and $U(1)_B$ currents
in the vertices.

The presence of fermions from the  meson multiplet ${\cal M}$ is   
absolutely
crucial for this  matching. Specifically, one finds for the one-loop
anomalies in both theories \cite{Sei-conf}: 
\begin{eqnarray} 
SU(N_f)^{3}  &\rightarrow & N_{c}D_{\rm fund},
\nonumber \\  
SU(N_f)^{2}U(1)_R  & \rightarrow &   -\frac{N_c^2}{N_f}
T_{\rm fund}, \nonumber \\ 
SU(N_f)^{2}U(1)_B  &\rightarrow &  N_c
T_{\rm fund}, \nonumber \\ 
U(1)^2_B U(1)_R  &\rightarrow &  -2N_c^2,
\nonumber \\ U(1)_R^3  & \rightarrow &  N_c^2 - 1 - 
2\frac{N_c^4}{N_f^2} ,\nonumber \\ 
U(1)_R  & \rightarrow &  -N_c^2 - 1 \, . 
\label{seibergmatching}
\end{eqnarray}

For example, in the $U(1)_R^3$ anomaly in the electric  theory the
gluino  contribution is proportional  to $N_c^2 - 1$ and  that of 
quarks to $-(N_c/N_f)^3 2N_fN_c = - 2N_c^4/N_f^2$; altogether  
$N_c^2 -
1- 2N_c^4/N_f^2$ as in (\ref{seibergmatching}). In the dual theory 
one 
gets from gluino and quarks another
contribution, $(N_f-N_c)^2 - 1- 2(N_f-N_c)^4/N_f^2$. Then  the  
fermions
$\chi$ from the meson multiplet ${\cal M}$
add extra $[(N_f-2N_c)/N_f]^3 N_f^2$, which is  precisely the
difference. 

The last line in Eq. (\ref{seibergmatching}) corresponds to the 
anomaly
of the $R$ current in the background gravitational field.
In the electric $SU(N_c)$ theory  the corresponding
coefficient is 
$$ N_c^2-1 -
 2(N_c/N_f)N_c N_f = - N_c^2 - 1
$$ 
 while in the magnetic  theory
 it is $-(N_f-N_c)^2 - 1$ from quarks and gluinos and
 $[(N_f-2N_c)/N_f] N_f^2$ from the ${\cal M}$ fermions, i.e.  in the 
sum
 again $- N_c^2 - 1$.

It is not difficult to check the matching of other triangles from Eq. 
(\ref{seibergmatching}).
The dependence on $N_c$ and $N_f$ is rather sophisticated, and it is 
hard  to
imagine that  this is an accidental coincidence. The fact that the 
electric and magnetic theories described above have the same global 
symmetries is an additional  argument in favor of their (infrared)  
equivalence. Of course, they have different gauge symmetries: 
$SU(N_c)$ in the first case and $SU(N_f-N_c)$ in the second.
The gauge symmetry, however, is not a regular symmetry;
in fact, it is not a symmetry at all. Rather, it is a redundancy in the 
description of the theory. One introduces first more degrees of 
freedom than actually exist, and then the  redundant variables are 
killed by the gauge freedom. That's why the gauge symmetry has no 
reflection in the spectrum of the theory. Therefore, distinct gauge 
groups
do not preclude the theories from being dual, generally speaking.
On the contrary, the fact that such dual pairs are found is very 
intriguing; it
allows one to look at the gauge dynamics from a new angle.

I have just said that various dual pairs of supersymmetric gauge 
theories  are found. To avoid misunderstanding I hasten to add that
although Seiberg's line of reasoning is very compelling it still falls
short of {\em proving} the infrared equivalence. The theory in the 
strong coupling regime is not directly solved, and we are hardly any 
closer now to the solution than we were a decade ago. The infrared 
equivalence has the status of a good solid conjecture substantiated 
by 
a number of various indirect arguments we have in our disposal 
(Sect. 3.4).

If we accept this conjecture we can make a remarkable step forward 
compared to the conformal limit of QCD studied in the
weak coupling regime in the 70-ies and 80-ies. Indeed, if $N_f$ is 
close to $3N_c$ (but slightly lower), i.e. we are near the right edge of 
the conformal window, the weakly coupled electric theory is in the 
conformal regime. Since it is equivalent (in the infrared) to the 
magnetic theory, which is strongly coupled at these values of
$N_f$ we, thus, establish the existence of a  {\em strongly coupled}
superconformal gauge theory. Moreover, when  $N_f$ is slightly 
higher than $3N_c/2$, i.e. near the left edge of the conformal 
window,
the magnetic theory is weakly coupled and in the conformal regime.
Its dual, the electric theory, which is strongly coupled near the 
left edge of the conformal window, must then be in the conformal 
regime too. In the middle of the conformal window, when both 
theories are strongly coupled, strictly speaking we do not know 
whether or not they stay superconformal. In principle, it is possible 
that they both leave the conformal regime. This could happen, for 
instance, if $\alpha_*$, the solution of the equation $\gamma 
(\alpha_*) = 1 - 3 N_cN_f^{-1}$, (temporary) becomes larger than 
$2\pi /T(G)$,
the position of the zero of the {\em denominator} of the NSVZ 
$\beta$ function, as we go further away  from the point
$N_f=3N_c$ in the direction of $N_f=3N_c/2$, and then
$\alpha_*$ becomes smaller than $2\pi /T(G)$ again, as we
approach $N_f=3N_c/2$. Such a scenario, although not ruled out, does 
not seem likely, however. 

\subsection{Traveling along the valleys}

So far, the dual pair of theories was considered at the origin of the 
vacuum valley. 
Both theories, electric and magnetic, have the vacuum valleys,
and a natural question arises as to what happens if we move away 
from the origin \footnote{This question was suggested to me by C. 
Wetterich. Note that if in the electric theory the vacuum degeneracy 
manifests itself in arbitrary vacuum expectations of $Q$ and $\tilde 
Q$, in the magnetic theory the expectation values of ${\cal Q}$ and 
$\tilde{\cal  Q}$ vanish.  The flat direction corresponds
to arbitrary expectation value of ${\cal M}$.
}. As a matter of fact, this question is quite crucial,
since if the theories are equivalent in the infrared, a certain 
correspondence between them should persist not only at the origin,
but at any other point belonging to the vacuum valley.
If a correspondence can be found, it will only strengthen the
conjecture of duality.

Thus, let us start from the electric theory and move away from the 
origin. Consider for simplicity a particular direction in the
moduli space, namely, 
\beq
Q=\tilde Q= \left( \begin{array}{c}
a_1 \\
0 \\
... \\
0 \\
\end{array} \right)\, ,
\label{edol}
\eeq
where $Q, \tilde Q $ are the superfields comprising, say,  the first
flavor. 
Moving along this direction we break the gauge symmetry 
$SU(N_c)$ down to $SU(N_c-1)$; $2N_c - 1$ chiral superfields
are eaten up in the (super)-Higgs mechanism providing
masses to $2N_c - 1$ $W$ bosons. Below the mass scale of 
these $W$ bosons the effective theory is SQCD with $SU(N_c-1)$
gauge group and $N_f -1$ flavors. (Additionally there is
one $SU(N_c-1)$ singlet, but it plays no role in the gauge dynamics.)
It is not difficult to see that decreasing both $N_f$ and $N_c$ by one 
unit in the electric theory
we move rightwards along the axis $N_f/3N_c$. In other words,
we move towards the right edge of our conformal window,
making the electric theory weaker. From what we already know, we 
should then expect
that the magnetic theory becomes stronger.

Let us have a closer look at the magnetic theory. The vacuum 
expectation value (\ref{edol}) is reflected in the magnetic theory
as the expectation value of the $(1,1)$ component of the meson
field ${\cal M}$. No Higgs phenomenon takes place, but, rather, 
${\cal M}_{1}^1\neq 0$. Then, thanks to the superpotential 
(\ref{msp}), the magnetic quark ${\cal Q}_1, \tilde{\cal Q}_1$
gets a mass, and becomes irrelevant in the infrared limit.
The gauge group remains the same, $SU(N_f-N_c)$,
but the number of active flavors reduces by one unit
(we are left with $N_f-1$ active flavors). This means that the first
coefficient of the Gell-Mann-Low function of the magnetic theory 
becomes more negative and the critical value $\alpha_{*\rm magn}$
increases. The theory becomes coupled stronger, in full accord
with our expectations.

Let us now try the other way around. What happens if we introduce 
the mass term to one of the quarks in the electric theory, say the 
first flavor? The gauge group remains, of course, the same, 
$SU(N_c)$. However, in the infrared domain the first flavor 
decouples, and we are left with $N_f-1$ active flavors.
The first
coefficient in the $\beta$  function of the electric theory becomes 
more negative; hence, the critical value $\alpha_{*\rm electr}$
increases. We move leftwards, towards the left edge of the
conformal window. Correspondingly, the electric theory becomes
stronger coupled, and we expect that the magnetic one
will be coupled weaker.

What is the effect of the mass term in the magnetic theory?
It is rather obvious that the corresponding impact reduces to
introducing a mass term in the superpotential (\ref{msp}),
\beq
{\cal W}' = f {\cal M}^i_j {\cal Q}_i\tilde {\cal Q}^j
+ m {\cal M}^1_1\, .
\eeq
Extending the superpotential is equivalent to changing the vacuum 
valley. Indeed, the expectation values of ${\cal Q}_1$
and ${\cal Q}_1$ do not vanish anymore.
Instead, the condition of the vanishing of the $F$ term implies
\beq
\frac{\partial{\cal W}}{\partial{\cal M}_1^1} =
{\cal Q}_1\tilde {\cal Q}^1 + m = 0\, .
\label{pmag}
\eeq
If $m$ is large, Eq. (\ref{pmag}) implies, in turn,
that the magnetic squarks of the first flavor develop a vacuum 
expectation value, the magnetic theory turns out to be in the Higgs 
phase,
the gauge group $SU(N_f-N_c)$ is spontaneously broken down to
$SU(N_f-N_c-1)$, and one magnetic flavor is eaten up in the 
super-Higgs mechanism. We end up with a theory with the gauge 
group $SU(N_f-N_c-1)$ and $N_f-1$ flavors. The ${\cal M}_1^i$ and
${\cal M}^1_i$ components of the meson field become sterile in the
infrared limit. 
In this theory the
first coefficient of the $\beta$ function is less negative,
$\alpha_{*\rm magn}$ is smaller, we are closer to the left edge of the 
conformal window, as was expected.

Summarizing, we see that Seiberg's conjecture of duality is fully 
consistent
with the vacuum structure of both theories. As a matter of fact, this 
observation may serve as an additional evidence in favor of duality.
Simultaneously it makes perfectly clear the fact that, if duality
does take place, it can be valid only in the infrared limit; by no 
means
the two theories specified above are fully equivalent.

One may ask what happens if we continue adding mass terms
to the electric quarks of the first, second, third, {\em etc.} flavors.
Adding large mass terms we eliminate flavors one by one.
In other words, we launch a cascade taking us back to smaller  
values of $N_f$.
The electric theory becomes stronger and stronger coupled.
Simultaneously the dual magnetic theory is coupled weaker and 
weaker.
When the number of active flavors reaches $3N_c/2$,
$\alpha_{*\rm magn} = 0$. Eliminating the quark flavors further
we leave the conformal window -- the magnetic theory looses 
asymptotic freedom and becomes infrared-trivial, with the 
interaction switching off at large distances.   We find ourselves in the
free magnetic phase, or the Landau phase. By duality, the correlation 
functions in the 
electric theory (which is superstrongly coupled in this domain of 
$N_f$) must 
have the same trivial behavior at large distances. To be perfectly
happy we would need to know the relation between all operators
of the electric and magnetic theory,
so that given a correlation function in the electric theory
we could immediately translate it in the language of essentially free 
magnetic 
theory. Alas ...  As was already mentioned, this relation is 
basically unknown. It can be explicitly found only  for some 
operators of a geometric nature.
Even though the general relation was not found, the achievement is 
remarkable. For the first time ever the gauge bosons of the weakly 
coupled theory (magnetic) are shown to be ``bound states"
of a strongly coupled theory (electric).

The last but one step in the reduction process
is when the number of active flavors is $N_c+2$. The gauge group of 
the electric theory is $SU(N_c)$ while that of the magnetic one is 
$SU(2)$. Under the duality conjecture the large distance behavior
of the superstrongly coupled electric theory is determined by the
massless modes of the essentially free magnetic theory: 
three ``photons",  three ``photinos",  $2(N_c+2)$ fields of the type
${\cal Q}$ and $2(N_c+2)$ fields of the type
$\tilde{\cal Q}$. We can return to the remark made in the very 
beginning of this section -- it is explained now. It becomes clear why 
all attempts to describe the infrared behavior of the theory
in terms of the variables $M, B, \tilde B$ failed in the $N_f=N_c+2$ 
case: these are not proper massless degrees of freedom.

The last step in the cascade that still can  be done is  reducing one 
more flavor, by adding a 
large mass term in the electric theory, or the corresponding entry of
the matrix 
${\cal M}$ in the superpotential of the magnetic theory.
At this last stage the super-Higgs mechanism in the magnetic theory
completely breaks the remaining  gauge symmetry. We end up with
$N_c+1$ massless flavors interacting with $(N_c+1)\times
(N_c+1)$ meson superfield ${\cal M}^i_j$ (plus a number of sterile 
fields inessential for our consideration). 
This remaining meson superfield ${\cal M}^i_j$ is
assumed to have no (or small) expectation values, so that we
stay near the origin of the vacuum valley.
Moreover, it is not difficult to see
that the instantons of the broken $SU(2)$ generates
the superpotential of the type det${\cal M}$.
This is nothing but a supergeneralization of the 't Hooft 
interaction \cite{tHooftPRD}. Indeed, 
the instanton generates fermion zero modes, one mode
for each $\psi_{\cal Q}$ and  $\psi_{\tilde{\cal Q}}$.
In the absence of the Yukawa coupling $f{\cal M}{\cal Q}\tilde{\cal 
Q}|_F$, there is no way to contract these zero modes, and their
proliferation results in the vanishing of the would-be 
instanton-induced superpotential. The Yukawa coupling
$f\phi_{\cal M}\psi_{\cal Q}\psi_{\tilde{\cal Q}}$ lifts the zero 
modes,
much in the same way as the mass term does. (Supersymmetrization 
of the result is achieved through the vertex 
$f\psi_{\cal M}\phi_{\cal Q}\psi_{\tilde{\cal Q}}$
where the boson field $\phi_{\cal Q}$ is induced
by the zero mode of $\psi_{\cal Q}$ and the gluino zero mode).
In this way we arrive at the instanton-induced superpotential
det${\cal M}$.
The full superpotential of the magnetic theory, describing the 
interaction of massless (or nearly massless) degrees of freedom
which are still left there, is
\beq
{\cal W} = f{\cal M}{\cal Q}\tilde{\cal 
Q} + \mbox{det}\, {\cal M}\, .
\eeq
Compare it with Eq. (\ref{spot}) which was derived for the 
interaction of the massless degrees of freedom of the $N_f=N_c+1$
model by exploiting a totally different line of reasoning. 
Up to a renaming of the fields involved, the coincidence is absolute!
Thus, the duality conjecture allows us to rederive
the $s$-confining potential in the ``electric" theory,
SQCD with the gauge group $SU(N_c)$ and $N_f=N_c+1$.
The fact that two different derivations lead to one and the same 
result further strengthens the duality conjecture, making one to 
think that actually it is more than a conjecture:
the full infrared equivalence of Seiberg's ``electric" and ``magnetic" 
theories does indeed take place. 

Patterns of Seiberg's duality in more complicated gauge theories
than that discussed above, including non-chiral matter sectors,
were studied in a number of publications. The dual pairs
proliferate! By now a whole zoo of dual pairs is densely populated.
Finding a ``magnetic" counterpart to the given ``electric"
theory remains an art, rather than science --
no general algorithm exists which would allow one to generate dual 
pairs automatically, although there is
a collection of  some helpful hints and recipes. Systematic searches 
for dual partners to every given supersymmetric theory 
is an intriguing and fascinating topic. At the
present stage it is too technical, however, to be included in this
lecture course. Even a brief discussion of the corresponding
advances would lead us far astray. The interested reader is referred 
to the original literature, see e.g. \cite{DU1,DU2,DU3}.

\newpage

\section{Lecture 4. New Phenomena in Other Gauge Groups}

\renewcommand{\theequation}{4.\arabic{equation}}
\setcounter{equation}{0}

After 1994 many followers worked on dynamical aspects of 
non-Abelian SUSY gauge theories. In many instances the 
development 
went not in depth but, rather, on the surface. Various ``exotic" gauge 
groups and matter representations were considered, supplementing 
the list of models considered in the previous section  by many new 
examples with essentially the same dynamical
behavior. The corresponding discussion
might be interesting to experts but is hardly appropriate here. 
Along with thorough explorations of the previously
known patterns 
some interesting nuances in dynamical scenarios were revealed {\em 
en route}. In this section we will focus on these new dynamical
phenomena.
Below we will briefly review some exciting findings ``after 
Seiberg". Preference will be 
given to results of general interest -- multiple 
inequivalent 
branches, 
 oblique confinement and
electric-magnetic triality.

New  phenomena takes place in  $SO(N_c)$ SUSY gauge theories with 
matter
in the
vector representation of the gauge group.
Unlike the $SU(N_c)$
theories with the fundamental quarks, where there is
no invariant distinction between the Higgs and confinement regimes, 
in $SO(N_c)$ these phases (as well as the oblique confinement
phase)
can be distinguished.  Indeed, the  Wilson loop in
the spinor representation of $SO(N_c)$ cannot be screened by the 
dynamical quarks and,
therefore, presents  a gauge invariant order parameter for 
confinement. 

As was explained in Sect. 3, technically the most fruitful idea
is considering $N_f$, the number of the quark flavors, as a free
parameter. Note, that in the orthogonal groups, we do not need 
subflavors -- one chiral superfield in the {\em vector} representation 
presents one flavor. It is useful to note that the adjoint
representation in the orthogonal groups has dimension 
$N_c (N_c-1)/2$ and $T(G) = N_c -2$. The vector
representation has dimension $N_c$ and $T(\mbox{vect}) =1$.
The group $SO(3)$ is exceptional:
the vector representation is the same as adjoint, and $T(G) = 
T(\mbox{vect}) =2$.

Below we will consider $SO(N_c)$ theories with $N_f = N_c - 4$, $N_c 
- 3$
and $N_c - 2$ which exhibit inequivalent phase branches,
massless glueball/exotic states at the origin of the  moduli space
and oblique confinement \cite{KINS} 
(see also \cite{KINS1}). A brief digression in $SO(3)$
theories will provide us with the first example of the 
electric-magnetic
 triality.

All results obtained for the orthogonal groups can be
readily adapted for the symplectic groups by formally extrapolating
the parameter $N$ in $SO(N)$ to negative values \cite{NMSK},
so that the dynamical behavior of the $Sp(N)$ theories
\cite{SPSUSY} merely 
parallel that of $SO(N)$.

\subsection{Two branches of $N_f=N_c-4$ theory}

Assume that $N_c > 4$ and $N_f=N_c-4$.
The structure of the vacuum valleys is very similar to that
we discussed in detail in Sects. 2 and 3 for the unitary groups.
The moduli matrix is
\beq
M^{ij} = Q^iQ^j\, ,
\eeq
where the summation over the color index (running from 1 to
$N_c$) is implicit. A generic point from the vacuum valley
$\langle M^{ij}\rangle \neq 0$
corresponds to  spontaneous breaking of the gauge symmetry
$SO(N_c)$ down to $SO(4)$. Since $SO(4)= SU(2)\times SU(2) $
(we will mark one of these subgroups by subscript L and another by 
R),
at low energies, below the masses of  those gauge bosons that
became heavy $W$'s, we have two $SU(2)$ theories of  
gluons/gluinos
coupled to massless axion/dilaton fields. These two low-energy
 theories are in the strong coupling regime;
 the corresponding scale parameters
$\Lambda_{L,R}$  are related to the fundamental parameter
$\Lambda$ of the high energy theory as
\beq
\Lambda_{L,R}^6 = \frac{\Lambda^{2N_c - 2}}{\mbox{det}M}\, ,
\label{orthlam}
\eeq 
where I omitted an irrelevant numerical constant.
Equation (\ref{orthlam}) is most easily verified
by matching the running $\alpha$ 
from the NSVZ $\beta$ functions of the original $SO(N_c)$
theory and the low-energy $SO(4)$ in the limiting case when
$M^{ij}$ is proportional to the unit matrix.

The gluino condensate can develop independently in the
$SU(2)_L$ and $SU(2)_R$ theories, 
\beq
\langle\lambda\lambda\rangle_{L,R} =\pm \Lambda_{L,R}^3\, .
\label{orthgc}
\eeq
The existence of the gluino condensate implies
the following superpotential
\beq
{\cal W}(M) = \langle\lambda\lambda\rangle_{L}
+\langle\lambda\lambda\rangle_{R} = 
(\varepsilon_L +\varepsilon_R)\left(
\frac{\Lambda^{2N_c - 2}}{\mbox{det}M}\right)^{1/2}\, ,
\label{orthsp}
\eeq
where $\varepsilon_{L,R}$ are the phase factors,
$\varepsilon_{L,R}=\pm 1$, corresponding to two possible signs of
the gluino condensates. Here I used Eq. (\ref{orthlam})
for the low-energy scale parameter. The relation between the
gluino condensate in the low-energy theory
and the emerging  superpotential is perfectly the same as that
observed in  the $SU(N_c)$ theories long ago \cite{ADS1}. 

The crucial novel element is the occurrence of physically distinct
solutions: (i) $\varepsilon_{L} = \varepsilon_{R} = +1$;
(ii) $\varepsilon_{L} = \varepsilon_{R} = -1$;
 (iii) $\varepsilon_{L} =+1,\,  \varepsilon_{R} = -1$;
and (iv)  $\varepsilon_{L} =-1,\,  \varepsilon_{R} = +1$.
The solutions (i) and (ii) are related by a discrete
symmetry of the model and are equivalent. By the same token,
(iii) and (iv) are equivalent. It is quite evident, however,
that the first pair leads to a familiar picture, with
a superpotential destroying the vacuum valley,
while the second pair of solutions corresponds to no superpotential
-- the classical valley persists as the quantum moduli space.
In other words, one and the same fundamental theory
has two branches, two drastically different realizations.

Let us have a closer look at the second branch. The most interesting
point is the origin, $M=0$. At this point the full $SO(N_c)$
gauge symmetry is unbroken. Classically all $N_c(N_c-1)/2$
gauge bosons are massless -- the theory is singular at the origin.
The singularity might manifest itself in the form 
of the kinetic terms
of the fields $M^{ij}$.
In the quantum theory, due to confinement,
 this singularity must  be smoothed out (much 
in the same way as it happens in the $SU(N_c)$ theory
with $N_f=N_c$). The only
apparent reason why the singularity might
still survive is the emergence of some additional massless
bound states. 

The conclusion that there are no extra massless particles at 
$\langle M\rangle = 0$,
 beyond those residing in the moduli fields $M^{ij}$,
is based, as usual, on the 't Hooft matching condition.
In the absence of condensates the global (unbroken) symmetry
of the model 
\footnote{Additional discrete symmetries exist. They are irrelevant
in this consideration and will be briefly discussed later.}
is $SU(N_f)\times U(1)_R$. One can check
that the anomalous $AVV$ triangles calculated at the fundamental 
level are exactly saturated by the massless fermions from 
$M^{ij}$ \cite{KINS}. Thus, following the standard logic, we believe
that the only massless particles are represented by the
moduli fields, and their kinetic terms are
non-singular. If so, we encounter here
another example of confinement without
the spontaneous breaking of the chiral symmetry.

\subsection{$N_f = N_c - 3$ (massless glueballs and exotic states)}

Adding one more flavor, i.e. considering 
$N_f=N_c-3$, allows one to break the gauge symmetry
down to $SO(3)$. It is convenient to consider first a 
special point from the vacuum valley where the expectation values
of $N_c-4$ flavors are large, and the VEV of the last flavor is small.
Then the pattern of the gauge symmetry breaking is
two-stage. At the first stage the symmetry is
broken down to $SU(2)_L\times SU(2)_R$, as in Sect. 4.1;
then it is further broken down to a
diagonally embedded  $SO(3)$.
Omitting details of the derivation let me quote
the superpotential generated due to the gaugino condensation
in the low-energy theory 
\cite{KINS},
\beq
{\cal W}(M) = (1+\varepsilon ) 
\frac{\Lambda^{2N_c - 3}}{\mbox{det}M}\, ,\,\,\, \varepsilon =\pm 
1\, .
\label{orthsp3}
\eeq
Again, as in Sect. 4.1, we have 
  two physically 
inequivalent 
phase
branches (corresponding
to $\varepsilon = 1$ and
$\varepsilon =-1$), 
one with a dynamically generated superpotential and the 
other
with a moduli space of the quantum vacua.
The latter is of most interest.
What can be said about the low-energy spectrum on this branch?

If a mass term is given to the last flavor we want
 the solution to smoothly pass into the $\varepsilon_L
\varepsilon_R = -1$ branch of the $N_f = N_c - 4$ theory
as the mass parameter  becomes large.
If the kinetic term of the fields $M^{ij}$
is non-singular, and there are no other massless
states (other than those in $M^{ij}$), the smooth transition
to this branch is impossible. Indeed, adding
the tree level superpotential 
\beq
{\cal W} = mM^{i_\ell i_\ell}
\label{R1}
\eeq
($i_\ell$ is the last flavor) and integrating out $M^{i_\ell i_\ell}$
under the assumption that the kinetic term is non-singular
we end up with no supersymmetric solution \cite{ISS}.

A possible scenario ensuring the desired  smooth transition
is the emergence of additional massless fields at the origin.
Since additional massless fields are absent at  generic points
from the vacuum valley, they must have a superpotential
giving them a mass at $M\neq 0$. A simplest possibility is
 a massless 
$N_f$-plet $q_i$  with the superpotential
\beq
\Delta {\cal W} = - M^{ij}q_iq_j\, .
\label{R2}
\eeq	

Combining now Eqs. (\ref{R1}), (\ref{R2}) and integrating out
the heavy superfield $M^{i_\ell i_\ell}$ we do obtain
two physically equivalent solutions with no superpotential.
The two-fold ambiguity is due to the fact that
$\langle q_{i_\ell}\rangle = \pm \sqrt{m}$. This is precisely the
 $\varepsilon_L
\varepsilon_R = -1$ branch of the $N_f = N_c - 4$ theory.

From the superpotential (\ref{R2}) we infer that the
$U(1)_R$ charge of the superfield $q_i$ is $1/N_f$. It is
not difficult to check that with this particular 
 massless sector ($M^{ij}$ plus $q_i$) 
and this charge assignment all anomalous $AVV$ triangles 
corresponding
to the conserved global symmetries $SU(N_f)\times U(1)_R$
are matched \cite{KINS}. 

One can ask whether it is possible
to build, from the fundamental fields of the theory, an 
interpolating local gauge invariant  product with the
external  quantum numbers
coinciding with those of $q_i$. The answer to this question is
positive,
\beq
q_i \sim (Q)^{N_c-4}W^2\, .
\label{glue}
\eeq
In other words, we are free to interpret the massless states $q_i$
as exotic quark-gluon bound states. 
The emergence of these states is remarkable by itself.
At $N_c = 4$ we deal with massless glueballs (cf. Ref. \cite{Kovner}).

\subsection{ $N_f = N_c -2$ (massless monopoles/dyons; oblique
confinement)}

Theories with $N_f=N_c-2$ 
turn out to have much in common with the extended-SUSY  $N=2$ 
model
whose solution was found by Seiberg and Witten
\cite{SEIWIT} (see also \cite{Seib2,SWFOL}).
In particular, massless monopoles and dyons
appear at  certain points on the moduli space.
Their condensation causes confinement (oblique
confinement). To be able to understand
the corresponding dynamics, and details of relevant
derivations, in full, one must master the results of
Ref. \cite{SEIWIT}. Certainly, we cannot submerge
in this topic now, nor do I see any reasons
why  should we  undertake this endeavor.
 The 
$N=2$ model, specific features of the
Seiberg-Witten  solution and 
specific tools used to reveal them, is a very vast topic
 extensively covered in numerous dedicated reviews. 
The reader is referred to the 
 list of recommended literature  at the end. We will
 settle here for a descriptive discussion 
of the $SO(N_c)$ model at hand
which, hopefully,
gives undistorted general idea of the basic ingredients.

We start from the following obvious observation:
the $SO(N_c)$ model with $N_f=N_c-2$ flavors
has a moduli space parametrized by the matrix of the expectation
values
$\langle M^{ij} \rangle$; a generic point from the vacuum
valley corresponds to the spontaneous
breaking of the original $SO(N_c)$
gauge symmetry down to $SO(2)$. Since $SO(2)$
is the same as $U(1)$ the theory has a {\em massless photon}.
Hence, the theory is in the Coulomb 
phase.
The sector of massless states consists of all moduli fields, the
massless photon and all its superpartners. In other words, at low
energies we deal with QED; the moduli fields are electrically
neutral. Those states that are electrically charged
(e.g. the gauge bosons from $SO(N_c)/SO(2)$)
are generically massive.

The value of the QED coupling constant depends
on where exactly we are  on the vacuum valley.
Denote the inverse (complexified) coupling constant by $T$,
\beq
T\equiv \frac{1}{g^2} + i\frac{\vartheta}{8\pi^2}\, .
\label{ccc}
\eeq
The low-energy electromagnetic coupling constant
is a holomorphic function of the moduli. Because of the
$SU(N_f)$ flavor symmetry of the model
it can actually depend only on a specific combination
of the moduli, namely,
\beq
U\equiv \mbox{det}\, M\,  , \,\,\, M^{ij} = Q^i Q^j\, .
\label{defu}
\eeq
  If det$M\gg \Lambda^{2N_c - 4}$ the relation between
$T$ and the high-energy coupling constant of the
original $SO(N_c)$ theory $T_0$ is given by a familiar
one-loop formula,
\beq
T = T_0 -\frac{1}{8\pi^2}\,  \ln\frac{M_0^{2N_c-4}}{U}\, .
\label{looporth}
\eeq
This expression is perturbatively exact. It does receive
nonperturbative corrections, however, which modify
the $U$ dependence of $T$ at small $U$.

To see how this happens we, first, observe that
the matter field $R$ charge with respect to the
anomaly-free $U(1)_R$ symmetry
vanishes. (The reader is invited to check this
statement, as well as all other numerical factors
mentioned in this section.) In other words, the $R$ charge
of $M^{ij}$ is zero. This implies, in particular, 
that no superpotential can be generated along the valley --
the reason is the same as in the $SU(N_c)$ model with $N_f = N_c$.
By the same reason the $F$ terms of
the form 
$$
\left[ W^2 \left( \frac{\Lambda^{2N_c - 4}}{U}\right)^k\right]_F\, ,
$$
where $k$ is integer, can be generated by instantons; $k=1$
corresponds to the one-instanton correction, $k=2$ to 
two-instanton and so on. These $F$ terms are equivalent
to 
 the instanton corrections in $T$
of the type
\beq
\left( \frac{\Lambda^{2N_c - 4}}{U}\right)^k\, ;
\label{instorth}
\eeq
and, as a matter of fact,  they do
 appear. Instead of relying on the
 $R$-charge arguments one could just count the number of the 
fermion zero modes in the instanton background, with the
 same conclusion.
 
Summation of  all instanton terms (\ref{instorth}), one by one,
would be a brute force approach to calculating $T(U)$.
Nobody attempted to follow this road, of course.
A roundabout way based on subtle considerations
of analytical and other general properties \cite{KINS}
leads to an exact formula for $T(U)$ in terms of
elliptic integrals. To write the formula
we first introduce the curve
\beq
y^2 = x^3 + x^2 (-U +8\Lambda^{2N_c-4}) +
16 \Lambda^{4N_c-8} x\, ,
\label{elcu}
\eeq  
which obviously has branch points at
$x = 0, \infty $ and 
\beq
x_\pm = \frac{1}{2}\left[ U -8 \Lambda^{2N_c-4}
\pm\sqrt{U(U-16\Lambda^{2N_c-4})}\right]\, .
\label{xorth}
\eeq
The exact dependence $T(U)$
is given by the ratio
\beq
T(U) =\frac{1}{16\pi i}\, \,
\frac{\int_{x_-}^{x_+} dx/y(x)}{\int_{0}^{x_-} dx/y(x)}\, .
\label{exacT}
\eeq
The branches of the square roots are defined in such a way
that for positive (real) $a$ the square root $\sqrt{a}$ is positive.
Then at large $U$ Eq. (\ref{exacT}) reproduces
the perturbative logarithmic $U$ dependence (\ref{looporth}).
It allows us to go further, however, and examine
the behavior of the complexified coupling constant
in the entire complex plane. From Eq. (\ref{xorth})
it is perfectly clear that at $U$ tending to zero and to
$16\Lambda^{2N_c +4}$ the denominator of
Eq. (\ref{exacT}) develops a logarithmic singularity
since  $x_+ \ra x_-$. The singularities in the effective
gauge coupling constant can  appear only if there
are massless particles in the physical spectrum
which carry electric and/or magnetic charges and are
 coupled to our photon. We, thus, conclude that in
two points on the moduli space, $U=0$
and $16\Lambda^{2N_c +4}$ the sector of the massless states
is extended: apart from the photon and the moduli fields
it includes some massless electrically/magnetically charged
particles.

More exactly,  we define   two submanifolds, ${\cal M}_1$
and ${\cal M}_2$, of the vacuum manifold 
(both  are non-compact)
\beq
\langle M^{ij}\rangle = M_*^{ij}\, , \,\,\, \mbox{det}M_* = 
0\,\, \mbox{on}\,\,  {\cal M}_1\,\,\mbox{or}\,\,\,  \mbox{det}M_* 
=16\Lambda^{2N_c +4} \,\,\,  \mbox{on}\,\,  {\cal M}_2\, .
\eeq
Examining the character of the singularity in $T$ allows one
to say
which particular massless particles contribute. 
In this way it was found \cite{KINS} that
on the second manifold ${\cal M}_2$ dyons are
massless. They have both electric and magnetic charges $\pm 1$. 
Since outside ${\cal M}_2$  they are massive, a superpotential of 
a special form (vanishing on ${\cal M}_2$) must be generated.
If the massless dyons are denoted by $E^\pm$,
the superpotential must obviously have the form
\beq
{\cal W} = (U-16\Lambda^{2N_c +4} )E^+E^-\left[1+{\cal 
O}\left(\frac{U-
16\Lambda^{2N_c +4} }{\Lambda^{2N_c +4}}\right)\right]\, .
\eeq

As soon as we leave ${\cal M}_2$ and pass to ${\cal M}_1$, the dyons 
 $E$ become massive. Instead, other particles loose their mass
on ${\cal M}_1$, so that their contribution to $T$
ensures a proper singularity.  Equation (\ref{R2}), which takes place 
in the
$N_f = N_c-3$
theory, gives us a hint that the number
of distinct species of massless monopoles on ${\cal M}_1$ is larger 
than one and is 
related to $N_f$. The hint is based on anticipation of a smooth 
transition from $N_f = N_c -2$ to the case of $N_f = N_c -3$, see 
below. 

Let us introduce $2N_f$ chiral superfields $q_i^\pm$,
$i =1,2, ..., N_f$. The superfields $q_i^+$ describe monopoles with the 
magnetic charge $+1$, $q_i^-$ monopoles with 
 the magnetic charge $-1$, with a superpotential term roughly 
speaking of the form
\beq
{\cal W} = M^{ij}q^+_iq^-_j \, .
\label{moninter}
\eeq
The moduli field $M^{ij}$ is supposed to belong to ${\cal M}_1$.
If the rank of this matrix is $r$, where $r$ is obviously less than 
$N_f$, then Eq. (\ref{moninter})  corresponds to $N_f - r$
massless supermultiplets. The electric and magnetic charges of
$q$, $E$ and the original ``quarks" $Q$ are related. Namely, the 
relation is such that one can think of $E$ as of a bound state of 
$Qq$,
$$
E^\pm \sim q_i^\pm Q^i\, .
$$

At the origin of the vacuum valley, when $M^{ij}=0$, all $N_f$ species 
$q^\pm_i$ are massless. At this point, the full global symmetry of the 
model,
$SU(N_f)\times U(1)_R$, is presumably unbroken.
As usual, we check the self-consistency of this hypothesis by 
matching
the 't Hooft triangles. To this end we calculate four triangles,
$U(1)_R$, $U(1)_R^3$, $SU(N_f)^3$ and $SU(N_f)^2U(1)_R$,
at the fundamental level ($\lambda$ and $\psi_Q$) and at the 
composite level. The massless fermions at the composite
level are those residing in $M^{ij}$, $q^\pm_i$ and the photon
supermultiplet. As was expected, all four AVV triangles do match 
\cite{KINS}!

The last question to be addressed in this section is
the issue of transition from the $N_f = N_c -2$ theory to $N_f = N_c 
-3$ upon adding the mass term to one of the original quarks.  
The tree level superpotential is the same as in Eq. (\ref{R1}).
It is not difficult to see that with  the mass term  added
the whole vacuum valley shrinks to 
the submanifolds ${\cal M}_1$ and ${\cal M}_2$. In other words,
at $m\neq 0$ (see Eq. (\ref{R1})) the zero vacuum energy is achieved
on two solutions: $M^{ij}\in {\cal M}_1$ and $M^{ij}\in {\cal M}_2$.

The second solution corresponds to condensation of dyons,
$\langle E^+E^-\rangle \sim m$. The non-vanishing expectation value 
of the dyon field entails confinement of the electric charges, i.e. the 
original quarks of the theory are confined, in accordance with the 
arguments 
presented in Sect. 1. As a matter of fact, since the 
condensed objects are dyons, we deal here with the {\em oblique }
confinement, see Sect. 1.4. Simultaneously, the former massless 
photon becomes massive. The would-be moduli $M^{ij}$, where
$i,j = 1,2, ... N_f -1$, survive as light fields with the superpotential
(\ref{orthsp3}) with positive $\varepsilon$.  Thus, this solution 
smoothly passes into the $\varepsilon = 1$ branch of the
$N_f = N_c -3$  theory. 

For the first solution, $M^{ij}\in {\cal M}_1$, adding the mass term
(\ref{R1}) for the quarks of the last flavor entails condensation of 
monopoles, $\langle q^+_{N_f}q^-_{N_f}\rangle \neq 0$. The fact that
they do condense immediately follows from minimization of
the superpotential
$$
M^{ij} q_i^+q_j^- + mM^{N_fN_f}\, .
$$
The first term in this superpotential,  Eq. (\ref{moninter}),
is actually somewhat simplified; I have omitted details
irrelevant for qualitative conclusions but important
in quantitative analysis. Since the monopoles condense, all fields
carrying the electric charges -- in particular,
the original quarks -- are confined. The massless unconfined fields
are the moduli $M^{ij}$ with $i,j=1,2,..., N_f-1$, and the  chiral 
superfields
\beq
q_i \sim \left( q_i^+q_{N_f}^- - q_i^-q_{N_f}^+ \right)\, ,\,\,\, 
i = 1,2, ..., N_f - 1 \, .
\label{MM}
\eeq
A superpotential is generated coupling the above moduli
to $q_iq_j$, cf. Eq. (\ref{R2}). Thus, the solution
$M^{ij}\in {\cal M}_1$ does indeed
pass into the $\varepsilon =-1$ branch of the $N_f =N_c-3$ theory
upon adding the mass term to one of the quarks.
Comparing the results described above with the dynamical portrait
of the $N_f = N_c - 3$ theory we conclude that $N_c-3$
monopoles (\ref{MM}) that remain massless can actually be
interpreted as massless exotics or glueballs in terms of the original 
degrees of freedom of the fundamental Lagrangian, cf. Eq. 
(\ref{glue}). 

Most of the phenomena we discuss in this section were originally 
discovered
in another model, the extended $N=2$ SUSY theory 
with the $SU(2)$ gauge group \cite{SEIWIT}. In some of the 
examples \cite{SEIWIT}, where the
matter hypermultiplets were present, there are monopoles and 
dyons too, whose condensation leads to confinement and
oblique confinement of the quarks belonging to the
fundamental representation of $SU(2)$. We know already, however,
that in the presence of the scalar fields in the fundamental 
representation there is no invariant distinction between the
Higgs phase and the phase of confinement. One can speak only of the
strong coupling {\em versus} weak coupling regimes,
which are smoothly connected (Sect. 1). The $SO(N_c)$ example
considered here presents a picture of three distinct physically
inequivalent phases: Higgs, confinement and oblique confinement.

\subsection{Electric-magnetic-dyonic triality}

Traveling further along the $N_f$ axis, towards higher values of 
$N_f$, we find ourselves in a situation where the original
$SO(N_c)$ theory is dual in the infrared domain
to another
theory,  with  the gauge group $SO(N_f-N_c +4)$,
the same number of the dual quarks as in the original theory, 
and an additional gauge singlet field $M^{ij}$. 
Leaving aside certain nuances, we find a full
parallel to Seiberg's duality for the unitary groups, see Sects. 3.3 and 
3.4. In particular, the
interval  ${3
\over 2} (N_c-2)< N_f < 3 (N_c-2)$ is the conformal window of the 
$SO(N_c)$ theories, where both the electric and  magnetic
theories flow to the same non-trivial infrared fixed point 
(superconformal, or non-Abelian Coulomb phase). To the left of the 
conformal window, at $N_c-2< N_f \leq {3 \over 2} (N_c-2)$, the
magnetic degrees of freedom are free in the infrared
(the magnetic Landau phase).
They can be regarded as superstrongly bound states of the original 
(electric) gluons and matter, just in the same way as the
corresponding domain in the 
$SU(N_c)$ theories.

Instead of continuing this excursion in the direction of higher $N_f$ 
for the generic values of $N_c$
presenting a dynamical  picture which is
already pretty familiar, we turn to an exceptional case
of the gauge group $SO(3)$,
 where a new phenomenon occurs.  We will limit ourselves to a 
specific choice of the
matter sector -- two triplet ``quark" fields,
$Q^1_a$ and $Q^2_a$ where $a =1,2,3$. This particular problem 
is nicely reviewed in Ref. \cite{KINS1,ins}. Many other
instructive  examples are considered  in Refs.
\cite{KINS,Elitzur}. 

The remarkable new phenomenon just mentioned  is 
the occurrence of {\em two} theories dual to the original one.
If the original theory is ``electric", one of the dual theories is 
``magnetic" and the other is ``dyonic".  Thus, instead of duality,
one can speak of {\em triality}. Moreover, a symmetry which is
explicit in the original electric theory is realized
in the magnetic and dyonic theories as a
 {\em quantum 
symmetry}.  
Such (quantum)  symmetries are not easily visible in the   
Lagrangian 
because
they are implemented by nonlocal transformations of the fields.

Let us have a closer look at the model we will be dealing with.
First of all, at the classical level the vacuum valley is
obviously parametrized by three gauge invariant moduli
$M^{ij}$ (note that $M^{12}= M^{21}$). The expectation value of the
first triplet $Q^1$ breaks the gauge symmetry $SO(3)$
down to $SO(2)$, which is then further broken
by the expectation value of the second triplet $Q^2$. Thus generically,
if det$M^{ij}\neq 0$, the gauge group is completely broken and the 
theory is in the Higgs phase.  

If the values of all moduli
are large, all gauge bosons are very heavy, and we are in the 
quasiclassical (weak coupling) regime. Nonperturbative effects can 
come only from instantons. Do instantons generate a superpotential 
which lifts
the vacuum degeneracy?

The theory at hand has a global flavor symmetry $SU(2)$
associated with the presence of two quark triplets. Moreover, it has a 
conserved $R$ current. The first fact implies that, if there is a 
superpotential, it can depend only on det$M$. Moreover, the 
$U(1)_R$
charges of the matter fields are such that det$M$
could enter the superpotential only linearly. It is clear then
that no superpotential is generated by instantons. The same 
conclusion
follows from a straightforward analysis of the
zero modes in the instanton background. Thus, the classical vacuum 
valley remains intact upon inclusion of the quantum effects.

As long as we stay away from the origin of the three-dimensional
complex manifold of the moduli (i.e. det$M\neq 0$), we are in the 
Higgs phase. Far away from the origin the theory is weakly
coupled. Needless to say that the emerging dynamical portrait  is 
quite boring.

As usual, surprises await us near the origin of the moduli space 
where the strong coupling regime is attainable. At det$M=0$ we 
expect to find more varied dynamics. Let us see whether our 
expectations come true.

At first we consider a part of the vacuum valley where det$M=0$ but 
$M\neq 0$. This part obviously corresponds to spontaneous breaking 
of the gauge $SO(3)$ down to $SO(2)=U(1)$.  The massless-state 
sector consists of one photon, a pair of electrically charged fields plus 
some neutral fields. In other words, in this ``corner" we deal 
with massless SQED. According to Landau, the electric charge is
totally screened at large distances, so we find ourselves in the free 
electric phase. This dynamical regime has been already discussed 
previously; therefore, our finding is not too exciting.

If all moduli are small, $\langle M^{ij}\rangle \ra 0$, the $SO(3)$ 
gauge group is 
unbroken and the strong coupling regime sets in. It is natural that 
the most nontrivial situation can and does take place here 
\cite{KINS}. 

The scale of strong interactions is determined by $\Lambda$. The 
question one begins with is what fields are light in this scale,
if at all. In the theory under consideration we 
do know that the moduli fields $M$ are light.  If we approach from 
the large $M$ side, from the Higgs phase, the moduli fields have no 
superpotential.  This does not mean, however, that no superpotential 
is generated in the strong coupling regime in {\em other}
phases of the theory.  The situation reminds that in supersymmetric 
gluodynamics where the gluino condensate may or may not develop
depending on the phase of the theory (see Sect. 2.6 and Ref. 
\cite{Kovner}). Another rather close parallel exists with the 
$s$-confining theories, where a superpotential for light degrees of 
freedom is generated at the origin, while no superpotential is allowed 
in the  Higgs weak coupling regime (Sect. 3.2). 
If the superpotential is generated by strong interactions then, from  
the $R$ charge counting and dimensional arguments, we conclude 
\cite{KINS} that 
it must have the form 
\beq
{\cal W} = \frac{\eta}{8\Lambda} \mbox{det}M + \frac{1}{2}
\mbox{Tr} \, mM \, ,
\label{spso3}
\eeq
where the second (tree-level) mass term is added by hand,
in anticipation of future applications. It presents 
 a mass deformation
$$
\Delta{\cal L}_m = \frac{1}{2}
 m_{ji}Q^iQ^j
$$
 of the original massless theory; $m_{ji}$ is the quark mass matrix.
So far we were considering $m=0$. The parameter $\eta$
in Eq. (\ref{spso3}) is a numerical constant. In the Higgs phase
$\eta = 0$. As we will see shortly, there are two other solutions,
$\eta =\pm 1$; the first solution corresponds to oblique confinement, 
the second to confinement. 

The occurrence of the nontrivial solutions with $\eta=\pm 1$ can be 
detected in the following way.  Assume that the mass matrix is 
diagonal,
$\{ m_{ij}\} =$ diag$(m_1,m_2)$, 
and 
\beq
m_{1,2}\neq 0\, , \,\,\,  m_1 / m_2 \ra 0\, .
\label{so3limit}
\eeq
Since the second quark is much heavier than the first one, we can
integrate out $Q^2$, reducing the theory to that with one
triplet quark $Q^1$ which has a tiny mass $m_1$. In the limit
$m_1 \ra 0$ one deals with the $N=2$ model 
 solved by Seiberg and Witten \cite{SEIWIT}.  As well-known, 
this model possesses a single moduli $M^{11}$ which is locked
by the mass term $m_1$  at
$\pm 4\tilde\Lambda^2 $  where the monopoles/dyons become 
massless 
and condense. Here $\tilde\Lambda $ is the scale of the strong 
interactions in the low-energy $SO(3)$ theory with one triplet quark. 
Using the NSVZ $\beta$ function it is easy to obtain
an exact relation between the both scale parameters,
$\tilde\Lambda^2 = m_2\Lambda$. The monopole/dyon condensate
is proportional to $m_1$. Thus, in the limit (\ref{so3limit})
we expect $M^{11} = \pm 4 m_2\Lambda$ and $M^{22}\ra 0$.
The positive value of $M^{11}$ corresponds to the monopole
condensation (confinement), the negative value to the dyon
condensation (oblique confinement).

Now, we observe  that Êexactly this
pattern of the $M^{ij}$ condensates emerges from  Eq. (\ref{spso3}). 
Namely, minimizing the 
superpotential (\ref{spso3})
one gets \cite{KINS}
\beq
\langle M^{ij}\rangle = -4 \eta \Lambda \, \mbox{det} m\,  (m^{-
1})^{ij}\, .
\label{so3cond}
\eeq

Of special interest is the point $m_2\ra 0$ 
(along with $m_1/m_2\ra 0$). In this limit, on the one hand,
all moduli vanish and, on the other hand, the monopoles and dyons
become massless simultaneously. These objects are mutually 
non-local. Their presence at the point $\langle M^{ij}\rangle
=0$ is most naturally interpreted as the onset of the
conformal behavior, cf. Sect. 3.3 and Refs. \cite{Seib2,Sei-conf}
\footnote{Another example of a similar dynamical scenario
was found recently in the $SU(N_c)$ theory with one
adjoint matter field and several fundamentals and antifundamentals 
\cite{Kapu}.}. In other words, at $\langle M^{ij}\rangle
=0$ the theory is, presumably, in the non-Abelian  Coulomb phase.

The fact that the first term in the superpotential (\ref{spso3})  is
generated by strong interactions of the original ``electric" quarks and 
gluons
is established above on the basis of indirect arguments. A question 
which immediately comes to one's mind is whether one can get this 
superpotential in a more direct way. In a sense, the answer is 
positive.

As we already know, one of the most powerful tools from the
magic tool-kit of supersymmetry is Seiberg's duality.
If we are able to identify a dual partner equivalent to the
original ``electric" theory in the infrared, which is simpler
than the original model (for instance, the dual partner
can be weakly coupled while the original theory is strongly coupled),
then the problem is solved. Although the procedure of building dual
models is not formalized, there is a standard strategy based on the 't
Hooft matching and additional self-consistency checks. Implementing 
this strategy one finds \cite{KINS} that the original theory has 
two duals in the case at hand! Thus, as a matter of fact, we deal with 
a triplet of theories. For the reasons which will become clear shortly
the first dual theory will be called $T_1$ and the second $T_{-1}$. 

Both dual theories dubbed $T_{\pm 1}$ have the gauge groups 
$SO(3)$ and the matter sectors including two triplet quarks $q_i$,
$(i=1,2)$. Additionally, they include three gauge singlets
$M^{ij}$ where $i,j=1,2$ and $M^{ij}=M^{ji}$. I stress
that the field $M^{ij}$ is to be treated as elementary in the theories
$T_{\pm 1}$; it has mass dimension two. The gauge singlets
 $M^{ij}$ interact with $q_i$ through a superpotential,
much in the same way as in the ``magnetic" theory of Sect. 3.3.
The theories $T_1$ and $T_{-1}$ differ one from another
by the form of the superpotential,
\beq
{\cal W}_\varepsilon 
= \frac{1}{12\Lambda} M^{ij} (q_iq_j) +
\varepsilon \left(
\frac{1}{24\Lambda} \mbox{det}\, M +
\frac{1}{24\Lambda} \mbox{det}\, \{q_iq_j\}
\right)
\label{trialsp}
\eeq
where $\varepsilon =1$ for $T_1$ (``magnetic" dual) and 
$\varepsilon =-1$ for $T_{-1}$ (``dyonic" dual).
For simplicity the scale parameters of the original theory 
and its duals are set equal. This assumption is inessential
and can be readily lifted. The superpotential (\ref{trialsp})
is non-renormalizable. This is unimportant too, since the theories
$T_{\pm 1}$ are effective low-energy theories anyway. 
They are suitable only for describing the large distance behavior. 

The theories  $T_{\pm 1}$ have the $D$ 
flat directions, which are just  the same as in the original
``electric" model. They are parametrized by three
gauge invariant chiral products $N_{ij} =q_iq_j$. The composite fields
$N_{ij}$ are would-be moduli in the dual theories.  Interaction with 
the elementary fields $M^{ij}$ makes them massive.
The coefficients in Eq. (\ref{trialsp}) are fixed by duality.
One can also check that the relative weight
of various terms in  Eq. (\ref{trialsp}) is correct by
flowing down from other theories \cite{KINS}. 

The simplest phase of the original ``electric" theory
is the Higgs phase. In this phase superpotential is not generated,
$\eta = 0$. Our goal is exploring the
confinement/oblique confinement phases of the ``electric" theory.
To this end we start with the simplest (Higgs) phase of the
theories $T_{\pm 1}$. In this phase interaction of the dual quarks 
with the dual
gluons generates no superpotential, so that the interaction of the 
would-be moduli is exhausted by Eq. (\ref{trialsp}). If we integrate
out the massive composite fields $N$ by using equations of motion
\footnote{The equations of motion imply that
Tr$MN = - 4\varepsilon$det$M$ and det$N$ = 4 det$M$.}
the superpotential for the gauge singlet elementary field
$M$ obviously takes the form
$$
\frac{-\varepsilon}{8\Lambda} \mbox{det}\, M\, .
$$
This is exactly the first term in Eq. (\ref{spso3}), with
$\eta = -\varepsilon$.  Thus, the Higgs phase in the ``magnetic" 
theory 
($\varepsilon = 1$) corresponds to the confinement phase
of the original ``electric" theory. By the same token,
the Higgs phase in the ``dyonic" theory 
($\varepsilon = -1$) corresponds to the 
oblique confinement phase
of the  ``electric" theory. We see that the first term
in the superpotential Eq. (\ref{spso3}), which  in the 
``electric" theory appears as a result of complicated strong-
interaction 
dynamics, is present essentially at the tree level in the
``magnetic" and ``dyonic" theories. 

Without going into further details let me note that
the Higgs phase of the ``electric" theory corresponds
to oblique confinement in $T_{1}$ and to confinement in
$T_{-1}$. Moreover, one can check
that dualizing $T_{\pm 1}$  -- each of these theories has two
dual partners -- one obtains permutations of the same three theories.

In summary, the phase structure of the $SO(3)$ theory with $N_f =2$
is so rich that it exhibits virtually all phases of the gauge theories 
considered so far. If the quark mass term in the
original model is set to zero, in a 
 generic point from the vacuum 
valley, det$M\neq 0$,  we 
find ourselves in the Higgs phase. On the non-compact 
two-dimensional subspace,
det$M= 0$, $M\neq 0$, there exists an unbroken $U(1)$, with the 
massless photon. Since the electrically charged fields 
are massless, the theory is Landau-screened in the infrared.
We deal here with the free electric phase.
Finally, if $M=0$ we are in the conformal (non-Abelian
Coulomb) phase. Adding a small quark mass term
we force the monopoles (dyons) to condense pushing the theory in 
the confinement or oblique confinement phase. 
 The  ``electric"
theory gives a weak coupling description of the Higgs branch of the
theory, $T_1$  gives a weak coupling description of the
confining branch of the original theory and $T_{-1}$  gives a weak 
coupling 
description
of the oblique confinement branch of the original theory.

\subsection{Quantum symmetry}

The electric theory we discussed above has a discrete symmetry
$Z_8$. This symmetry is a remnant of the anomalous 
chiral rotations of the matter fields. (I remind that the Konishi 
current
is anomalous). The fact that a discrete subgroup $Z_8$ survives 
becomes quite evident e.g. from the instantons. The number of the 
zero modes of the matter fermions is eight. 

Although the $Z_8$ symmetry is obvious in the original ``electric" 
Lagrangian, in the Lagrangian of the dual magnetic theory we see 
only
$Z_4$. The full $Z_8$ is not visible at the Lagrangian level;
it must (presumably) be realized as a non-local symmetry of the
quantum states \cite{KINS}. Such a situation is familiar in
the string context where it is referred to as {\em quantum 
symmetries}. More sophisticated patterns of the quantum 
symmetries of the dual pairs are treated in Ref. \cite{DU2}.

 The confining and the oblique confinement
branches of the $SO(3)$ theory discussed in Sect. 4.4. are related by a 
spontaneously broken global
discrete symmetry.  Therefore, the magnetic and the dyonic theories 
are
similar. 
This particular  example of the dyonic theories is not unique. More 
dyonic 
theories were revealed in Ref. \cite{KINS}. 
In these more complicated  examples  there is no global
symmetry which makes the theories similar.  The electric, magnetic 
and
dyonic theories are really distinct.    

\section{Lecture 5. Towards QCD}

\renewcommand{\theequation}{5.\arabic{equation}}
\setcounter{equation}{0}

In spite of discouraging developments of the recent years I still 
 believe that high energy physics is an empiric science
whose ultimate task is describing and understanding the laws of 
Nature.  For those who share this opinion the major question
in studying various dynamical scenarios in supersymmetric gauge 
theories   is 

\vspace{0.1cm}

{\em What lessons can be drawn for non-supersymmetric theories, 
first 
and foremost 
QCD?} 

\vspace{0.1cm}

From   this standpoint one can view the  previous 
sections of this lecture course as  an extended 
introduction.

 Although there are  few quantitative achievements in the 
direction of actual QCD, if at all, 
some qualitative insights were obtained. One   useful lesson
has been already discussed: we have seen that the MAC approach
 does not necessarily lead
to correct answers for the chiral symmetry breaking condensates
developing in the strong coupling regime. This lesson is negative, 
however. Below we will consider several issues where 
SUSY might provide with positive answers.

QCD is {\em the} theory of strong interactions 
in Nature,
describing three light (but not massless) quarks coupled to the octet 
of gluons. The number of colors is three, not two or four,
and the light quark masses are such as they are, not smaller or 
larger.  From the early days of QCD it became clear, however, that
it
is extremely advantageous to treat the theory in a more flexible way,
as a ``laboratory".  The ``laboratory" aspect is the second face of QCD.
For, instance, setting the quark masses to zero allows one to
develop the chiral perturbation theory. Considering the multicolor 
limit,
$N_c\ra\infty$, gives crucial insights in phenomenologically 
important problems (e.g. the Zweig rule, or factorization in the weak
non-leptonic decays), and serves as a consistent basis for the Skyrme 
model of baryons and many other approaches. 

The recent advances in supersymmetric gauge theories teach us 
the same story, confirming the old wisdom: 
whatever parameters we have in complicated and messy theories 
describing our world, it is worth
trying treat them as free parameters. Changing them may reveal
novel features of the model, hidden in more conservative 
approaches, and provide important insights. Analysis of various 
gauge groups and matter sectors reveals a rich spectrum of different
dynamical scenarios in SUSY gauge theories, some of which, in this or 
that form, may be relevant to QCD.

The conformal window was discussed in detail in Sect. 3.3. In QCD it 
exists too.
The right edge of the window lies at $N_f=16$ (if $N_c = 3$). Unlike 
SUSY QCD, however, we do not know exactly where the left edge lies.
One could only dream of analytic methods like those presented in 
Sect. 3.3. From numerical studies in Lattice QCD one may conjecture 
that the conformal window in QCD extends down to $N_f\approx 7$
\cite{Iwasaki}.

Towards the left side of the conformal window the critical value 
$\alpha_{s*}$ becomes large, and the theory is strongly coupled.

At $N_f=3$ experiment tells us that  the chiral symmetry
breaking takes place and the quarks are confined.
Does the chiral symmetry breaking sets in simultaneously with 
confinement?

Generally speaking,  these two phenomena are distinct, and it is 
conceivable 
that they occur not simultaneously.  For many years it was believed 
that confinement of color in QCD implies the chiral symmetry 
breaking \cite{Casher}. Are 
we sure today that that's the case? Supersymmetric examples teach 
us to be ready for
surprises.  
Confinement without the chiral symmetry breaking is still an open 
possibility in  the theory with, say, 5 or 6 massless quarks, although  
I hasten to  add that  no indications exist in favor of 
such a scenario at present. 
Moreover, the reverse -- the chiral symmetry breaking without 
confinement -- seems quite plausible.  {\em A priori} it is not  ruled 
out 
that the chiral 
symmetry 
breaking without 
confinement  occurs at the edge of the 
conformal 
window. 

Thus, we see that even the most global questions concerning the 
dynamical behavior in QCD remain unanswered. That's why new  
understanding of supersymmetric gauge dynamics gave rise to great 
expectations among QCD practitioners. The following 
strategy may prove to be fruitful. One starts from a supersymmetric 
theory where a solution  of a particular dynamical aspect of interest
is known. One then introduces explicit (soft) supersymmetry
breaking by adding mass terms to gluinos and/or squarks.
When these mass terms are sent to infinity
we find ourselves in a non-supersymmetric theory, with decoupled
gluinos/squarks.
Ideally, we would like to calculate in this limit.
Unfortunately, this is still beyond reach. What can be carried out, 
however, is the analysis of the theory with small enough
 SUSY breaking mass terms. One may hope 
that the trend revealed in this way continues to take place in the 
domain
of larger gluino/squark masses.  In this way we get a qualitative 
picture of
what is to be expected in QCD and other non-supersymmetric 
theories in the strong coupling regime. In some  cases, on the 
contrary, one can predict phase transitions in the gluino/squark 
masses. Although such an outcome is less interesting
than the possibility of solving issues in QCD, it still
provides us with  information about gauge dynamics
which may turn out useful. 

Below we will see how this strategy works in some simple
examples.

\subsection{$\vartheta$ dependence and the puzzle of ``wrong" 
periodicity}

Since the mid-seventies it is known \cite{V1} that there is a hidden 
parameter,
the vacuum angle $\vartheta$ in QCD and pure Yang-Mills theory (no 
quarks). The vacuum wave function is of the Bloch type.
Moreover,  in the absence of strictly  massless quarks  the physical 
quantities depend on $\vartheta$, and this dependence
must be periodic, with the period $2\pi$. 

On the other hand,
 various Ward identities  can be used in certain instances to show
that the $\vartheta$ parameter enters through the ratio
$\vartheta / N$ where $N$ is some integer, typically, the number of 
colors or the number of light flavors. For, instance, the 
Witten-Veneziano formula reads \cite{V2}
\beq
m_{\eta '}^2f_{\eta '}^2 = \left. 36 \frac{d^2\varepsilon_{\rm 
nlq}}{d\vartheta^2}\right |_{\vartheta =0}\, ,
\label{WV}
\eeq
where $\varepsilon$ is the vacuum energy density,
the subscript nlq indicates that it has to be ``measured" in pure 
Yang-Mills theory (no light quarks), $m_{\eta '}$ and $f_{\eta '}$
are the $\eta '$ meson mass and the coupling constant,
and the limit $N_c\ra\infty$ is implied. The second derivative on the 
right-hand side is nothing but the topological succeptability of the 
vacuum.
As well-known \cite{V2}
\beq
m_{\eta '}^2 = {\cal O}\left(\frac{1}{N_c}\right)\, , \,\,\,
f_{\eta '}^2 = {\cal O}\left({N_c}\right)\, , \,\,\,
\varepsilon_{\rm nlq} = {\cal O}\left({N_c^2}\right)\, ,
\label{NCD}
\eeq
which implies, in turn, that the left-hand side of Eq. (\ref{WV})
is ${\cal O}(N_c^0)$. To make the right-hand side of the correct order
in $N_c$ one is forced to assume that  the vacuum energy density in 
pure Yang-Mills theory depends on $\vartheta /N_c$ rather than
on $\vartheta $. 

A similar conclusion can be reached by analyzing the chiral Ward 
identities in the theory with light (but not exactly massless) quarks.
If the number of the light quarks is $N_f$, the topological 
succeptability $d^2\varepsilon_{\rm lq}/d\vartheta^2$ can be 
demonstrated \cite{CRE} to be proportional
to $1/N_f$. The subscript lq marks the light quarks. Since on  
general grounds one expects $\varepsilon_{\rm lq} $ to be a linear 
function of 
$N_f$, we conclude  that the vacuum energy density in the world 
with the 
light quarks switched on  must depend on  $\vartheta$ through 
$\vartheta /N_f$. In QCD the number of colors and the
number of light quarks is 3. So, both the Witten-Veneziano argument 
and that of Crewther point  to the  expected  $\vartheta $ 
dependence of 
 physical quantities of the type $f(\vartheta 
/3)$, making one suspect  that something is wrong with the $2\pi$ 
periodicity. 

These observations gave rise to numerous speculations 
that the standard construction \cite{V1}
of the vacuum wave function of the 
Bloch
type based on instantons was wrong; significant effort has
been invested in searches of  additional 
degenerate 
``pre-vacua" which should have been included in the construction of 
the ``correct" Bloch
wave function, but were actually overlooked. If these additional 
``pre-vacua" existed, a new 
superselection rule should  have been   imposed (a concise review of 
the topic
can be found in Ref. \cite{SMI}). The speculations were seemingly 
further supported by 't Hooft's torons -- field configurations defined 
in a finite volume for pure Yang-Mills $SU(N_c)$ theories and having  
fractional topological charges proportional to $1/N_c$ \cite{toron}.

This situation -- an apparently wrong $\vartheta$ dependence --
is already familiar to us. As we saw in Sect. 2.3.1, the gluino 
condensate in SUSY gluodynamics depends on $\vartheta$
through the factor $\exp (i \vartheta /N_c)$.  When $\vartheta$
continuously evolve from 0 to $2\pi$ we do {\em not} return to the 
same value of the gluino condensate.  The way out is obvious. One 
has $N_c$ degenerate and physically equivalent vacua, each of the 
Bloch type.
In order to restore the correct $2\pi$ periodicity one has just to 
rename these vacua after $\vartheta \ra \vartheta +2\pi$. Thus, all 
physically observable  quantities will 
be actually $2\pi$-periodic, in spite of the fact that the gluino 
condensate in the {\em given} vacuum is not.

The set of $N_c$ states intertwined under the $\vartheta$
evolution
 consists of distinct 
vacua, totally disconnected from each other in infinite volume,
rather than merely ``pre-vacua" to be included 
in the Bloch wave function. These distinct vacua reflect the 
spontaneous breaking of $Z_{2N_{c}}$ down to $Z_2$. 
Thus, the controversy ``distinct vacua {\em versus} superselection 
rule" is solved in favor of the former option \cite{DVSH}. 
A clear-cut signal confirming the distinct vacua scenario is the  
emergence of the   domain walls interpolating 
between the spatial regions of different vacua,
with a finite energy density \cite{DVSH}.  The existence of the 
domain walls rules
out the  superselection rule scenario.

In short, the problem is fully solved in SUSY gluodynamics, case 
closed.
Now, what can be said about non-supersymmetric  Yang-Mills 
theory, with no fermion 
fields?

To get a hint we will introduce the gluino mass term, $m_g$. As 
long as $m_g\ll\Lambda$,  theoretical analysis is reliable --
we merely slightly perturb the theory near its supersymmetric limit
and calculate the effects due to this perturbation
in the first order in $m_g$. 

The most drastic impact of $m_g\neq 0$ is lifting the vacuum 
degeneracy. If the gauge group is $SU(3)$  
the gluino mass term explicitly breaks $Z_6$ down to $Z_2$.
Correspondingly, 
we will have now
three non-degenerate states: one with a negative energy density
$\varepsilon =-m_g\Lambda_0^3$, and two degenerate states,
with  positive energy densities $\varepsilon =-
m_g\Lambda_0^3\cos(2\pi /3) =m_g\Lambda_0^3 /2$.
The parameter
$m_g$ is chosen to be real and positive; then the gluino condensate 
in the ground state is negative and equal to ($\vartheta =0$)
$$
\langle\lambda\lambda\rangle \equiv -\Lambda_0^3/2\, ,
$$
 where 
$\Lambda_0$ is a positive parameter of dimension of mass.
The sign of the gluino condensate can be established from a general 
analysis
parallelizing that of Ref. \cite{BANK}. The first state has the lowest 
energy and is the genuine vacuum.
Two other states are excited quasistable states, with the
spontaneously broken $CP$. For very small $m_g$
these false ``vacua" are almost stable. The lifetime decreases, 
however, as $m_g$ approaches $\Lambda$.  The family of states 
entangled under the $\vartheta$ evolution now consists of three 
physically {\em inequivalent} states. 

In the limit of small $m_g$ 
the expectation value of the operator $G^2 +iG\tilde G$ ($G$
is the gluon field strength tensor)
is proportional to that of $m_g\lambda\lambda$. So, both can be 
used as appropriate order parameters for the states discussed above.
In particular, the quasistable states are characterized by 
non-vanishing (and opposite) values of $G\tilde G$. When gluino 
becomes heavy and decouples, it is the gluon operator
that survives as the order parameter. 

Although at $m_g\neq 0$ the degeneracy of three states inherent to 
the supersymmetric limit  is gone, the $\vartheta$ evolution still 
intertwines these three states together. They  interchange their 
positions when $\vartheta$ varies from 0 to 
$2\pi$.
The one with the negative energy density, the true vacuum at
$\vartheta = 0$,  takes place of one of the two  false vacua, which 
formerly had 
positive energy density, and {\em vice versa}.  The  definition of 
the genuine vacuum (the lowest-energy state) has to be changed 
{\em en rout}, at $\vartheta = \pi$. All physical quantities (which for 
every given $\vartheta $
must be defined in the genuine vacuum) are $2\pi$-periodic,
although by naively inspecting the gluon condensate at
small $\vartheta$ we could have easily drawn a false conclusion
of a false $\vartheta$ dependence through $\exp
(i\vartheta /3)$. 

Now, what happens when $m_g$ becomes comparable to
$\Lambda$, and, eventually, much larger than $\Lambda$,
so that the gluino decouples leaving us with   pure Yang-Mills 
theory? There are two logical possibilities. The extra minima might 
just disappear. However, 
to match the Witten-Veneziano formula it is natural to think that 
that's not what happens. Two false ``vacua" may survive in QCD as 
quasistable states,
separated by a barrier from the genuine vacuum. They still form a 
triplet of the states entangled with each other under the
$\vartheta$ evolution. 
Certainly, there is no small parameter which might ensure
a large lifetime; one can only hope for an interplay of numerical 
factors. 

If the above picture is correct, the extra quasistable states may show 
up, as droplets of the false vacuum, in the
nuclear-nuclear collisions, or in any environment where
the hot quark-gluon plasma is formed and then cools down,
much in the same way as the droplets of the disoriented chiral 
condensate
\cite{DCC}. 

The fact that $N_c$ distinct vacua in supersymmetric gluodynamics
form a family whose members interchange their positions each time
$\vartheta$ reaches $2\pi$, $4\pi$ and so on is known
for over a decade \cite{SV4}. If $m_g =0$ the $\vartheta$ 
dependence of the gluino condensate is unobservable.  As soon as 
$m_g$ becomes non-vanishing the vacuum energy exhibits a 
$\vartheta$
dependence through the condensates  $G^2$ and $m_g\lambda^2$. 
The $\vartheta$ dependence  is now observable. Needless to say it 
remains observable in the limit $m_g\ra \infty$, i.e. in 
(non-supersymmetric) gluodynamics. 

Inclusion of the matter fields clearly makes the corresponding 
analysis more complicated. There exists a limit, however, 
in which one can reliably deal \cite{Evans} with supersymmetry 
breaking terms
in the same vein we dealt with in SUSY gluodynamics. Introduce 
$N_f$ flavors with a supersymmetric mass 
term, 
$m_{\rm SUSY}\neq 0$. The vacuum structure remains intact --
we still have $N_c$ degenerate vacua at zero, intertwined in one 
family. The $\vartheta$ dependence  is unobservable
since there exists a classically conserved anomalous current
(built from the gluino and squark fields). Assume now
that a SUSY breaking term is introduced as an $F$ term of a spurion 
field $\mu$,
\beq
\Delta {\cal L} = \mu Q\tilde Q|_F\, , \,\,\,  F_\mu \neq 0 \, ,
\label{mabr}
\eeq
so that
$$
F_\mu \ll m_{\rm SUSY}\, .
$$
It is pretty obvious that the SUSY breaking effects in the
$D$ terms will be proportional to $|F_\mu |^2$. The quark-squark 
mass splitting will be linear in $F_\mu$ and is fully controllable.
The term (\ref{mabr}) does not allow the squark field to be rotated.
Correspondingly, the classically conserved anomalous current
is gone, and gone with it is the $\vartheta$ independence of the
physical quantities.

The issue of the $\vartheta$ dependence in appropriately deformed
$N=2$ theories is discussed in \cite{EvansKonishi}. 

\subsection{Questions and
Êlessons   in supersymmetric QCD}

To be as close as possible to the real world we have to incorporate
light quarks. The number of the  light quarks is  three, 
and so is the number of colors, so that
the case $N_f=N_c$ is of most interest. SQCD with $N_f=N_c$
was thoroughly discussed in Sect. 3.2. If the mass term of the matter 
superfields is set to zero the  vacuum valley of the theory is 
parametrized by the moduli $M^i_j$, $B$ and $\tilde B$ subject to 
constraint (\ref{qmodnf}). Any point from this manifold is a 
legitimate vacuum.  If all moduli have large values,
the gauge symmetry is completely broken, all gauge bosons are very 
heavy (in the scale 
 $\Lambda$), and the theory is obviously
in the weak coupling regime. Since the scalar fields (squarks)
are in the fundamental representation we deal with a unified
Higgs/confinement phase.  In Sect. 3.2 we discussed 
what happens when one enters the part of the valley where the 
moduli become small and the strong coupling regime sets in. In this 
part of the valley two points were obviously singled out:
(i) the one corresponding to the standard pattern of the 
chiral symmetry breaking, $B=\tilde B = 0\, , \,\, \,  
M^i_j=\Lambda^2\delta^i_j$; and (ii) the point with the unbroken 
chiral symmetry and broken baryon charge, $B = -\tilde B 
=\Lambda^{N_f}\, ,\,\,\,  M^i_j = 0$, see Eqs. (\ref{bbvac}), 
(\ref{anpo}). Here we will briefly consider a SUSY-breaking 
deformation of this model 
analyzed in Ref. \cite{Peskin,Hoker}. To make  the expressions we 
will be dealing with more 
concise from now on $\Lambda$ will be put to unity.
All dimensionful quantities will be measured in the units of
$\Lambda$.

Following \cite{Peskin} we will assume that the squarks
and/or  gluinos get a mass term, added in the Lagrangian by hand.
These mass terms represent a soft supersymmetry breaking, 
\beq
\Delta{\cal L} = - m_Q^2 \left( |Q|^2+|\tilde{Q}|^2\right) 
- (m_g\lambda^2+\mbox{h.c.})
\label{ssb}
\eeq
where $Q$ and $\tilde Q$ are the lowest components of the 
corresponding superfields. This particular expression is not 
most general.
Other soft supersymmetry breaking  terms are possible.
The choice (\ref{ssb}) is singled out by the fact
that at $m_g =0$ all global symmetries of the
original supersymmetric model, see Eq. (\ref{AD5}), are preserved.
A non-vanishing gluino mass, $m_g \neq 0$, explicitly breaks the 
conservation of the $R$ current, leaving all other global symmetries 
intact. Correspondingly, the model with $m_g \neq 0$
will be referred to as the $\not\!\!{R}$ model, while that
with $m_g =0$ as the $R$ model. Tending $m_Q, m_g \ra\infty$
we return back to the chiral limit of QCD.
If $m_g$ is kept at zero while $m_Q\ra\infty$ the theory we arrive 
at is not QCD but, rather, the theory of quarks and
gluons plus one Majorana fermion in the adjoint representation.
It also exhibits the strong coupling behavior
and is interesting on its own.

The deformation of SQCD  caused by the gluino mass term $m_g\neq 
0$
switches on supersymmetry breaking in a smooth way.
The point $m_g =0$ is non-singular, so that one can study all
correlation functions of interest at small $m_g$ as an expansion
in $m_g$. In this way, the vacuum energy density was calculated to 
order ${\cal O}(m_g)$ in the previous section. 

Introduction of the squark mass term is more problematic.
Indeed, for the negative values of $m_Q^2$ the theory becomes 
unstable and tends to develop large VEV's of the scalar fields
\cite{Hoker} (see also \cite{Hoker2}). Strictly speaking, the theory 
with 
negative  $m_Q^2$ is not defined at all unless we add a quartic term 
for stabilization.  Therefore, at $m_Q^2= 0$
a phase transition must take place, and expansion near 
$m_Q^2=0$ is potentially dangerous. We will return
to this question later on. 

For the time being let us assume that $m_Q^2>0$
In the presence of the squark mass ($m_Q^2>0$)
the vacuum degeneracy is gone, the theory is locked in 
a concrete lowest-energy state which is,  generally speaking, 
non-degenerate.  If $m_Q^2\ll \Lambda^2$ one may hope that
the predictive power we had in the supersymmetric limit is not lost, 
and everything is calculable. 

Saying that everything is calculable is an  exaggeration.
Although we do know what happens with the  $F$ terms
under the deformation specified above, there is an ambiguity in the
$D$ terms of the relevant fields. These terms were unknown even 
in the limit of exact supersymmetry. Adding a soft supersymmetry
breaking does not help to pin them down, of course.  To be able
to predict the impact of the   squark/gluino masses
one has to accept certain assumptions about the kinetic terms.
Since we are aimed at a qualitative picture  anyway, it
 is natural  to make a simplest possible choice still capturing 
general features that must be inherent to any sensible kinetic term.
As long as our conclusions do not strongly depend
on details of the kinetic terms at hand, they seem to be trustworthy.

At first  we will focus on the $R$ model.  A few comments on the 
$\not\!\!{R}$ will be given at the end of this section. 

In the supersymmetric limit the 
particles residing in $M, B,\tilde B$ are massless. When a mass term
$m_Q^2$  is switched on some of the particles still remain massless,
while others acquire a mass. If $m_Q^2\ll \Lambda^2$
the acquired mass is small, and one can limit oneself
to the linear approximation. To answer physically interesting 
questions we must 
build an effective Lagrangian for the light degrees of freedom, 
find  the location of the vacuum state, and the spectrum 
\footnote{Eventually we will proceed to the second stage --
 what happens when
$m_Q$ and $ m_g$ become large. Needless to say that this
question can be 
addressed only at the qualitative level.}. 

Intuitively it is clear that the squark  mass pushes the vacuum
towards small values of $M,B$ and $\tilde B$.  
As a matter of fact, if it were not for the constraint (\ref{bbvac})
the lowest-energy state would be achieved at $M=B=\tilde B = 0$.
The constraint (\ref{bbvac}) does not allow $M,B$ and $\tilde B$
to vanish simultaneously. 
Upon a brief reflection, two points (\ref{bbvac}) and 
(\ref{anpo}), which were singled out even in the absence of 
supersymmetry breaking, become prime suspects.

The effective low-energy Lagrangian for the light composite fields
depends on a number of constants in the $D$ terms and in the 
SUSY breaking part. The latter constants appear when one rewrites
the mass term (\ref{ssb}), introduced at the fundamental level, in 
terms
of the composite fields $M,B$ and $\tilde B$.  The values of the 
constants remain undetermined. Depending on the relative weight of 
the $M$ and $B$ parts the minima at the points (\ref{bbvac}) and 
(\ref{anpo}) ``breathe".  One of them lies higher
than another. Although both logically possible versions
are analyzed in Ref. \cite{Peskin}, there are good reasons to
believe  that the 
minimum at $B=\tilde B = 0\, , \,\, \,  
M^i_j=\Lambda^2\delta^i_j$ must be  deeper, while  the minimum at 
$B = -\tilde B 
=1\, ,\,\,\,  M^i_j = 0$ is irrelevant (i.e. it is local and 
quantum-mechanically unstable). 

Why? Our starting point was supersymmetric QCD with the {\em 
massless} matter superfields. We could have started, instead, from 
the massive SQCD, with a small supersymmetric mass term for  all 
matter superfields.
Then, this new theory could have been deformed 
 by adding the SUSY-breaking term (\ref{ssb}) implying  heavier
squarks. 
Such an approach is even more realistic, since the light quarks in 
QCD, although light, are not exactly massless. If we start, however, 
from the {\em massive} SQCD, there is no vacuum degeneracy
from the very beginning; the only solution for the vacuum is given 
by
Eq. (\ref{bbvac}).

I remind that this 
vacuum solution, $B=\tilde B = 0\, , \,\, \,  
M^i_j=\Lambda^2\delta^i_j$, corresponds to the conventional 
dynamical pattern
with the spontaneously broken axial chiral symmetry and unbroken 
baryon charge. This is just the pattern we deal with in QCD. The 
``wrong"  vacuum (\ref{anpo}) corresponds to the unbroken
chiral symmetry and spontaneously broken baryon charge. 

In the $R$ model the $U(1)_R$
symmetry is not broken explicitly (since $m_g=0$);
the condensate $\langle M^i_j\rangle =\delta^i_j$ does not break it 
either.
The success of the 't Hooft matching 
in SQCD (Sect. 3.2) implies that
there is no spontaneous breaking of $U(1)_R$ in the limit of {\em 
exact} supersymmetry.  When the SUSY-breaking mass term 
$m_Q^2$  is introduced, a phase transition  may occur,
leading to a spontaneous breaking of $U(1)_R$. 
As we already know, the point $m_Q^2=0$ is singular anyway.
What we do not know is whether the $U(1)_R$ breaking
phase transition  occurs right at $m_Q^2=0$, or it takes place  at a 
finite value 
of $m_Q^2$. The former  option, disregarded in Ref. \cite{Peskin},  is 
somewhat 
exotic but  seemingly   cannot be ruled out on general grounds.

To begin with, we may assume, following  \cite{Peskin}, that  the $R$ 
symmetry is not  spontaneously broken 
in some finite range of $m_Q^2$, 
 $$
0 <  m_Q^2\leq (m_{Q}^2)_* \, .
$$
Then the light particle spectrum 
at $m_Q^2$ from this interval is  almost evident. All fermions,
$\psi_M, \psi_B$ and $\psi_{\tilde B}$, must remain massless.
As for their boson counterparts,  some of them ($N_f^2 -1$
pseudoscalar bosons) have to remain massless since they are the 
Goldstone bosons of the spontaneously
broken global $SU(N_f)$ symmetry. All others get masses generated 
by  the 
SUSY-breaking term (\ref{ssb}).  The constraint (\ref{bbvac}) 
leads to a subtlety which deserves mentioning.
The matrix $M$ has $N_f^2$ complex elements. It can be
conveniently 
 parametrized  as
\beq
M = \exp \left( \frac{\pi_V+i\pi_A}{\sqrt{2N_f}}\right) \exp
(\pi_V^a t^a ) \exp
(i \pi_A^a t^a )
\label{paramp}
\eeq
where $t^a$ are the generators of $SU(N_f)$, all $\pi$'s are real,  and  
$a = 1,2, ..., N_f^2 -1$.  The fields $B$ and $\tilde B$ provide two 
extra complex
parameters, so that altogether we deal with $N_f^2+ 2$ complex 
fields. The constraint (\ref{bbvac})  eliminates one complex field,
namely,
\beq
\pi_V+i\pi_A  = \sqrt{(2/N_f)} B\tilde B 
\eeq
plus cubic and higher order terms.  The fermionic
partner of $\pi_V+i\pi_A$ is also removed from the theory. Since the 
order
parameter of the spontaneously broken $SU(N_f)_L\times SU(N_f)_R
\ra SU(N_f)_V$
is $\langle M^i_j\rangle =\delta^i_j$, it is easy to see that the
corresponding Goldstone bosons are $\pi_A^a$. 
The masses of the remaining $N_f^2 +3$ (real)  bosons are 
proportional
to $m_Q$.

How do we learn that the composite fermions are massless
even though $m_Q^2 \neq 0$?
The global (unbroken) symmetry of the 
model at hand is
$SU(N_f)_V\times U(1)_B\times U(1)_R$. Since  the $R$ current
has the axial component, one has to match the 't Hooft triangles.
In the limit of the exact supersymmetry they do match (Sect. 3.2) --
the quark and gluino contribution at the fundamental level is 
matched by that of the massless composite fermions $\psi_M, 
\psi_B$ and $\psi_{\tilde B}$. To ensure the matching
persists 
at $m_Q^2\neq 0$ (with the unbroken $R$ symmetry),
we are forced to conclude that the masses of $\psi_M, \psi_B$ and 
$\psi_{\tilde B}$ remain at zero even though the squarks have a 
mechanical mass.

The masslessness of $\psi_M$ is  surprising, to put it mildly.  Indeed,  
$\psi_M$
is a composite state built of $\psi_Q$ and ${\tilde Q}$. The mass 
term of ${\tilde Q}$ is a free parameter, and it is hard to 
imagine that the strong interaction fine-tunes itself in such a way
that the mass of the composite state does not depend on this free 
parameter at all. 

A possible way out of this hurdle was indicated in Ref. \cite{Peskin}.
It was noted that the squarks $Q$ and $\tilde{Q}$ have exactly the 
same quantum numbers (color, flavor and Lorentz) 
as the gluino-quark composites,
\beq
 Q\ra \lambda^\alpha (\psi_Q)_\alpha\, , \,\,\,
 \tilde{Q}\ra \lambda^\alpha (\psi_{\tilde Q})_\alpha\, ,
\eeq
where $\alpha$ is the Lorentz index, and the color and flavor indices 
are
suppressed. Therefore, one may think of $\psi_M$ as of
a state built from $\psi_Q$ and $(\lambda\psi_{\tilde Q})$.
Since neither $\psi$'s nor $\lambda$'s have mechanical masses in 
the $R$ model, the masslessness of  $\psi_M$ seems less 
counterintuitive.  An extremist evolution of this idea leads
to a hypothetical scenario \cite{Peskin}, according to which there is 
no phase transition in $m_Q^2$ at all, and massless composites
$(\psi_Q\lambda\psi_{\tilde Q})$ survive in the limit
$m_Q^2\ra \infty$, when the theory at hand flows to QCD 
supplemented by  one extra (Majorana) spinor in the adjoint 
representation. 

The possibility which seems preferable to me is
the $R$-breaking phase transition at $m_Q^2=0$. If we postulate that 
at any positive non-vanishing value of $m_Q^2$ an $R$ violating 
condensate,
say, $\langle \lambda\lambda\rangle$ develops,
there will be no need in saturating the 't Hooft triangles by massless
fermions. Then they will be saturated by a Goldstone meson,
the ``ninth" Goldstone, built from a mixture of $\bar 
\lambda\lambda$
and $\bar\psi\psi$. The theory emerging in this way is very similar
to the conventional QCD. 

Thus, we see that the deformation of SQCD by the squark mass term 
does not take us too far. Too many unknowns remain, precluding us 
from making definite conclusions about the dynamical behavior of
the non-supersymmetric theory even for small values of $m_Q^2$.
The issue is clearly not ripe enough, further effort is needed
before insights in QCD dynamics can be obtained. 

As far as the $\not\!\! R$ model is concerned,
here the route to QCD is much smoother. The $R$ symmetry is
explicitly broken by $m_g\neq 0$ from the very beginning.
At small $m_g$ the  composite fermions can be shown to have 
masses proportional to $m_g$.  If we take the limit $m_g\ra\infty$
first, the gluino field decouples while  the mass of the composite 
fermions presumably is frozen at $\Lambda$.  $N_f^2 - 1$ 
bosons survive as massless Goldstones; the remaining $N_f^2 + 3$ 
are still light. Increasing now the value of $m_Q^2$
in the positive direction we decouple all squarks, and make $N_f^2 + 
3$  light bosons heavy (i.e. of order $\Lambda$). The Goldstones that 
survive in this limit are familiar QCD pions. 

Concluding, let me note that the potential of this approach is far from 
being exhausted. It seems that this range of ideas 
-- SUSY-breaking deformations of supersymmetric models whose 
solution we know -- 
can bring lavish fruits. Various supersymmetric models
can be used as a starting point. In particular, one can start from 
$N=2$ supersymmetry \cite{Alvarez}. No matter where
the starting point is, the destination is the same, QCD. 

Will we reach this destination in the foreseeable future?

\newpage

\section{Conclusions}

This lecture course summarizes advances in theoretical 
understanding of nonperturbative phenomena in the strong 
coupling regime. If before the  SUSY era,
the number of exact nonperturbative results in four-dimensional 
field theory could be counted on one hand, 
with the advent of  supersymmetry a wide spectrum 
of problems relevant to the most intimate aspects of strong  gauge 
dynamics
found exact solutions. Mysteries unravel. Our understanding
of gauge theories is dramatically deeper now than it was a decade 
ago. When preparing these lectures, I intended to share with you,
all the excitement and joys associated with the continuous advances 
in 
this field spanning over 15 years. Hopefully, the message I tried to
convey will be appreciated in full. 

Supersymmetric gauge dynamics is very rich, but life is richer, still.
The world surrounding us is not supersymmetric. It remains to be 
seen whether the remarkable discoveries and elegant, powerful 
methods developed in supersymmetric gauge theories
will prove to be helpful in solving the messy  problems of  real-life 
particle physics.  So far, not much has been done in this direction. 
In today's climate it is rare that the question of practical applications 
is even posed.  I hope that we reached a turning point: high-energy 
theory will return  to its empiric roots. The 
 command we obtained of
supersymmetric gauge theories will be 
a key
which will open to us Pandora's box of problems  of 
Quantum Chromodynamics, {\em the} theory of our world. 
Pandora opened the jar that contained all human blessings, and they 
were gone.  Will the achievements obtained in supersymmetric gauge 
theories be lost in the Planckean nebula?

\vspace{0.3cm}

\section{Acknowledgments}

I would like to thank B. Chibisov, I. Kogan, A. Kovner, 
G. Veneziano and especially A. Vainshtein  for numerous 
discussions and helpful comments. I am grateful to S. Ferrara, F. 
Fucito,
K.  Intriligator,
V. Miransky, M. Sato, and  P. West, for pointing out  misprints and 
omissions in the first version, and to 
B. Chibisov
for his kind assistance in preparation of the manuscript. Parts of 
these
lecture notes were written during my stay at  CERN Theory
Division in fall 1996 and at Isaac Newton Institute for 
Mathematical Sciences, Cambridge University, in March 1997.
I am grateful to colleagues from the above groups for
 hospitality. 

This work was supported in part by DOE under the grant number
DE-FG02-94ER40823.

\newpage

\section{Appendix: Notation, Conventions, Useful Formulae}

\renewcommand{\theequation}{A.\arabic{equation}}
\setcounter{equation}{0}

In this Appendix  the key elements of the formalism
used in supersymmetric gauge theories are outlined. Basic formulae 
are collected for convenience.

The notation we follow is close to that of the canonic text book of 
Bagger and Wess \cite{6}. There are some distinctions, though.
The most important of them is the choice of the metric.
Unlike Bagger and Wess, we use the standard metric $g_{\mu\nu}
=(+---)$.  There are also distinctions
in normalization, see Eq. (\ref{intnorm}). 

The left-handed 
spinor is denoted by undotted indices, e.g. $\eta_{\beta}$. The 
right-handed spinor is denoted by dotted indices, e.g. 
$\bar{\xi}^{\dot\beta}$. (This convention is standard in 
supersymmetry
but is opposite to one accepted in the text-book \cite{LL}).
The Dirac spinor $\Psi$ then takes the form 
\beq
 \qquad \Psi =  \left( \begin{array}{l}
                          \bar{\xi}^{\dot\beta} \\
                          \eta_{\beta}
                          \end{array}
                   \right)\, .
\eeq
 Lowering and raising of the spinorial indices is done by
applying the Levi-Civita tensor from the left,
\beq
\chi^{\alpha} = \epsilon^{\alpha\beta} \chi_{\beta} \;, \quad 
\chi_{\alpha} = \epsilon_{\alpha\beta} \chi^{\beta},
\eeq
and the same for the dotted indices, where
\beq
\epsilon^{\alpha\beta} = - \epsilon^{\beta\alpha}\;,\quad 
\epsilon^{12} =\; -\;\epsilon_{12} = 1\, . 
\eeq
The products of the undotted and dotted spinors are defined as 
follows:
\beq
\eta\chi = \eta^{\alpha}\chi_{\alpha} = - 
\eta_{\alpha}\chi^{\alpha}\;,\quad 
\bar{\eta}\bar{\chi} = 
\bar{\eta}_{\dot\alpha}\bar{\chi}^{\dot\alpha}\, .
\eeq
Under this convention $(\eta\chi)^+ =\bar{\chi}\bar{\eta}$. 
Moreover,
\beq
\theta^{\alpha}\theta^{\beta} =  - \frac{1}{2}\epsilon^{\alpha\beta}  
\theta^2\, ,\quad 
\theta_{\alpha}\theta_{\beta} =  \frac{1}{2}\epsilon_{\alpha\beta}  
\theta^2\, , \quad 
\bar\theta^{\dot\alpha}\bar\theta^{\dot\beta} =   
\frac{1}{2}\epsilon^{\dot\alpha\dot\beta} \bar\theta^2\, ,\quad
\bar\theta_{\dot\alpha}\bar\theta_{\dot\beta} =  - 
\frac{1}{2}\epsilon_{\dot\alpha\dot\beta}
\bar\theta^2\, .
\eeq
The vector quantities (representation $(\frac{1}{2}\;,\frac{1}{2}))$
are obtained in the spinorial formalism by multiplication by
\beq
\quad (\;\sigma^{\mu}\;)_{\al\db} 
\;=\;\{\;1\;,\vec{\tau}\;\}_{\al\db}
\eeq
where $\vec{\tau}$ stands for the Pauli matrices, for instance,
\beq
\quad {A}_{\al\db} = {A}_{\mu} ( \sigma^{\mu} )_{\al\db}.
\eeq
Note that
\beq
{A}_{\mu}{B}^{\mu}\;=\; \frac{1}{2}{A}_{\al\db}{B}^{\al\db}\;,
\quad {A}_{\al\db}{A}^{\ga\db}=\delta^{\ga}_{\al}\;{A}_{\mu}{ 
A}^{\mu}.
\eeq
The square of the four-vector is understood as
\beq
{A}^2 \equiv  {A}_{\mu}{A}^{\mu} = \frac{1}{2}{A}_{\al\db}{ 
A}^{\al\db}\, .
\eeq
If the matrix  $(\sigma^{\mu} )_{\al\db}$ is ``right-handed'' it is
convenient to introduce its ``left-handed'' counterpart,
\beq
( \bar{\sigma}^{\mu} )^{\db\al}= \;\{\;1\;,-\vec{\tau}\;\}_{\db\al} \, .
\eeq 
The matrices that appear in dealing with representations $(1\;,\;0)$
 and $(0\;,\;1)$ are 
\beq
( \vec{\sigma}  )^{\al}_{\be} = \; \vec{\tau}_{\al\be}\;,\quad
( \vec{\sigma}  )^{\al\be}= \eps^{\be\delta} 
\vec{\sigma}^{\al}_{\delta} \, ,
\label{matrix1}
\eeq
and the same for the dotted indices. The matrices $( \vec{\sigma}  
)^{\al\be}$ are
symmetric, $( \vec{\sigma}  )^{\al\be}\;=\;( \vec{\sigma}  )^{\be\al}$.
In the explicit form
\beq
( \vec{\sigma}  )^{\al\be} = \{ \;\tau^3\;,\;-i{\bf 1}\;,\;-
\tau^1\;\}_{\al\be} \, .
\label{matrix}
\eeq
Note that with our definitions
$$
( \vec{\sigma}  )_{\al\be} = \{ \;-\tau^3\;,\;-i{\bf 
1}\;,\;\tau^1\;\}_{\al\be} \, .
$$
The left (right) coordinates ${\it x}_{L,R}$ and covariant derivatives 
are
\begin{eqnarray}
&&({\it x_{L}})_{\al\da} = {\it x}_{\al\da} - 
2\;i\;\th_{\al}\bar{\th}_{\da}\;,
\quad 
({x_{R}})_{\al\da} = {\it x}_{\al\da} + 2\;i\;\th_{\al}\bar{\th}_{\da}\;,
\nonumber\\
&&{D}_{\al} =   \frac{\partial}{\partial\th^{\al}} - i\; 
\partial_{\al\da}\bt^{\da}\;,
\quad
{\bar{D}}_{\da} = - \frac{\partial}{\partial\bt^{\da}} + i\;\th^{\al} 
\partial_{\al\da} ,
\end{eqnarray}
so that
\begin{eqnarray}
&\{ {D}_{\al}\;{\bar{D}}_{\da} 
\}\;=\;2i\partial_{\al\da}\;,&\nonumber\\
&{ \bar{D}}_{\db}({\it x}_{L})_{\al\da}=0\;,\quad
&{D }_{\be}({\it x}_{R})_{\al\da}=0\;,\nonumber\\
&{\bar{D}}_{\db}({\it x}_{R})_{\al\da}=-
4\;i\;\th_{\al}\ep_{\db\da}\;,\quad
&{D }_{\be}({\it x}_{L})_{\al\da}=4\;i\;\bt_{\da}\ep_{\be\al} \, .
\end{eqnarray}
The law of the supertranslation is
\begin{eqnarray}
&&\th\rightarrow\th +\ep\;,\quad \bt\rightarrow\bt +\bep 
\;,\nonumber\\
&&{\it x}_{\al\db} \rightarrow {\it x}_{\al\db} - 
2\;i\;\th_{\al}\bep_{\db}
+ 2\;i\;\ep_{\al}\bt_{\db} \;,\nonumber \\
&&({\it x}_L)_{\al\db} \rightarrow ({\it x}_L)_{\al\db} - 
4\;i\;\th_{\al}\bep_{\db}\;,\nonumber\\
&&({\it x}_R)_{\al\db}\rightarrow ({\it x}_R)_{\al\db} + 
4\;i\;\ep_{\al}\bt_{\db}\, .
\end{eqnarray}

The integrals over the Grassmann variable are normalized as follows
\beq
\int d^2\th\;\th^2\;=\;2\;,\quad \int d^4\th \;\th^2\;\bt^2\;=\;4\;,
\eeq
and we define
\beq
\{\dots\}_F\;=\;\frac{1}{2}\;\int d^2 \th \{\dots\}\;,\quad
\{\dots\}_D\;=\;\frac{1}{4}\;\int d^4 \th \{\dots\}.
\label{intnorm}
\eeq

A generic non-Abelian SUSY gauge theory has the Lagrangian
$$
{\cal L} = \left\{\frac{1}{4 g_0^2}{\rm Tr}\,\int d^2\th\;W^2\, +\, 
\mbox{H.c.}\right\}\, +
$$
\beq
+\frac{1}{4}\, \sum_i\,  \int  d^4\th \bar{Q}_ie^V Q_i
+\left\{ \frac{1}{2}\int d^2 \th {\cal W}(\{Q_i\}) \, + \, \mbox{H.c.} 
\right\}
\, ,
\label{gensusylag}
\eeq
where
$$
\frac{1}{g^2_0} = \frac{1}{g^2} -\frac{i\vartheta}{8\pi^2}\, ,
$$
is the (complexified) gauge coupling constant, the sum in Eq. 
(\ref{gensusylag}) runs over all matter superfields $Q_i$ present in 
the theory, and ${\cal W}(\{Q_i\})$ is a generic superpotential.
Most commonly one deals with the superpotential corresponding to 
the mass term of the matter fields. In many models,   cubic terms are 
gauge invariant; then  they are  allowed too (and do not spoil 
renormalizability of the theory).

Furthermore,
the superfield $W_{\al}$, which 
includes the
 gluon strength tensor, is defined as follows:
\beq
{W}_{\al} = \frac{1}{8}\;\bar{D}^2\left({\rm e}^{-V}\;D_{\al} \;{\rm 
e}^{V}   
\right)
\eeq
where $V $ is the vector superfield. In  the Wess-Zumino gauge
\beq
V = -2\th^{\al}\bt^{\da}A_{\al\da} - 2i\bt^2(\th\la) + 2i 
\th^2(\bt\bar{\la}) + 
\th^2\bt^2 D\, ,
\eeq    
 $V=V^{a}T^{a}$
and $T^{a}$
stands for the generators of the gauge group $G$. In the fundamental 
representation of $SU(N)$, a case of most practical interest, 
$$
{\rm Tr}\left( T^a T^b \right) = \frac{1}{2}\;\delta^{ab}\, .
$$
For a general representation $R$ of any group $G$ we define
$$
{\rm Tr}\left( T^a T^b \right)_R = T(R)\, \delta^{ab}\, .
$$
If $R$ is the adjoint representation, $T(\mbox{adjoint})
\equiv T(G)$.

The supergauge transformation
has the form
\beq
Q_i\rightarrow e^{i\La}\, Q_i\, ,   \quad e^V\rightarrow 
e^{i\bar{\La}}e^V 
e^{-
i\La}\, ,    \quad 
W_{\al}\rightarrow e^{i\La} W_{\al} e^{-i\La}   \, ,
\eeq
where $\La$ is an arbitrary chiral superfield ($\bar{\La}$ is 
antichiral). 
In components
\beq
{W}_{\al} = i\left( \la_{\al} + i\th_{\al}D - \th^{\be}\;G_{\al\be} - 
i\th^2{\cal D}_{\al\da}\bar{\la}^{\da} \right)
\eeq
where $\la_{\al}$ is the gluino (Weyl) field, ${\cal D}_{\al\da}$ is the  
covariant 
derivative, and $G_{\al\be}$ is the gluon field strength tensor in the
 spinorial notation.

The standard gluon field strength tensor transforms as $(1,0)+(0,1)$ 
with respect 
to the
Lorentz group. Projecting out pure $(1,0)$ is achieved by virtue of 
the 
$(\sigma)_{\al\db}$ matrices,
\beq
G_{\al\be}\;=\;-\frac{1}{2} 
G_{\mu\nu}(\sigma^{\mu})_{\al\db}(\sigma^{\nu})_{\be\dot{\delta}}
\ep^{\db\dot{\delta}} = (\vec{E} 
-\,  i\, \vec{B})(\vec{\sigma})_{\al\be}\, .
\eeq
Then 
$$
G^{\al\be}G_{\al\be}\; =\;2(\vec{B}^2 - \vec{E}^2 + 2\,  i\,
\vec{E}\vec{B}) = 
G_{\mu\nu}G_{\mu\nu} - i\, G_{\mu\nu}\tilde{G_{\mu\nu}} 
$$
where
\beq
\tilde{G}_{\mu\nu} = 
\frac{1}{2}\ep_{\mu\nu\al\be}G^{\al\be}\, ,  \quad 
(\ep_{0123} = -1)\, .
\eeq
  
The supercurrent supermultiplet has the following general form
$$
J_{\al\da} = R^0_{\al\da} - \frac{1}{2}\left\{ i\th^{\be}( J_{\be\al\da} 
- 
\frac{2}{3}\eps_{\be\al}\eps^{\ga\delta}J_{\delta\gamma\da}) + 
\mbox{H.c.}    \right\} - 
$$
\beq
  \th^{\be}\bt^{\db}  
    \left( {J_{\al\da\be\db}} -
     \frac{1}{3}\eps_{\al\be}\eps_{\da\db}\eps^{\ga\delta}
     \eps^{\dg\dot{\delta}} J_{\ga\dg\delta\dot{\delta}} \right)+\dots
\label{decom2}
\eeq
where $R^0_{\al\da}$ is the $R_0$ current,  $J_{\be\al\da} $ is the 
supercurrent, 
and ${J_{\al\da\be\db}}$ is related to the energy-momentum tensor, 
$\th_{\mu\nu}$,  in the
following way
\beq
J_{\al\da\be\db} = - (\sigma^i)_{\al\be}(\sigma^j)_{\da\db} 
\left\{
\th^{ij} + \th^{00}g^{ij} - \eps^{ijk}\th^{0k} 
\right\} + 
\frac{1}{2}\eps_{\al\be}\eps_{\da\db} \th^{\mu}_{\mu} \, ;
\label{enmt}
\eeq 
here $i,j,k = 1,2,3$ ,  $g_{\mu\nu}$ is the metric tensor and 
matrices
$(\sigma^i)_{\al\be}$ are defined in Eq.(\ref{matrix1}).

The general anomaly relation (three ``geometric" anomalies) is
\beq
\bar{D}^{\da}J_{\al\da} = \frac{1}{3} D_{\al}\left\{
\left[ 3{\cal W } - \sum_i Q_i \frac{\partial{\cal W }}{\partial Q_i}
\right] - 
\left[ \frac{3T(G)- \sum_i T(R_i)}{16\pi^2}{\rm Tr}W^2 + 
\frac{1}{8}\sum_i\gamma_i 
\bar{D}^2
(\bar{Q}_i e^{V} Q_i) \right]
\right\}\, ,
\label{geom}
\eeq
where $\gamma_i$ are the anomalous dimensions of the matter 
fields $Q_i$. The general Konishi anomaly has the form
\beq
\frac{1}{8} \bar{D}^2(\bar{Q}_i^+ e^{V} Q_i) = 
\frac{1}{2}Q_i \frac{\partial{\cal W }}{\partial Q_i} +
\frac{\sum_i T(R_i)}{16\pi^2}{\rm Tr}W^2\, .
\label{ka1}
\eeq 

In conclusion let us present the full component expression for the 
simplest SU(2) model with one flavor (two subflavors), assuming that 
the superpotential in the case at hand reduces to the mass term of 
the quark (squark) fields. This model was discussed in detail in Sects. 
1.3.2 and 1.3.5. If the index $f$ denotes the subflavors, $f=1,2$,
\begin{eqnarray}
{\cal L} = \frac{1}{g^2}
\left\{ -\frac{1}{4}G^a_{\mu\nu}G^a_{\mu\nu} + \la^{\al,a}i{\cal 
D}_{\al\da}\bar{\la}^{\da,a}
+ \frac{1}{2} D^a D^a  \right\}
+\nonumber\\
\psi^{f\al}i{\cal D}_{\al\da}\bar{\psi}^{f\da} + 
({\cal D}_{\mu}\phi^{+f})({\cal 
D}_{\mu}\phi^{f})
+ F^{+f}F^{f} + 
\nonumber\\
 i\sqrt{2}( \phi_1^+\la\psi_1 + \phi_1\bar{\la}\bar{\psi}_1 +
\phi_2^+\la\psi_2 + \phi_2\bar{\la}\bar{\psi}_2 ) +\nonumber\\
 \frac{1}{2} D^a\;(\phi^+_1 T^a \phi_1 + \phi^+_2 T^a \phi_2) +
\nonumber\\
m\;( \phi_1 F_2 + \phi_1^+ F_2^+  + \phi_2 F_1 + \phi_2^+ F_1^+
+ \psi_1\psi_2 + \bar{\psi}_1\bar{\psi}_2 )\, .
\end{eqnarray}
In this model the component expression for the supercurrent is 
$$
{ J_{\al\be\db}}= 2 \left\{\frac{1}{g^2}\left[
i G^a_{\be\al}\bar{\la}^a_{\db} -
3\eps_{\be\al}D^a\bl^a_{\db}\right]  \right.
+
$$
\beq
\left.
\sqrt{2}\left[  (\partial_{\al\db}\phi^+)\psi_{\be} -i\;\eps_{\be\al} F 
\bar{\psi}_{\db}\right]-
\frac{\sqrt{2}}{6}\left[ \partial_{\al\db}(\psi_{\be}\phi^+) + 
\partial_{\be\db}(\psi_{\al}\phi^+)
- 3 \eps_{\be\al}\partial_{\db}^{\ga}(\psi_{\ga}\phi^+) \right]  
\right\}\, .
\label{sqcdcur} 
\eeq

\newpage

\section{Recommended Literature}

It is assumed that the reader is familiar
with the  text-books on supersymmetry:

\vspace{0.1cm}

J. Bagger and J. Wess, {\em Supersymmetry and Supergravity}, 
(Princeton 
University Press, 1983).

\vspace{0.1cm}

P. West, {\it Introduction to Supersymmetry and Supergravity}
(World Scientific, Singapore, 1986).

\vspace{0.1cm}

 S.J. Gates, M.T. Grisaru, M. Ro\v{c}ek and W. Siegel, {\em Superspace}
(The Benjamin/ Cummings, 1983).

\vspace{0.1cm}

D. Bailin and A. Love,  {\it Supersymmetric Gauge Field Theory
and String Theory} (IOP Publishing, Bristol, 1994).

\vspace{0.3cm}

A solid introduction to supersymmetric instanton calculus is given in

\vspace{0.1cm}

{\em Instantons in Gauge Theories}, ed. M. Shifman,
(World Scientific, Singapore, 1994), Chapter VII.

\vspace{0.1cm}

A brief survey of those  aspects of supersymmetry which are most 
relevant to the recent developments can be found in 

\vspace{0.1cm}

J. Lykken, {\em Introduction to Supersymmetry},
hep-th/9612114.

\vspace{0.3cm}

\begin{center}

{\bf Reviews on Exact Results in SUSY Gauge Theories and Related 
Issues}

\end{center}

\vspace{0.1cm}

N. Seiberg, {\em  The Power of Holomorphy --
Exact Results in 4D SUSY Field Theories}, in {\em Proc. VI
International Symposium on  Particles, Strings, and Cosmology
   (PASCOS 94)},   Ed.  K. C. Wali,  (World
   Scientific, Singapore, 1995) [hep-th/9408013].

\vspace{0.1cm}

K. Intriligator and N. Seiberg, {\em Lectures on Supersymmetric 
Gauge Theories and 
Electric - Magnetic
Duality},  {\it Nucl. Phys. Proc. Suppl.}  {\bf 45BC} (1996) 1
[hep-th/9509066].

\vspace{0.1cm}

K. Intriligator and N. Seiberg, {\em Phases of N=1 Supersymmetric
Gauge Theories and Electric - Magnetic Triality},
in Proc. Conf.
  {\it Future Perspectives in String Theory (Strings '95)}
     Eds. I. Bars, P. Bouwknegt, J. Minahan, D.
   Nemeschansky, K. Pilch, H. Saleur, and. N. Warner (World Scientific,
Singapore, 
   1996) [hep-th/9506084]. 

\vspace{0.1cm}

D. Olive, {\em Exact Electromagnetic Duality}, 
{\it Nucl. Phys. Proc. Suppl.} {\bf 45A} (1996) 88 [hep-th/9508089].

\vspace{0.1cm}

P. Di Vecchia, {\em Duality in Supersymmetric Gauge Theories},
 {\it Surveys High Energ. Phys.} {\bf 10} (1997) 119
[hep-th/9608090].

\vspace{0.1cm}

L. Alvarez-Gaum\'{e} and S.F. Hassan,
{\em Introduction to S-Duality  in N=2 Supersymmetric 
Gauge Theories}, {\it Fortsch.Phys.} {\bf 45} (1997) 159 [hep-
th/9701069].

\vspace{0.1cm}

W. Lerche, {\em Notes on N=2 Supersymmetric
Yang-Mills Theory}, {\it Nucl. Phys. Proc. Suppl.} {\bf 55B} (1997) 83 
[hep-th/9611190].

\vspace{0.1cm}

A. Bilal, {\em Duality in  N=2 SUSY
Yang-Mills Theory: A Pedagogical Introduction to the Work of 
Seiberg and Witten}, hep-th/9601007.

\vspace{0.1cm}

S. Ketov, {\em Solitons, Monopoles, and Duality:
From Sine-Gordon to Seiberg-Witten}, {\it Fortsch. Phys.} {\bf 45} 
(1997) 237 [hep-th/9611209]. 

\vspace{0.1cm}

M. Peskin, {\em Duality in Supersymmetric Yang-Mills Theory},
hep-th/9702094.


\newpage

\begin{thebibliography} {99}

\bibitem{polyakov}
A. Polyakov, {\it Nucl. Phys. } {\bf 120} (1977) 429.

\bibitem{mandelstam}
S. Mandelstam, {\it Phys. Reports} {\bf 23} (1976) 245.

\bibitem{thooft}
G. 't Hooft, in {\it 1981 Carg\'{e}se Summer School Lecture Notes on 
Fundamental Interactions, NATO Adv. Study Inst. Series B: Phys.,}
Vol. 85, ed. M. L\'{e}vy {\it et al.}  (Plenum Press, New York, 1982)
[reprinted in G. 't Hooft, {\it Under the Spell of the Gauge Principle}
(World Scientific, Singapore, 1994), page 514];
{\it Nucl. Phys. } {\bf B190} (1981) 455.

\bibitem{GL}
Yu. Golfand and E. Likhtman, {\it Pis'ma ZhETF}, {\bf 13} (1971)
452 [{\it JETP Lett.} {\bf 13} (1971) 323]; see also in
{\it Problems of Theoretical Physics}, I.E. Tamm
Memorial Volume (Nauka, Moscow, 1972), page 37;\\
E.P. Likhtman, {\it Irreducible representations, extensions of the 
algebra of 
Poincar\'{e} generators by bispinor generators},
{\it Kratk. Soob. Fiz. -- Short Comm. Phys. FIAN}, 1971, No. 5, page 
197.


\bibitem{Higgs}
P.W. Higgs, {\it Phys. Lett.} {\bf 12} (1964) 132; {\bf 13} (1964) 508;
 {\it Phys. Rev.} {\bf 145}
(1966) 1156;\\
F. Englert and R. Brout, {\it Phys. Rev. Lett.} {\bf 13} (1964) 321;\\
G.S. Guralnik, C.R. Hagen and T.W.B. Kibble, {\it Phys. Rev. Lett.} {\bf 
13} (1964) 585.

\bibitem{Mattis}
M. Mattis, {\it Phys. Reports} {\bf 214} (1992) 159;\\
V.A. Rubakov and  M.E. Shaposhnikov, 
{\it  Usp. Fiz. Nauk} {\bf 166} (1996)
493 [{\it Phys. Uspekhi} {\bf 39} (1996) 461]. 

\bibitem{GeG}
H. Georgi and S. Glashow, {\it Phys. Rev. Lett.} {\bf 28} (1972) 1494.

\bibitem{THP}
A. Polyakov, {\it JETP Lett.} {\bf 20} (1974) 194;\\
G. 't Hooft, {\it Nucl. Phys.} {\bf B79} (1974) 276.

\bibitem{RAJ}
R. Rajaraman, {\it Solitons and Instantons}, (North-Holland,
Amsterdam, 1987), Chapter 3.

\bibitem{Bog}
E. Bogomol'nyi, {\it Sov. J. Nucl. Phys.} {\bf 24} (1976) 449.

\bibitem{PrS}
M.K. Prasad and C.H. Sommerfield, {\it Phys. Rev. Lett.} {\bf 35}
(1976) 760.

\bibitem{Harv}
J. Harvey, in Proc. {\em 1995 Summer School in High-Energy Physics 
and Cosmology},  Eds. E. Gava, A. Masiero, K.S. Narain, S.
   Randjbar-Daemi, Q. Shafi,  (World Scientific, Singapore, 1997) 
[hep-th/9603086]; \\
L. Alvarez-Gaum\'{e} and S.F. Hassan,
{\it Fortsch. Phys.} {\bf 45} (1997) 159.


\bibitem{DirM}
P.A.M. Dirac, {\it Proc. Roy. Soc.} {\bf A133} (1931) 60.\\
The classical source on the Dirac monopole is
T.T. Wu and C.M. Yang, {\it Phys. Rev.} {\bf D12} (1975) 3845.

\bibitem{SchM}
J. Schwinger,  {\it Phys. Rev.} {\bf 144} (1966)  1087.\\
This paper as well as all classical Dirac's papers on the monopole
are reprinted in Russian in  {\em The Dirac Monopole},
Eds. B. Bolotovskii and Yu. Usachev,  (Mir Publishers,
Moscow, 1970).

\bibitem{JZ}
B. Julia and A. Zee, {\it Phys. Rev.} {\bf D11} (1975) 2227.

\bibitem{SEIWIT}
N. Seiberg and E. Witten,
{\it Nucl. Phys.} {\bf  B426} (1994) 19; (E) {\bf B430} (1994) 485;
{\it Nucl. Phys.} {\bf B431} (1994) 484.

\bibitem{BRFS}
T. Banks and E. Rabinovici, 
{\it Nucl. Phys.} {\bf B160} (1979) 349;\\
E. Fradkin and S. Shenker, {\it Phys. Rev.} {\bf D19} (1979) 3682.

\bibitem{DRuj}
A. De Rujula,   R.C. Giles, and R.L. Jaffe,
{\it Phys. Rev. }{\bf D17} (1978) 285.

\bibitem{CR}
J. Cardy and E. Rabinovici, {\it Nucl. Phys. } {\bf B205} (1982)1;\\
J. Cardy, {\it Nucl. Phys. } {\bf B205} (1982)17.

\bibitem{Ed}
E. Witten, {\it Phys. Lett.} {\bf B86} (1979) 283.

\bibitem{moninst}
P. Rossi, {\it Nucl. Phys.} {\bf B149} (1979) 170.

\bibitem{moninst1}
Yu. Simonov, {\it Yad. Fiz.} {\bf 42} (1985) 557 [{\it Sov. J. Nucl. 
Phys.} {\bf 42} (1985) 352].

\bibitem{moninst2}
M. Chernodub and F. Gubarev, in {\it Nonperturbative Approaches
to Quantum Chromodynamics}, Proc. Int. Workshop of the European 
Center for
Theoret. Studies in Nucl. Phys. and Related Areas, Trento, 1995, Ed. D. 
Diakonov (Gatchina, 1995), page 217;\\
H.  Suganuma, K. Itakura, H. Toki, and O. Miyamura,
{\it ibid.}, page 224;\\
R.C. Brower, K.N. Orginos and C.-I. Tan, {\it Phys. Rev.} {\bf D55} 
(1997) 6313.

\bibitem{Simonov}
Yu. Simonov, in {\it Selected Topics in Nonperturbative
QCD},  Proc. 
International School of Physics, ``Enrico Fermi", Course 80,
Varenna, Italy, 1995, Eds. A. Di Giacomo and D. Diakonov
(IOS Press, Oxford, 1996), page 339
 [hep-ph/9509403]; 
\\
M. Polikarpov,
{\it Nucl. Phys. Proc. Suppl.} {\bf 53} (1997) 134.

\bibitem{monlatun}
L. Del Debbio, M. Faber, J. Greensite and S. Olejnik,
{\it Nucl. Phys. Proc. Suppl.} {\bf 53} (1997) 141.

\bibitem{monlat}
A. Kronfeld, G. Schierholz and U.-J. Wiese,
{\it Nucl. Phys.} {\bf B293} (1987) 461;\\
S. Hioki, S. Kitahara, S. Kiura, Y. Matsubara, S. Ohno
and T. Suzuki, {\it Phys. Lett.} {\bf B272} (1991) 326;\\
S. Kitahara, Y. Matsubara, and T. Suzuki, {\it Prog. Theoret. Phys.} 
{\bf 93} (1995) 1.

\bibitem{monlat1}
H. Suganuma, S. Umisedo, S. Sasaki, H. Toki and O. Miyamura,
{\it Aust. J. Phys.} {\bf  50} (1997) 233;\\
A. Hart and M. Teper, {\it Nucl. Phys. Proc. Suppl.} {\bf 53}
(1997) 497;\\
T. Suzuki {\em et al.}, {\it Nucl. Phys. Proc. Suppl.}
{\bf 53} (1997) 531.

\bibitem{Pauli}
W. Pauli, {\it Pauli Lectures on Physics}, Vol. 6,
{\it Selected Topics in Field Quantization} (MIT Press, Cambridge, 
1973), page 
33.

\bibitem{SS}
A. Salam and J. Strathdee, {\it Nucl. Phys. } {\bf B76} (1974) 477;
{\it Phys. Rev. } {\bf D11} (1975) 1521.

\bibitem{WessZ}
J. Wess and B. Zumino, {\it Nucl. Phys.} {\bf B70} (1974) 39;
{\it Phys. Lett.} {\bf B49} (1974) 52.

\bibitem{Sergio}
S. Ferrara and B. Zumino, {\it Nucl. Phys.} {\bf B79} (1974) 413.

\bibitem{NSVZ1}
V. Novikov, M. Shifman, A. Vainshtein and V. Zakharov, {\it Nucl. 
Phys.} {\bf 
B229} (1983) 407.

\bibitem{SV4}
M. Shifman and A. Vainshtein, {\it Nucl. Phys.} {\bf B296} (1988) 
445.

\bibitem{Kovner}
A. Kovner and M. Shifman, {\it Phys. Rev. } {\bf D56} (1997) 2396.

\bibitem{VENY}
G. Veneziano and S. Yankielowicz, {\it Phys. Lett.} {\bf B113} (1982) 
321;\\
T. R. Taylor, G. Veneziano and S. Yankielowicz, 
{\it Nucl. Phys.} {\bf B218} (1983) 
493.

\bibitem{NSVZ2}
V. Novikov, M. Shifman, A. Vainshtein and V. Zakharov, {\it Nucl. 
Phys.} {\bf 
B229} (1983) 381; {\it Phys. Lett.} {\bf B166} (1986) 334.

\bibitem{SSF}
A. Salam and J. Strathdee, {\it Nucl. Phys.} {\bf B87} (1975) 85;\\
P. Fayet, {\it Nucl. Phys.} {\bf B90} (1975) 104.

\bibitem{Grisa}
M. Grisaru, in
{\it Recent Developments in Gravitation} (Carg\'{e}se Lectures, 1978),
Eds. M. Levy and S. Deser (Plenum Press, New York, 1979), page 577,
and references therein.

\bibitem{Witten}
E. Witten, {\it Nucl. Phys.} {\bf B202} (1982) 253.

\bibitem{BDSF}
F. Buccella, J.-P. Derendinger, C. Savoy and S. Ferrara,
{\it Phys. Lett.} {\bf B115} (1982) 375, and 
in Proc. Europhys. Study Conf. {\it Unification of the Fundamental 
Particle 
Interactions. II}, Eds. J. Ellis and S. Ferrara (Plenum Press, New York, 
1983),
page 349. 

\bibitem{ADS1}
I. Affleck, M. Dine and N. Seiberg,
{\it Nucl. Phys.} {\bf B241} (1984) 493.

\bibitem{ADS2}
I. Affleck, M. Dine and N. Seiberg,
{\it Nucl. Phys.} {\bf B256} (1985) 557.

\bibitem{Luty1}
M. Luty and W. Taylor, {\it Phys. Rev. } {\bf D53} (1996) 3399.

\bibitem{6}
 J. Bagger and J. Wess, {\it Supersymmetry and Supergravity},
2nd Edition (Princeton University Press, Princeton, 1992).

\bibitem{7}
E. Witten, {\it Nucl. Phys.} {\bf B403} (1993) 159.

\bibitem{SVZ1}
A. Vainshtein, V. Zakharov and M. Shifman, {\it Usp. Fiz. Nauk},
{\bf 146} (1985) 683 [{\it Sov. Phys. - Usp.} {\bf 28} (1985) 709].

\bibitem{Koga1}
I. Kogan, A. Morozov, M. Olshanetsky and M. Shifman,
{\it Yad. Fiz.} {\bf 43} (1986) 1587 [{\it Sov. J. Nucl. Phys.}
{\bf 43} (1986) 1022].

\bibitem{Seib1}
N. Seiberg, {\it Phys. Rev. } {\bf D49} (1994) 6857;
{\it Nucl. Phys.} {\bf B435} (1995) 129.

\bibitem{Seib2}
K. Intriligator and N. Seiberg, {\it Nucl. Phys.} {\bf B431} (1994) 551;
K. Intriligator, R. Leigh,  and N. Seiberg, {\it Phys. Rev. } {\bf D50} 
(1994) 
1092.

\bibitem{Seib3}
K. Intriligator, R. Leigh,  and M. Strassler, {\it Nucl. Phys.} {\bf B456} 
(1995)
567.

\bibitem{Popp1}
E. Poppitz and L. Randall, {\it Phys. Lett.} {\bf B336} (1994) 402;
J. Bagger, E. Poppitz and L. Randall, {\it Nucl. Phys.} {\bf B426} 
(1994) 3.

\bibitem{Gidd1}
S. Giddings and J.M. Pierre, {\it Phys. Rev. } {\bf D52} (1995) 6065.

\bibitem{Gher1}
T. Gherghetta, C. Kolda and S. Martin, {\it Nucl. Phys.} {\bf B468} 
(1996) 37.

\bibitem{GRS}
J. Wess and B. Zumino, {\it Phys. Lett.} {\bf B49} (1974) 52;\\
J. Iliopoulos and B. Zumino, {\it Nucl. Phys.} {\bf B76} (1974) 310;\\
P. West, {\it Nucl. Phys.} {\bf B106} (1976) 219;\\
M. Grisaru, M. Ro\v{c}ek, and W. Siegel,
{\it Nucl. Phys. } {\bf B159} (1979) 429.

\bibitem{SU5}
Y. Meurice and G. Veneziano, {\it Phys. Lett.} {\bf B141} (1984) 69;\\
I. Affleck, M. Dine and N. Seiberg,  {\it Phys. Lett.} {\bf B137} (1984) 
187.

\bibitem{Amati}
D. Amati, K. Konishi, Y. Meurice, G. Rossi and G. Veneziano,
{\it Phys. Rep.} {\bf 162} (1988) 557.

\bibitem{SU52}
I. Affleck, M. Dine and N. Seiberg,  {\it Phys. Rev. Lett.} {\bf 52} 
(1984) 1677;\\
Y. Meurice and G. Veneziano, {\it Phys. Lett.} {\bf B141} (1984) 69.

\bibitem{FEZU}
S. Ferrara and B. Zumino, {\it Nucl. Phys. } { \bf  B87} (1975) 207.

\bibitem{SV2}
M. Shifman and A. Vainshtein,
{\it Nucl. Phys.} {\bf B277} (1986) 456.

\bibitem{KSV}
I. Kogan, M. Shifman, and A. Vainshtein, {\it Phys. Rev.} {\bf D53}
(1996) 4526.

\bibitem{WZM}
J. Wess and B. Zumino,
{\it Phys. Lett.} {\bf B49} (1974) 52.

\bibitem{SV3}
M. Shifman and A. Vainshtein,
{\it Nucl. Phys.} {\bf B359} (1991) 571.

\bibitem{JJW}
I. Jack, D.R.T. Jones, and P. West, {\it Phys. Lett.}
{\bf B258} (1991) 382. 

\bibitem{Lisa}
E. Poppitz and L. Randall, {\it Phys. Lett.}
{\bf B389} (1996) 280.

\bibitem{Konishi1}
T.E. Clark, O. Piguet, and K. Sibold, {\it Nucl. Phys.} {\bf B 159} (1979) 
1;\\
K. Konishi, {\it Phys. Lett.} {\bf B135} (1984) 439;\\
K. Konishi and K. Shizuya, {\it Nuov. Cim.} {\bf A90} (1985) 111.

\bibitem{NSVZ3}
V. Novikov, M. Shifman, A. Vainshtein and V. Zakharov, {\it Nucl. 
Phys.} {\bf 
B260} (1985) 157; for a review see \cite{SVZ1}.

\bibitem{VZS}
A. Vainshtein, V. Zakharov and M. Shifman,
{\it Yad. Fiz.} {\bf 43} (1986) 1596 [{\it Sov. J. Nucl. Phys.}
{\bf 43} (1986) 1028], Sect. 3; see also {\it Pis'ma ZhETF}
{\bf 42} (1985)  182 [{\it JETP Lett.} {\bf 42} (1985) 224];
for a recent discussion see M. Shifman,
{\it Int. J. Mod. Phys.} {\bf  A11} (1996) 5761.

\bibitem{JJN}
I. Jack, D.R.T. Jones and C.G. North,
{\it Nucl. Phys.} {\bf B486} (1997) 479.

\bibitem{Nathan}
N. Seiberg, {\it Phys. Lett.} {\bf 206B} (1988) 75.

\bibitem{seibhol}
N. Seiberg, {\it Phys. Lett.} {\bf B318} (1993) 469.

\bibitem{ADS3}
I. Affleck, M. Dine and N. Seiberg,
{\it  Phys. Rev. Lett.} {\bf 51} (1983) 1026.

\bibitem{ADS4}
I. Affleck, M. Dine and N. Seiberg,
{\it  Phys. Lett.} {\bf B137} (1984) 187;
{\it  Phys. Rev. Lett.} {\bf 52} (1984) 493; {\it  Phys. Lett.} {\bf 
B140} (1984) 
59.

\bibitem{DKL}
L. Dixon, V. Kaplunovsky and J. Louis,
{\it Nucl. Phys.} {\bf  B355} (1991) 649.

\bibitem{LM}
H. Li and K. Mahanthappa, {\it Phys. Lett.} {\bf B319} (1993) 152;
{\it Phys. Rev.} {\bf D49} (1994) 5532.

\bibitem{instvol}
M. Shifman (Ed.), {\it Instantons in Gauge Theories}
(World Scientific, Singapore, 1994), Chapter VII.

\bibitem{Zum}
B. Zumino, {\it  Phys. Lett.} {\bf B69} (1977) 369.

\bibitem{DADI}
A. D'Adda and P. Di Vecchia, {\it  Phys. Lett.} {\bf B73} (1978) 162.

\bibitem{ROVE}
G.C. Rossi and G. Veneziano, {\it Phys. Lett.} {\bf B138} (1984) 195.

\bibitem{Khoze}
N. Dorey, V. Khoze, and M. Mattis, {\it Phys. Rev.}
{\bf D54} (1996) 2921; {\it Phys. Rev.} {\bf D54} (1996) 7832;  {\it 
Phys. Lett.} {\bf B388} (1996) 324; {\it Phys. Lett.} {\bf B390} (1997) 
205;\\
 H. Aoyama, T. Harano, M. Sato, and S. Wada, {\it Phys. Lett.} {\bf 
B388} (1996) 331;\\ 
K. Ito and  N. Sasakura, {\it Nucl. Phys.} {\bf B484} (1997) 141;
{\it  Mod. Phys. Lett.} {\bf A12} (1997) 205;\\
F. Fucito and T. Travaglini, {\it Phys. Rev.} {\bf D55} (1997) 1099;\\
T. Harano and M. Sato, {\it Nucl. Phys.} {\bf B484} (1997) 167;\\
 Y. Yoshida, hep-th/9610211;\\
M. Slater,  {\it Phys. Lett.} {\bf B403} (1997) 57.
 
\bibitem{Yung}
 A. Yung,   {\it Nucl. Phys.} {\bf B485} (1997) 38.  

\bibitem{MAC}
S. Dimopoulos, {\it Nucl. Phys.} {\bf B168} (1980) 69;\\
M. Peskin,  {\it Nucl. Phys.} {\bf B175} (1980) 197. 

\bibitem{antiMAC}
M. Vysotsky, I. Kogan, and M. Shifman,  {\it Yad. Fiz.}
{\bf 42} (1985) 504 [{\it Sov. J. Nucl. Phys.} {\bf 42} (1985) 318]. 

\bibitem{DDS}
A.C. Davis, M. Dine and N. Seiberg, {\it Phys. Lett.} {\bf B125} (1983) 
487.

\bibitem{thooft1}
 G. 't Hooft,  in  {\it Recent Developments in Gauge Theories},
 Eds. G. 't Hooft {\em et al.},  (Plenum Press, New York, 1980).

\bibitem{dolgov} 
 A. Dolgov and V. Zakharov, {\it Nucl. Phys.}
 {\bf B12}  (1971) 68.

\bibitem{ShifPR}
M. Shifman, {\it Phys. Rep.} {\bf 209} (1991) 161.

\bibitem{VW}
C. Vafa and E. Witten, {\it Nucl. Phys.} {\bf B234} (1984) 173.

\bibitem{ColW}
S. Coleman and E. Witten, {\it Phys. Rev. Lett.} {\bf 45} (1980) 100.

\bibitem{s1}
E. Poppitz and S. Trivedi, {\it Phys. Lett.} {\bf B365} (1996) 125; 
P. Pouliot, {\it Phys. Lett.} {\bf B367} (1996) 151.

\bibitem{s2}
K. Intriligator and P. Pouliot, {\it Phys. Lett.} {\bf B353} (1996) 471.

\bibitem{s3}
P. Cho and P. Kraus, {\it Phys. Rev.}
{\bf D54} (1996) 7640.

\bibitem{s4}
C. Cs\'{a}ki, W. Skiba and M. Schmaltz, {\it Nucl. Phys. } {\bf B487} 
(1997) 128.

\bibitem{s5}
K. Intriligator and N. Seiberg, {\it Nucl. Phys. } {\bf B444} (1995) 125.

\bibitem{s6}
P. Pouliot, {\it Phys. Lett.} {\bf B359} (1995) 108;
P. Pouliot and  M. Strassler, {\it Phys. Lett.} {\bf B370} (1996) 76;
{\it Phys. Lett.} {\bf B375} (1996) {175}.

\bibitem{s7}
I. Pesando, {\it Mod. Phys. Lett.} {\bf A10} (1995) {1871};
S. Giddings and J. Pierre {\it Phys. Rev.} {\bf D52} (1995) {6065}.

\bibitem{s8}
C. Cs\'{a}ki,   M. Schmaltz, and W. Skiba,
{\it  Phys. Rev. Lett.} {\bf  78} (1997) 799; 
{\it Phys. Rev.} {\bf D55} (1997) 7840.

\bibitem{PRD}
For a concise review and relevant references see e.g. I. Hinchliffe, 
{\it Phys. Rev.} {\bf D54} (1996) 77. 

\bibitem{MB}
A. Belavin and A. Migdal, {\it Pis'ma ZhETF} {\bf 19} (1974) 317 
[{\it JETP Lett.}
{\bf 19} (1974) 181]; {\em  Scale Invariance and Bootstrap in the 
Non-Abelian Gauge
Theories}, Landau Institute  Preprint-74-0894,  1974 (unpublished).

\bibitem{BZ}
T. Banks and  Zaks, {\it Nucl. Phys.} {\bf B196} (1982) 189.

\bibitem{Sei-conf}
N. Seiberg, {\it Nucl. Phys.} {\bf B435} (1995) 129.

\bibitem{Kut}
 D. Kutasov, A. Schwimmer, and  N. Seiberg, {\it Nucl. Phys.} {\bf 
B459} (1996) 455. 

\bibitem{tHooftPRD}
G. 't Hooft, {\it Phys. Rev.} {\bf D14} (1976) 3432; (E) 
{\bf D18} (1978) 2199.

\bibitem{DU1}
P. Pouliot, {\it Phys. Lett.} {\bf B359} (1995) 108;\\
D. Kutasov, {\it Phys. Lett.} {\bf B351} (1995) 230;\\
D. Kutasov and A. Schwimmer, {\it Phys. Lett.} {\bf B354} (1995) 
315;\\
P. Pouliot and M. Strassler, {\it Phys. Lett.} {\bf B370} (1996) 76;
{\bf B375} (1996) 175;\\
K. Intriligator, {\it Nucl. Phys.} {\bf B448} (1995) 187;\\
K. Intriligator, R. Leigh and M. Strassler, {\it Nucl. Phys.} {\bf B456} 
(1995) 567.

\bibitem{DU2}
J. Distler and A. Karch, hep-th/9611088.

\bibitem{DU3}
J. Brodie and M. Strassler, hep-th/9611197;\\
P. Cho,  {\it Phys. Rev.} {\bf D56} (1997) 5260;\\
and references therein. 

\bibitem{KINS}
K. Intriligator and  N. Seiberg, 
{\it Nucl. Phys.} {\bf B444} (1995) 125.

\bibitem{KINS1}
K. Intriligator and  N. Seiberg, in Proc. Conf.
  {\it Future Perspectives in String Theory (Strings '95)}
     Eds. I. Bars, P. Bouwknegt, J. Minahan, D.
   Nemeschansky, K. Pilch, H. Saleur, and. N. Warner (World Scientific,
Singapore, 
   1996) [hep-th/9506084]. 

\bibitem{NMSK}
N. Maru and S. Kitakado, {\it Mod. Phys. Lett.} {\bf A12} (1997) 691.

\bibitem{SPSUSY}
K. Intriligator and P. Pouliot,
{\it Phys. Lett.} {\bf B353} (1995) 471;\\
K. Intriligator, {\it Nucl. Phys.} {\bf B448} (1995) 187;\\
R. Leigh and M. Strassler, {\it Phys. Lett.} {\bf B356}
(1995) 492.

\bibitem{ISS}
K. Intriligator, N. Seiberg, and S. Shenker,
{\it Phys. Lett.} {\bf B342}
(1995) 152.

\bibitem{SWFOL}
A. Klemm, W. Lerche, S. Theisen, and S. Yankielowicz,
{\it  Phys. Lett.} {\bf B344} (1995) 169;\\
P. Argyres and  A. Faraggi, {\it  Phys. Rev. Lett.}
{\bf 74} (1995) 3931. 

\bibitem{ins}
K. Intriligator and N. Seiberg, {\it Nucl. Phys. Proc. Suppl.}  {\bf 45BC} 
(1996) 1
[hep-th/9509066].

\bibitem{Elitzur}
S. Elitzur, A. Forge, A. Giveon and E. Rabinovici,
{\it Nucl. Phys. Proc. Suppl.} {\bf 49} (1996) 174.

\bibitem{Kapu}
A. Kapustin, {\it Phys. Lett.} {\bf B398} (1997) 104.

\bibitem{Iwasaki}
Y. Iwasaki {\em et al.}, {\it Z. Phys.} {\bf C71} (1996) 343.

\bibitem{Casher}
A. Casher, {\it  Phys. Lett.} {\bf B83} (1979) 395.

\bibitem{V1}
V. Gribov, 1976, unpublished;\\
R. Jackiw and C. Rebbi, {\it Phys. Rev. Lett.} {\bf 37} (1976) 172;\\
C. Callan, R. Dashen, and D. Gross, {\it Phys. Lett.} {\bf B63} (1976) 
172.

\bibitem{V2}
E. Witten, {\it Nucl. Phys. } {\bf B156} (1979) 269;\\
G. Veneziano, {\it Nucl. Phys. } {\bf B159} (1979) 213.

\bibitem{CRE}
R. Crewther, {\it Phys. Lett.} {\bf B70} (1977) 349;
{\it Phys. Lett.} {\bf B93} (1980) 75. For a review see
 G.A. Christos, {\it  Phys. Rept.} {\bf 116} (1984) 251. 

\bibitem{SMI}
A. Smilga,  {\it Phys. Rev.} {\bf D54} (1996) 7757.

\bibitem{toron}
G. 't Hooft, {\it Comm. Math. Phys. } {\bf 81} (1981) 267.

\bibitem{DVSH}
G. Dvali and M. Shifman, hep-th/9612128.

\bibitem{BANK}
T. Banks and  A. Casher, {\it  Nucl. Phys.} {\bf B169} (1980) 103.

\bibitem{DCC}
A. Anselm, {\it Phys. Lett.} {\bf  B217} (1989) 169;
{\it Phys. Lett.} {\bf  B266} (1991) 482;\\
J.D. Bjorken, {\it Int. J. Mod. Phys.} {\bf A7} (1992) 4189;\\
 K. Rajagopal and  F. Wilczek, {\it  Nucl. Phys.} {\bf B399}
(1993) 395;\\
 J.-P. Blaizot and  A. Krzywicki, {\it  Phys. Rev.} {\bf D46} (1992) 
246.
 
\bibitem{Evans}
N. Evans, S. Hsu and M. Schwetz,
 {\it Phys. Lett.} {\bf B404} (1997) 77; see also earlier works
of these authors cited there. 

\bibitem{EvansKonishi}
K. Konishi, {\it Phys. Lett.} {\bf B392} (1997) 101;\\
N. Evans, S. Hsu and M. Schwetz, {\it  Nucl. Phys.} {\bf B484}
(1997) 124.

\bibitem{Peskin}
 O. Aharony, J. Sonnenschein, M. Peskin, and S. Yankielowicz,
{\it Phys. Rev. } {\bf D52} (1995) 6157. 

\bibitem{Hoker}
E. D'Hoker, Y. Mimura, and N. Sakai, {\it Phys. Rev.} {\bf D54} (1996) 
7724.

\bibitem{Hoker2}
E. D'Hoker, Y. Mimura, and N. Sakai,  hep-ph/9611458. 

\bibitem{Alvarez}
L. Alvarez-Gaum\'{e}, J. Distler, C. Kounnas, and M. Marino, {\it Int.  J. 
Mod. Phys. } {\bf  A11} (1996) 4745;\\
L. Alvarez-Gaum\'{e} and  M. Marino,  {\it Int. J. Mod. Phys.} {\bf 
A12} 
(1997) 975;\\
L. Alvarez-Gaum\'{e}, M. Marino, and  F.  Zamora,  hep-th/9703072. 

\bibitem{LL}
V.B. Berestetskii, E.M. Lifshits and L.P. Pitaevskii,
 {\it Quantum Electrodynamics},
(Pergamon Press, New York, 1982). 

\end{thebibliography}
\end{document}